\documentclass[twocolumn]{aastex7}

\usepackage{amsmath}
\usepackage{afterpage}
\usepackage[utf8]{inputenc}
\usepackage{colortbl}
\usepackage{graphicx}
\usepackage{subfig}
\usepackage{wrapfig}

\usepackage{hyperref}
\usepackage{amssymb}
\usepackage{pifont}

\usepackage[skip=0.333\baselineskip]{caption}
\usepackage{subcaption}

\received{\today}
\usepackage{graphicx}
\usepackage{subfig}
\graphicspath{{./}{Chapter2}}

\usepackage{amsmath}

\shortauthors{Burhenne et al.}
\shorttitle{Spatial and Temporal Starburst in the SMC and LMC}

\begin{document}

\title{Scylla V: Constraints on the spatial and temporal distribution of bursts and the interaction history of the Magellanic Clouds from their resolved stellar populations}

\author[0009-0005-0339-015X]{Clare Burhenne}
\email{cdb201@physics.rutgers.edu}
\affiliation{Department of Physics and Astronomy, Rutgers the State University of New Jersey, 136 Frelinghuysen Rd., Piscataway, NJ, 08854, USA}

\author[0000-0001-5538-2614]{Kristen B. W. McQuinn}
\email{kmcquinn@stsci.edu}
\affiliation{Department of Physics and Astronomy, Rutgers the State University of New Jersey, 136 Frelinghuysen Rd., Piscataway, NJ, 08854, USA}
\affiliation{Space Telescope Science Institute, 3700 San Martin Drive, 
Baltimore, MD 21218, USA}

\author[0000-0002-2970-7435]{Roger E. Cohen}
\email{rc1273@physics.rutgers.edu}
\affiliation{Department of Physics and Astronomy, Rutgers the State University of New Jersey, 136 Frelinghuysen Rd., Piscataway, NJ, 08854, USA}

\author[0000-0002-7743-8129]{Claire E. Murray}
\email{clairemurray56@gmail.com}
\affiliation{Space Telescope Science Institute, 
3700 San Martin Drive, 
Baltimore, MD 21218, USA}

\author[0000-0002-9820-1219]{Ekta Patel}
\email{ekta.patel@utah.edu}
\affiliation{University of Utah Department of Physics \& Astronomy, The University of Utah, 115 S 1400 E, Salt Lake City, UT 84112}
\affiliation{Department of Astrophysics and Planetary Sciences, Villanova University,  800 E. Lancaster Ave, Villanova, PA 19085, USA}

\author[0000-0002-7502-0597]{Benjamin F. Williams}
\email{benw1@uw.edu}
\affiliation{Department of Astronomy, University of Washington, Box 351580, U. W., Seattle, WA 98195-1580, USA}

\affiliation{The William H. Miller III Department of Physics \& Astronomy, Bloomberg Center for Physics and Astronomy, Johns Hopkins University, 3400 N. Charles Street, Baltimore, MD 21218, USA}

\author[0000-0003-0588-7360]{Christina W. Lindberg}
\email{christina.lindberg@live.com}
\affiliation{The William H. Miller III Department of Physics \& Astronomy, Bloomberg Center for Physics and Astronomy, Johns Hopkins University, 3400 N. Charles Street, Baltimore, MD 21218, USA}
\affiliation{Space Telescope Science Institute, 
3700 San Martin Drive, 
Baltimore, MD 21218, USA}

\author[0000-0002-9912-6046]{Petia Yanchulova Merica-Jones}
\email{petiay@gmail.com}
\affiliation{Space Telescope Science Institute, 
3700 San Martin Drive, 
Baltimore, MD 21218, USA}
\affiliation{University of Sofia, Faculty of Physics, 5 James Bourchier Blvd., 1164 Sofia, Bulgaria}
\affiliation{Institute of Astronomy and NAO, Bulgarian Academy of Sciences, 72 Tsarigradsko Chaussee Blvd., 1784 Sofia, Bulgaria}

\author[0000-0001-5340-6774]{Karl D.\ Gordon}
\email{kgordon@stsci.edu}
\affiliation{Space Telescope Science Institute, 3700 San Martin Drive, Baltimore, MD 21218, USA}

\author[0000-0003-1680-1884]{Yumi Choi}
\email{yumi.choi@noirlab.edu}
\affiliation{NSF National Optical-Infrared Astronomy Research Laboratory, 950 N. Cherry Avenue, Tucson, AZ 85719 USA}

\author[0000-0001-8416-4093]{Andrew E. Dolphin}
\email{adolphin@raytheon.com}
\affil{Raytheon Company, 1151 E. Hermans Road, Tucson, AZ 85756, USA}

\author[0000-0001-6326-7069]{Julia C. Roman-Duval}
\email{duval@stsci.edu}
\affil{Space Telescope Science Institute, 3700 San Martin Drive, Baltimore, MD 21218, USA}

\begin{abstract}
We measure the star formation histories (SFHs) from the Scylla survey in $\sim 98,000$~pc$^2$ and $\sim 75,000$~pc$^2$ of the SMC and LMC, respectively, using deep Hubble Space Telescope imaging (80$\%$ complete to $>1$~mag below the ancient main-sequence turnoff, $\sim 25.1$ and $26.0$ mag in F475W and F814W) from 74 pointings. We group the fields into eight sub-regions in the SMC and seven in the LMC. We use the birth rate parameter to identify bursts of star formation and measure their properties in each sub-region. Our methodology provides a standardized framework for burst identification and reveals both broad and fine burst characteristics. We identify global and local bursts, defined as those occurring in $\ge$ half or $<$ half of a galaxy's sub-regions, respectively. In the SMC we find two global ($\sim$ 5 and 1.5 Gyr ago) and one local burst ($\sim$ 3 Gyr ago). In the LMC we find one global burst ($\sim$ 3 Gyr ago). Comparing these findings with dynamical models of the LMC and SMC orbital histories, we find that when models predict a shared dynamical trigger for bursts across both galaxies, the burst begins earlier in the SMC with a greater enhancement in star formation rate than in the LMC. Finally, using age-metallicity relations (AMRs) and cumulative SFHs, we report that the Wing/Bridge region in the SMC resembles the southwestern LMC both chemically and in stellar mass assembly over the last $\sim$ 7 Gyr, possibly due to stellar material stripped from the LMC during their last interaction.

\end{abstract}

\section{Introduction}

The Small and Large Magellanic Clouds (S/LMC) are the two closest gas-rich dwarf galaxies to the Milky Way. Their proximity, 62 and 50 kpc, respectively \citep{degrijs2014, degrijs2015}, enables a detailed study of their properties and evolution. Crucially, the Clouds are interacting, which allows us to observe an ongoing interaction between two dwarf galaxies at close range.

Close pairs of interacting dwarfs are relatively rare, despite dwarf galaxies being the most common type of galaxy at all redshifts \citep{Karachentsev2013, binggeli1988}. Observations and simulations suggest that only $\sim 4\%$ of dwarf galaxies within redshifts $0.013 < z < 0.0252$ host a comparable low-mass companion, and $<1\%$ of LMC-analogs in the field have an SMC-analog companion \citep{Besla2018}. This rarity makes the Magellanic Clouds a uniquely valuable laboratory for studying in detail how interactions between dwarf galaxies can drive or regulate evolutionary processes in low-mass systems.

The most widely agreed-upon picture of the interaction history between the Clouds begins $\sim7$ gigayears (Gyr) ago, when they are thought to have become a binary pair (\cite{besla2012, diaz2012, pardy2018, Lucchini2021}, also see \cite{vasiliev2023, Sheng2024}). Since then, models predict that during the last 6 Gyr, there have been at least two pericentric passages between the SMC–LMC system \citep{patel2020}. Observational studies corroborate these interactions, suggesting that the most recent pericentric passage likely occurred within the past 200 Myr with an impact parameter $<$10 kpc, resulting in a direct collision between the two galaxies \citep{choi2022, zivick2018}.

In addition to interacting with each other, the Clouds are also interacting with the Milky Way. The Clouds likely fell into the Milky Way’s virial radius only 1–2 Gyr ago \citep{besla2007, besla2012, patel2020}, although some models suggest they are on their second infall \citep{vasiliev2023}. Other large-scale features of the Magellanic System, such as the Magellanic Stream and Bridge, are likely consequences of these interactions, either due to tidal stripping between the SMC and LMC or ram pressure stripping as the LMC entered the Milky Way’s halo \citep{nidever2008, besla2012, lucchini202}.

We used data from the Scylla program to study the SFH of the Magellanic Clouds as a function of position. Scylla is a multicycle pure-parallel HST campaign that imaged fields in the LMC and SMC in parallel to observations from the Ultraviolet Legacy Library of Young Stars as Essential Standards \citep[ULLYSES,][]{romanduval2025}. Previous space-based imaging of the LMC and SMC mainly focused on prominent star-forming regions \citep{sabbi2016}, deep pointings in the halos \citep{weisz2013, cignoni2013, dolphin2001}, or regions observed in parallel with targeted campaigns \citep[METAL,][]{roman2019}.  In contrast, the Scylla dataset includes 48 fields in the LMC, sampling the innermost $\sim 4.5$ kpc, and 48 fields in the SMC, sampling the innermost $\sim4$ kpc (in this work, we analyze 74 of these pointings). The spatial distribution of the Scylla fields allows us to sample a wide range of properties in the Clouds including a range in metallicities, gas column densities, radiation fields, and SFHs.

This paper, focusing on characterizing bursts of star formation across spatially distributed regions in the inner LMC and SMC, joins previous works from the Scylla project, including the overview survey paper \citep{Murray2024}, and three subsequent papers exploring the radial age gradient of the LMC \citep{Cohen2024a} and the SMC \citep{Cohen2024b}, and the intrinsic stellar and line of sight dust properties of resolved stars in the L/SMC \citep{Lindberg2024}.

\subsection{Interaction-driven bursts of star formation}

Interactions between galaxies are thought to play a significant role in triggering bursts of star formation, particularly in gas-rich systems. In the Magellanic Clouds, such interactions have heavily influenced their stellar mass assembly, as revealed by decades of star formation history (SFH) studies \citep[e.g.,][]{harris2004, Bekki_2007, harris2009, cioni2011, ruizlara2020, massana2022}. These bursts are often driven by processes like gas compression, tidal interactions, the redistribution of neutral hydrogen, and the funneling of gas into star-forming regions. In turn, the bursts drive stellar feedback, enrich the interstellar medium, and launch galactic winds, all of which are critical components of galaxy evolution \citep{Veilleux2005, raposo2016, McQuinn2019}.

There have been numerous investigations on how interactions between the Clouds may have impacted their SFHs. Ground-based programs such as the VISTA survey of the Magellanic Clouds system \citep[VMC;][]{cioni2011} and the more recent Survey of the MAgellanic Stellar History \citep[SMASH;][]{nidever2017}, offer wide spatial coverage and deep photometry across extended regions, making them ideal for mapping large-scale SFH trends. In contrast, archival Hubble Space Telescope (HST) imaging from studies like \citet{cignoni2013} and \citet{weisz2013} provides exquisite spatial resolution, allowing for the separation of individual stars even in densely populated regions, and reaching faint magnitudes below the oldest main-sequence turn-offs. Although HST’s narrow pencil-beam pointings limit continuous areal coverage, these observations are an excellent complement to ground-based surveys. 

Each of these studies measures SFHs that reveal broadly similar trends in star formation rate (SFR) over time. They seem to agree that there was an enhancement in the star formation of the SMC at least 8 Gyr ago, an enhancement in both galaxies beginning around 3 Gyr ago (which is especially prominent in the LMC), that there has been a more recent, distinct star formation event in the SMC around 1-1.5 Gyr ago, and that there is ongoing star formation in both galaxies. The widespread agreement in recent star formation is backed by theory. Modeling from \citet{diaz2012} suggests that the Magellanic Bridge could have formed during an interaction between the Clouds around 2 Gyr ago while modeling from \citet{Bekki_2007} suggests the Bridge may have been formed during the most recent interaction around 200 Myr ago. This most recent interaction may also account for the unusual morphology of the SMC and its Wing/Bridge region, which hosts a bimodal stellar distribution along the line of sight, the foreground component of which lies $\sim10$ kpc in front of the SMC \citep{nidever2013, subramanian2017, murray2023, Murray2024}.

Still, differences emerge in the detailed interpretation of these SFHs, particularly in identifying and dating enhancements in star formation activity. These differences are partly due to the non-uniform methods used to characterize enhancement features within the SFHs. In most studies, peaks in star formation are identified by local maxima in the SFR as a function of time, but the criteria used to define these peaks can vary, ranging from visual inspection to quantitative thresholds relative to baseline activity. While we do not attempt to reconcile these methods here, it is useful to recognize that differing definitions contribute to the differences in burst timing across studies, such as SMASH \cite{weisz2013} and others.

Moreover, differences in the underlying data, including the depth and coverage of photometry and the stellar evolution libraries used to measure the SFHs, along with other factors, can influence the recovered SFH and, consequently, the identification and timing of bursts. For example, a prominent star formation episode in the SMC has been dated to $\sim$9 Gyr \citep{weisz2013}, $\sim$8 Gyr \citep{massana2022}, and 8.3 Gyr \citep{rubele2018}. Similarly, enhancements reported by \citet{massana2022} occur synchronously across both galaxies at 0.45, 1.1, 2.0, and 3 Gyr ago show only partial overlap with features identified in \citet{rubele2018}, \citet{weisz2013}, and this work (see Section~\ref{sec:Bursts4}). Although some of these differences likely arise from the data themselves or the models used to measure the SFHs, the lack of a standardized quantitative framework for defining bursts further complicates direct comparisons between studies. 

To reconstruct the star formation rate as a function of time (SFR(t)) in a galaxy, one of the most robust techniques relies on fitting color–magnitude diagrams (CMDs) of resolved stellar populations. CMDs allow us to infer stellar ages and metallicities by comparing observed photometry to synthetic photometry built from stellar libraries of isochrones, enabling the reconstruction of detailed SFHs and subsequent calculation of the birthrate parameter. We also incorporate dynamical models from \citet{patel2020}, which allowed us to compare the timing of our measured bursts with key dynamical events in the SMC–LMC–Milky Way system.

In this work, we aim to standardize burst identification in the SFHs of the Magellanic Clouds by applying the birth rate parameter, $b = \mathrm{SFR} / \langle \mathrm{SFR} \rangle$, which compares the instantaneous SFR to some fiducial average \citep{Kennicutt1983, Scalo86}. However, the threshold value of $b$ used to define a burst varies. For example, the TiNy Titans (TNT) survey, a multiwavelength study of interacting dwarf galaxies, finds that interacting pairs typically exhibit SFRs $2$–$2.75 \times$ higher than the mean SFR of matched isolated counterparts \citep{Stierwalt2015}. A threshold of $b = 2$ is also used in other studies to identify elevated star formation \citep{Kennicutt2005, McQuinn2009, McQuinn2010}, while $b = 1$ is often used to define the beginning and end of an enhancement period. 

In Section \ref{sec:Data}, we present our observations and photometry. In Section \ref{sec:SFH3} we describe our methods for measuring the SFHs. In Section \ref{sec:Bursts4}, we present the results of our burst characterization. In Section \ref{sec:Trigger5.1} we introduce how we used the dynamical models from \citet{patel2020} to compare with our findings. In Section \ref{sec:Discussion5}, we discuss the implications of our dynamical analysis on bursts in the Clouds, and in Section \ref{sec:Conclusions} we summarize our findings. Finally, we include an appendix with the cumulative SFHs of each sub-region, and the SFRs versus time of all sub-regions with our burst metrics superimposed compared to the dynamical models from \citet{patel2020}.

\begin{figure*}
 	\centering
        \textbf{Scylla fields and regions}\par\medskip
\includegraphics[width=0.65\textwidth]{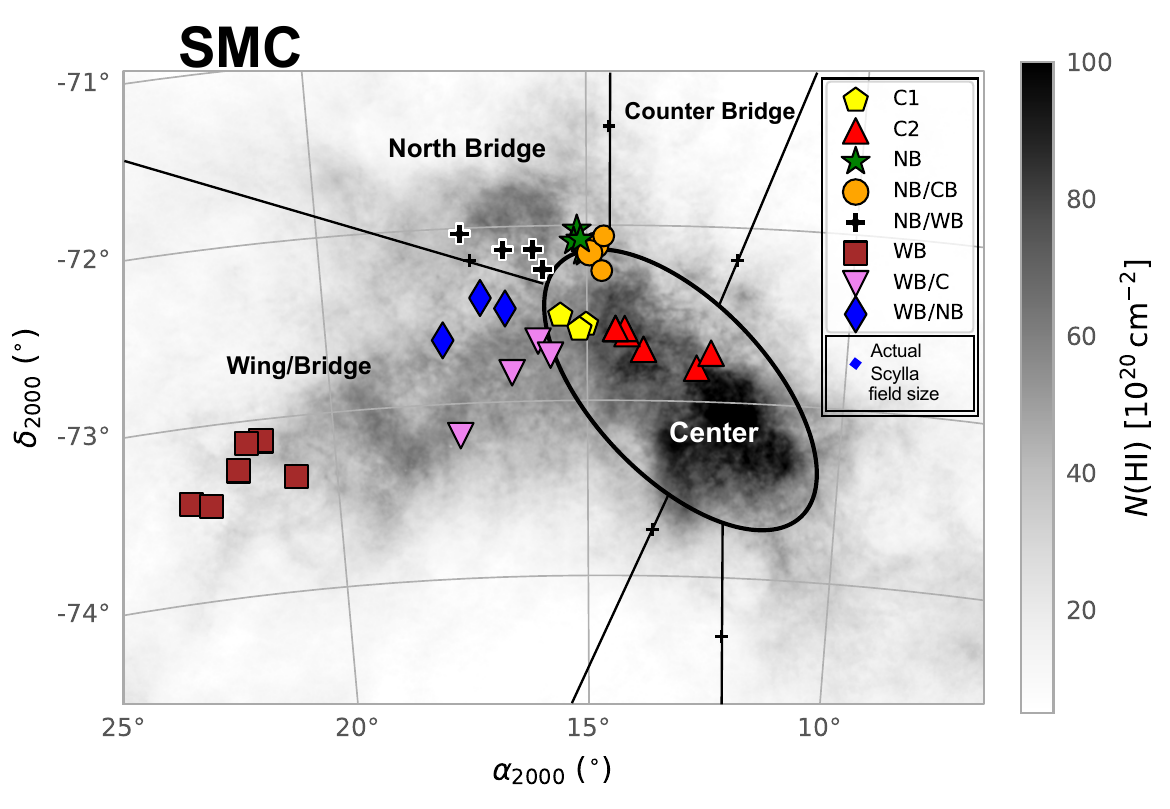}
\includegraphics[width=0.65\textwidth]{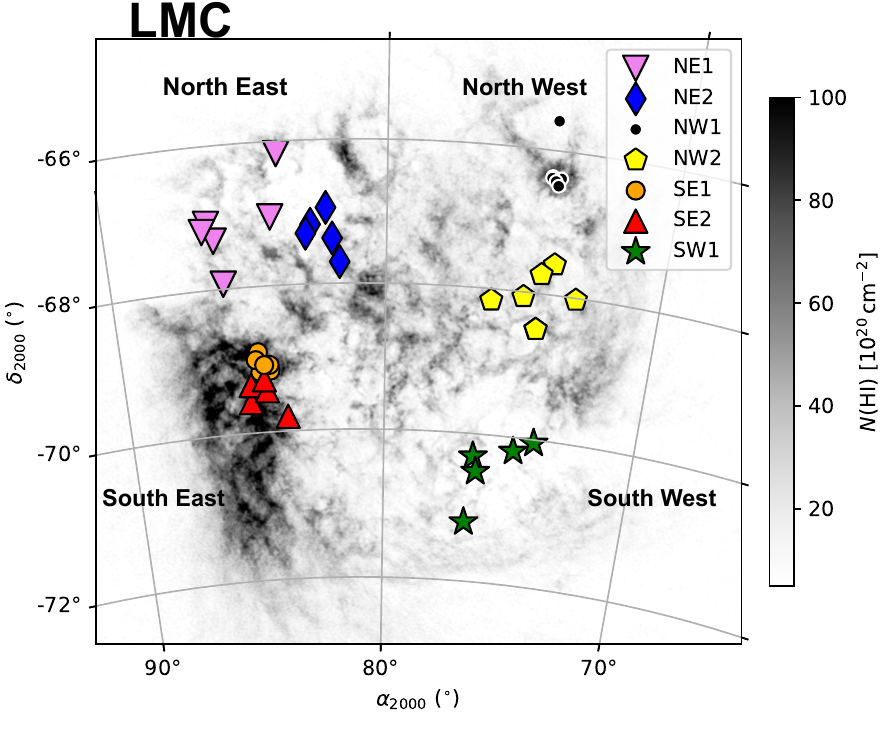}
	  \caption{The Scylla fields used in this analysis overplotted on an \ion{H}{1} map of the SMC (top) and the LMC (bottom) from the Galactic Australian Square Kilometre Array Pathfinder (GASKAP) survey \citep{Pingel_2022} for the SMC and \cite{kim2003} for the LMC. Each galaxy has been divided into four larger regions: for the SMC—North Bridge, Counter Bridge, Wing/Bridge, and Center; and for the LMC—North East, North West, South East, and South West. The regional divisions in the SMC are based on prior structural and kinematic groupings from the VISCACHA survey by \citet{dias2016}. The small black crosses in the SMC panel correspond to the approximate centers of substructures defined in \citet{Bica2020}. In the LMC, regional divisions are based on cardinal directions, using the dynamical center at (RA, Dec) = $(82.25, -69.5)$ as determined by \citet{vandermarel2001} as the central point. Within each large region, Scylla fields were grouped into smaller sub-regions based on spatial proximity. Fields belonging to the same sub-region share the same marker shape and color. There are eight sub-regions in the SMC and seven in the LMC, all listed in their respective legends. Details of the sub-region grouping strategy for each galaxy can be found in Sections \ref{sec:SMC3.2.1} and \ref{sec:LMC3.2.2}.}
  \label{fig:maps}
\end{figure*}

\section{Data} \label{sec:Data}
The data include imaging of 74 fields towards the LMC and SMC from HST in two optical filters, F475W and F814W. The observations were obtained via the Scylla pure-parallel imaging program (GO-15891, GO-16235, GO-16786; PI: C. E. Murray), in tandem with the spectroscopic HST survey ULLYSES.  Images were taken with the Wide Field Camera 3 (WFC3), and all fields used here have observations in F475W and F814W, which we utilize in the present work. A subset of fields in the full Scylla data set includes imaging in bands ranging from Ultraviolet (UV) to Infrared (IR), with up to seven WFC3 filters (F275W, F336W, F475W, F814W, F110W, F160W). Our catalog is limited to fields with at least two filters, as the sample of fields with greater than two filters is significantly smaller. For additional details on the observation strategy and the general objectives of Scylla, we refer the reader to the survey overview paper \citep{Murray2024}.

Figure \ref{fig:maps} presents the spatial coverage of 74 Scylla fields in the SMC and LMC, which cover $\sim$ $98,000$ pc$^2$ and $\sim$ $75,000$ pc$^2$, respectively. This coverage was achieved with 36 SMC and 38 LMC fields, which are overplotted on HI maps from the Galactic Australian Square Kilometer Array Pathfinder \citep[GASKAP,][]{Pingel_2022} for the SMC and \citet{kim2003} for the LMC. The Scylla fields in the LMC sample an area from right ascension (RA) = (70, 86) and declination (dec) = ($-$66, $-$70), while the fields in the SMC sample RA = (14, 21) and dec= ($-$72, $-$73.25). Fields in both galaxies provide excellent spatial sampling for the purposes of this study, which aims to identify bursts in spatially diverse regions across The Clouds. Because Scylla is a pure parallel survey, the fields are determined by the ULLYSSES pointings, which took spectroscopic data for O and B stars in the Magellanic Clouds using the Cosmic Origins Spectrograph (COS). Since O and B stars are preferentially located in areas with higher column densities of gas and dust, the Scylla dataset has fields that preferentially fall on or near star-forming regions and on areas with higher dust content.

% However, photometric coverage in the LMC is sparse in its central region and near the Southeast. In the SMC, coverage is skewed to the center and Southeast. Other Scylla collaboration papers improve spatial coverage by incorporating archival Hubble fields; see \citet{Cohen2024a}, \citet{Cohen2024b}. 

% or $0.04\%$ and $0.01\%$ of the total area, respectively

% For that reason, we checked for fields with severe internal differential extinction that could hamper SFH fits and excluded most of these fields from our analysis; the details of this process can be found in \citet{Cohen2024b}. Only one Scylla field used in this analysis, SMC$\_50$, met the cut criteria; however, this field was retained because its SFH solution was similar to the others used in this analysis. A more in-depth discussion of this can be found in Section \ref{sec:Fields3.2}.

In Figure \ref{fig:maps}, the fields are colored by their location on the sky and the similarity of their SFH solutions (among other criteria, which can be found in Sections \ref{sec:SMC3.2.1} and \ref{sec:LMC3.2.2} for SMC and LMC, respectively). The SFHs of these fields are combined in our final analysis, which increases stellar counts for robust SFHs with lower statistical uncertainties.

\begin{deluxetable*}{lcccc}
\tabletypesize{\scriptsize}
\tablecaption{Summary of $80\%$ Completeness Limits\label{table:completeness}}
\tablehead{ 
  \colhead{} & 
  \colhead{SMC F475W [mag]} & \colhead{SMC F814W [mag]} & 
  \colhead{LMC F475W [mag]} & \colhead{LMC F814W [mag]}
}
\startdata
Minimum & 26.01 & 24.06 & 25.03 & 23.25 \\
Maximum & 27.89 & 25.52 & 27.79 & 25.50 \\
Average & 26.93 & 24.78 & 27.07 & 24.88 \\ 
\enddata
\tablecomments{Minimum, maximum, and average $80\%$ completeness limits for the SMC and LMC in WFC3 filters F475W and F814W. A complete list of the $80\%$ completeness limits can be found in \citet{Cohen2024a} for the LMC and \citet{Cohen2024b} for the SMC.}
\end{deluxetable*}

\subsection{Photometry} \label{sec:Phot2.1}

The photometry reduction and overall data processing are described in detail in \citet{Cohen2024a}, \citet{Cohen2024b}. Here, we briefly summarize the steps. For each field, pre-processed and charge transfer efficiency (CTE) corrected images (flc.fits files) were combined to create a distortion-corrected, drizzled reference image using drizzlepac software \citep{avila2015}. Next, DOLPHOT \citep{dolphin2000, dolphin2016}, a stellar photometry package, masked bad pixels, applied a pixel area map, split the image into individual chips, and calculated the sky background. DOLPHOT has customized point spread functions (PSFs) for all filters on each HST imager, and applies the PSF corresponding to the camera and filter combination of each flc.fits file.  DOLPHOT uses a number of parameters to determine the PSF photometry procedure; we adopted values from \citet{williams2014, williams2021} determined by testing over a wide range of crowding conditions.

We filtered the DOLPHOT output using several quality metrics to create high-fidelity stellar catalogs. We start with a Scylla `.vgst' photometric catalog, in which all sources have at least one band with a signal-to-noise ratio (SNR) $>4$ and have had spurious, poorly measured, and non-stellar detections removed on a per filter basis, discussed further in \cite{Murray2024}.

We then imposed additional cuts to the .vgst file, where sources that are retained when they have an SNR $ \ge 5$, a $|$\texttt{sharp}$|$  $\ge 0.25$, a \texttt{crowd} parameter $\ge 0.25$, all in both filters, and flux flag = $1$ and photometric quality flag $= 0$ or $2$. Stars with a photometric quality flag of 1 are excluded because they contain too many bad and/or saturated pixels for reliable flux measurements. The \texttt{sharp} parameter measures how centrally concentrated the source flux is, with positive values being more concentrated sources (e.g., cosmic rays) and negative values being more extended sources (e.g., background galaxies). The \texttt{crowd} parameter measures how much brighter in magnitudes a source would have been if its neighbors were not subtracted when measaruing its flux. The flux flag identifies if a detection has been made; flux values for a detected source cannot be zero, but can be positive or negative. Further information regarding DOLPHOT parameters and their definitions can be found in the DOLPHOT manual\footnote{\url{http://americano.dolphinsim.com/dolphot/}}.

\subsection{ASTs} \label{sec:ASTs2.2}

We used artificial star tests (ASTs) to quantify photometric completeness, uncertainty, and bias, as well as facilitate the comparison of synthetic and observed CMDs. We injected a large number ($>10^5$) of fake stellar sources with known magnitudes into each of the images. We then performed photometry on the images to recover the artificial sources and applied the same quality cuts as those applied to the photometry.

The inputs for the ASTs were generated using the Bayesian Extinction and Stellar Tool \citep[BEAST;][]{gordon2016}, which models spectral energy distributions (SEDs) of stars in the UV to near-IR. The input star list was generated assuming a PARSEC evolutionary model \citep{Bressan2012}, a Kroupa \citep{Kroupa2001} initial mass function (IMF), and flat distributions over distance, age ($t$), metallicity ($Z$), extinction ($Av$) and average dust grain size ($Rv$), and the mixing parameter for dust extinction curves ($f_a$); the value ranges for these values may be found in \citet{Cohen2024a, Cohen2024b}. 

We quantified the completeness limits of the data using the ASTs. The completeness limit for each individual Scylla field in the SMC is reported in \cite{Cohen2024b} and in the LMC in \cite{Cohen2024a}. Importantly, these works found that Scylla photometry is $80\%$ complete down to $>1$ magnitude below the oldest main sequence turn-off (oMSTO). Our data are as deep or deeper than previous HST photometry used to measure the SFH of the Clouds \citep{weisz2013, cignoni2013}. Table \ref{table:completeness} lists the average $80\%$ completeness limit for the SMC and LMC fields used in this paper in the F475W and F814W filters.

\subsection{ Color Magnitude Diagrams } \label{sec:CMD2.3}

\begin{figure*}
 	\centering
        \textbf{Individual and combined Scylla CMDs with isochrones }\par\medskip
\includegraphics[width=0.76\textwidth]{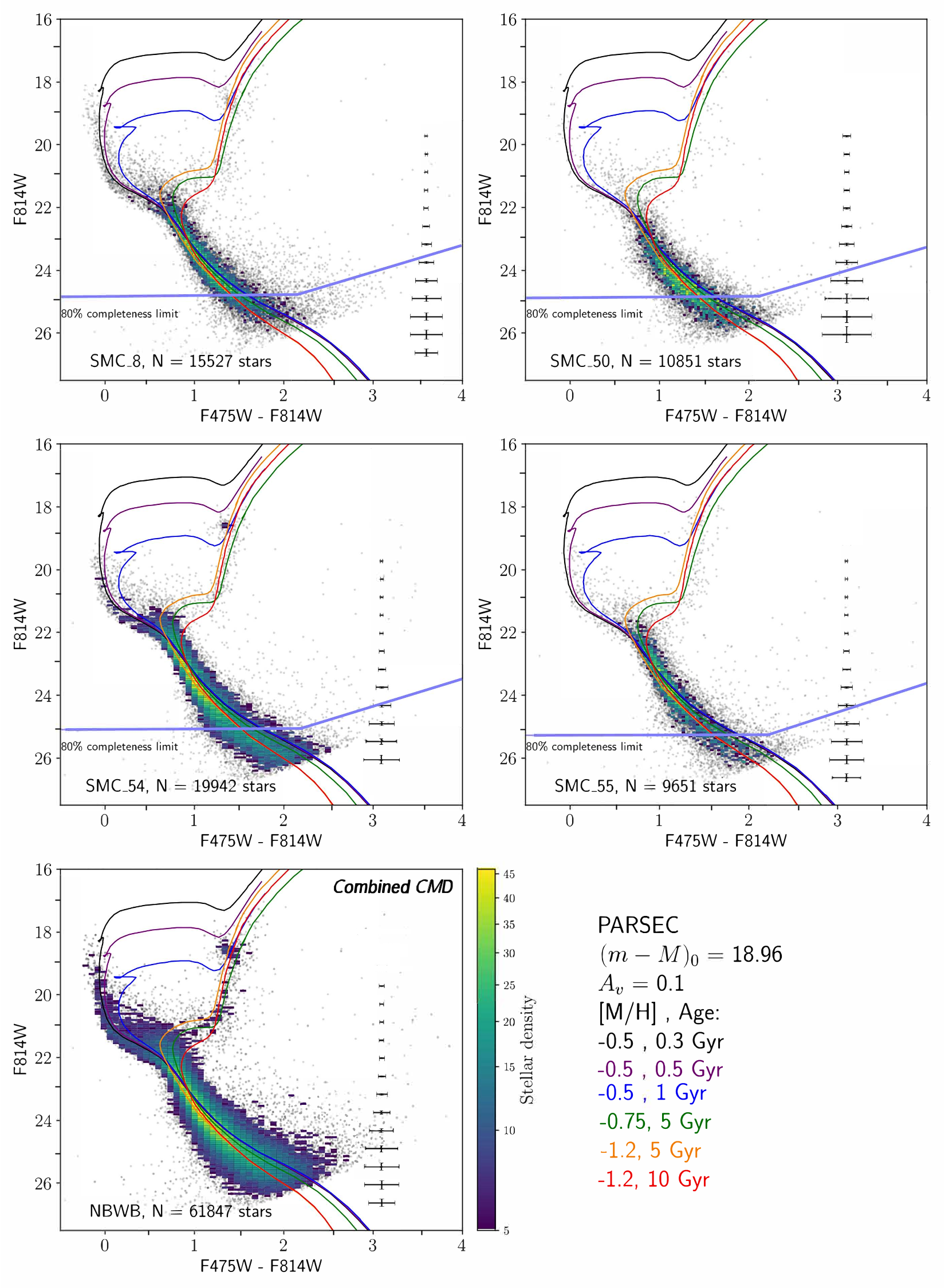}
	  \caption{Color-magnitude diagrams (CMDs) of the four individual Scylla fields (SMC 8, SMC 50, SMC 54, SMC 55) that comprise the sub-region North Bridge/Wing/Bridge (NBWB), along with a fifth CMD showing all four fields combined. While the CMDs are shown individually and combined for reference, the SFHs are derived from the individual fields and later combined to form the final SFH of the NBWB sub-region. This approach allows each field to be fit while improving statistical robustness in the SFH solutions by increasing the number of stars contributing to each age bin. This methodology is discussed in greater detail in Section \ref{sec:Bursts4}. Six isochrones are overplotted on all CMDs, with ages ranging from 0.3 to 10 Gyr and metallicities spanning $-1.2 < [M/H] < -0.5$. The $80\%$ completeness limits are indicated for all five CMDs with a solid blue line, demonstrating the exceptional depth of the Scylla data; the dataset remains $80\%$ complete to at least one magnitude below the oldest main sequence turnoff (oMSTO). Representative error bars are plotted on the right. Bright stars in the CMDs are saturated.}
      \label{fig:cmds}
\end{figure*}

The high-resolution, deep, HST imaging of stars in the Clouds produce well-populated ($\sim 10^4$ stars per field), high-fidelity CMDs. In Figure \ref{fig:cmds} we present the CMDs of four neighboring Scylla fields. Isochrones from the PARSEC stellar library \citep{Bressan2012} with a metallicity range of $ -1.2 \le [M/H] \le -0.5 $ and an age range of $10$ Gyr $\le$ Age $\le$  $0.3$ Gyr have been overplotted on the CMDs. This metallicity range is reasonably in line with the average value for the SMC \citep{Choudhury2018}. These individual CMDs illustrate a wide range of features found in the CMDs. For example,  the upper main sequence in SMC$\_{54}$ and SMC$\_8$ are more populated, indicating a higher degree of recent star formation in those fields. SMC$\_50$ and SMC$\_55$ both have fewer stars and more sparsely populated red clumps (RC) than SMC$\_8$ and SMC$\_{54}$. Across almost all CMDs, the brightest SMC sources are saturated. To determine if it would benefit the analysis to recover the magnitudes of saturated stars in our sample, we compared the SFHs of two fields with different exposures (SMC$\_36$ and SMC$\_37$, with 2 and 5 orbits, respectively). We found that the difference in exposure provided little gain in terms of either the number of dented stars or SFH precision. We chose not to recover magnitudes for saturated stars. Instead, we limited our detailed analysis to the ages of stars that are well-represented in the CMD and rely on the artificial star tests in the CMD-fitting technique to appropriately account for the saturated stars in the fields. The saturation is evident in the CMDs for the fields in all four panels and is a consequence of Scylla being a parallel program, as the Scylla exposures were constrained by the ULYSSES primaries.

In the bottom panel of Figure \ref{fig:cmds}, we show a CMD created by stacking the individual CMDs of the four neighboring fields (SMC 8, SMC 50, SMC 54, and SMC 55), which together define the North Bridge/Wing/Bridge (NBWB) sub-region (see Section \ref{sec:SMC3.2.1} for grouping rationale). While these combined CMDs are useful for visualization, the SFHs presented in this analysis were not measured from merged photometric catalogs. Instead, we measured the SFH of each field individually and statistically combined the resulting SFH solutions to generate a sub-region SFH. This approach increases the effective star counts contributing to each age bin, reducing statistical uncertainties, and helps to mitigate systematic effects from field-to-field variations. This method of SFH combination is applied consistently across our sample and is described in further detail in Section \ref{sec:Fields3.2}.

\section{Measuring Star Formation Histories} \label{sec:SFH3}

% The ages of older stars can be difficult to determine. This is partially a consequence of the age-metallicity degeneracy, which describes the interrelated nature of age and metallicity at the location of a star in a CMD \citep{Worthey1994}. One way to break this degeneracy is to obtain high-fidelity imaging that allows a CMD to reach below the old Main Sequence Turn off (oMSTO)  \citep{Gallart2005, Noel2009}. This is in part because below the oMSTO, isochrones have a stronger dependence along the color axis, which creates a greater spread between the isochrones, making it easier to differentiate between stars with the same ages and metallicities or visa versa. An example of this is given in Figure 2, which shows CMDs overlayed with isochrones from the PARSEC stellar library \citep{Bressan2012} with a metallicity range of $ -1.2 \ge [M/H] \ge -0.5 $ and an age range of $-10$ Gyr $\le$ Age $\le$  $0.3$ Gyr. 

We derive SFHs using the well-characterized CMD synthesis technique \citep{dolphin2002, aparicio2002}. Specifically, we use the software package MATCH, which fits an observed CMD with modeled CMDs and measures the best fit SFH using Poisson maximum likelihood statistics \citep{dolphin2002}. Each observed CMD is converted to a 2D density plot (Hess diagram) with a resolution in magnitude of 0.1 dex and a resolution in color of 0.05 dex. Synthetic photometry is then created using stellar evolution libraries and assuming an Initial Mass Function (IMF). MATCH combines the synthetic photometry with ASTs to simulate observational effects.

MATCH finds the best-fit SFR per time bin by comparing different combinations of synthetic stellar populations with the observed CMD. For SFH fitting, we assume the following values: a Kroupa IMF \citep{Kroupa2001}, a binary star fraction of 0.35, and a metallicity range of $ -2.0 \le [M/H]  \le - 0.2$ with a step size of 0.15 dex. \citet{Cohen2024a, Cohen2024b} detail how we assessed the impact of model choice on the SFH results by fitting each field three times, assuming a different stellar evolutionary model for each fit, namely: PARSEC stellar evolutionary library \citep{Bressan2012}, MIST \citep{choi2016}, and the updated BaSTI isochrones \citep{Hidalgo2018}. The SFH fits used for our analysis were those generated with the PARSEC evolutionary library. Previous testing has shown that the choice of model does not affect the SFH solution within our uncertainties, so we focused on a single stellar library to simplify our analysis. However, we do take into account systematic uncertainties in our SFH fits, which cover the range in solutions returned from different stellar libraries. See \citet{Cohen2024a, Cohen2024b}. We then designate a time-binning scheme to measure the SFH; details of this process can be found in Section \ref{sec:Timebin3.1}. 

\begin{figure*}
 	\centering
        \textbf{SFR testing on synthetic photometry}\par\medskip
\includegraphics[width=1.0\textwidth]{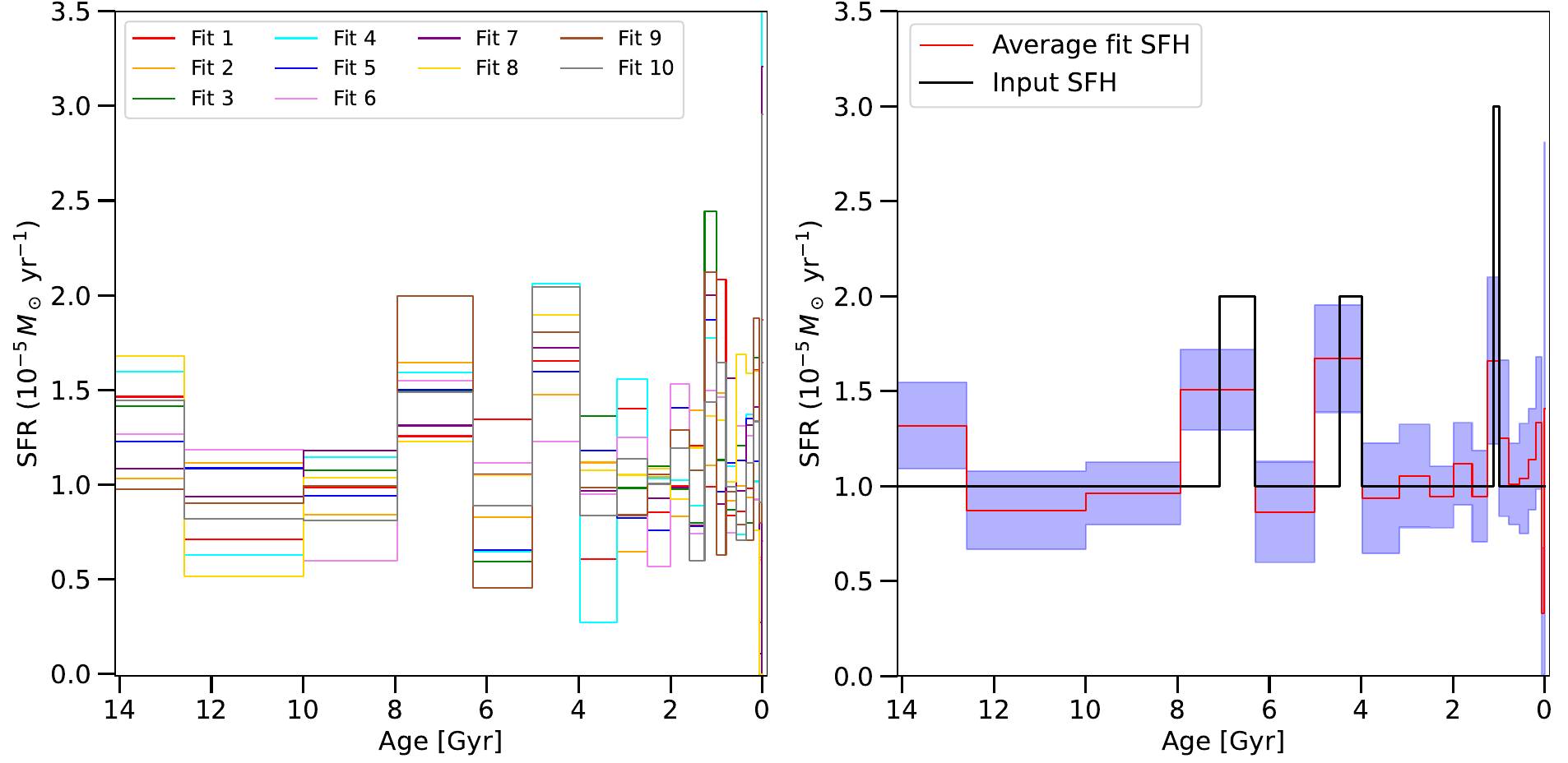}
	\caption{\textbf{Left}: Ten recovered SFRs as a function of time from synthetic photometry with simulated noise based on ASTs. \textbf{Right}: The average SFR as a function of time of the ten realizations (red) one standard deviation above and below (blue) and the input SFR as a function of time, in black. The input SFH has bursts at ages of: 1.0 – 1.1 Gyr, 3.9 – 4.5 Gyr, and  6.3 – 7.0 Gyr, which are shorter in duration than the width of the time bins used for the SFH recovery. We find that bursts are recovered that are smaller than the width of their bin at recent ($\sim 1$ Gyr) and old times ($\sim 7$ Gyr) with similar accuracy, and the peak of the bursts are lower as the star formation is averaged over a longer timescale.} 
        \label{fig:testing}
\end{figure*}

The SFH fits depend on additional free parameters, including the distance modulus $(m-M)_0$, the foreground extinction ($Av$) and the internal extinction ($dAv$). We determine the best-fit value for each parameter by allowing MATCH to search over a range of values for each parameter with 0.05 mag spacing. For LMC fields, we found good agreement between independent measurements of reddening and distance in the LMC \citep{Skowron2021, Choi2018a} and the MATCH fit values \citep{Cohen2024a}. Previous Scylla papers also find that the best-fit distances agree well across different sets of evolutionary models, with model-to-model differences of $ \le 0.05$ mag, and that those results agreed well between the three models tested \citep{Cohen2024a}. In the SMC, fitting for $dAv$, $Av$, and $(m-M)_0$ with MATCH has an additional level of complexity. MATCH assumes a single distance for the SFH fitting, but the distance distribution of stars in the SMC can be large ($\sim$ 14 kpc in central regions, \cite{subramanian2012} and up to 23 Kpc in towards the bridge \citep{nidever2013}) and dependent on the line-of-sight \citep{stanimirovic1999, Cioni2000, nidever2014}. \citet{Cohen2024b} used simulations to test the sensitivity of the SFHs to different line-of-sight distance distributions.

\begin{figure*}
 	\centering
        \textbf{Testing the dependence of combined SFHs on field groupings}\par\medskip
\includegraphics[width=1.0\textwidth]{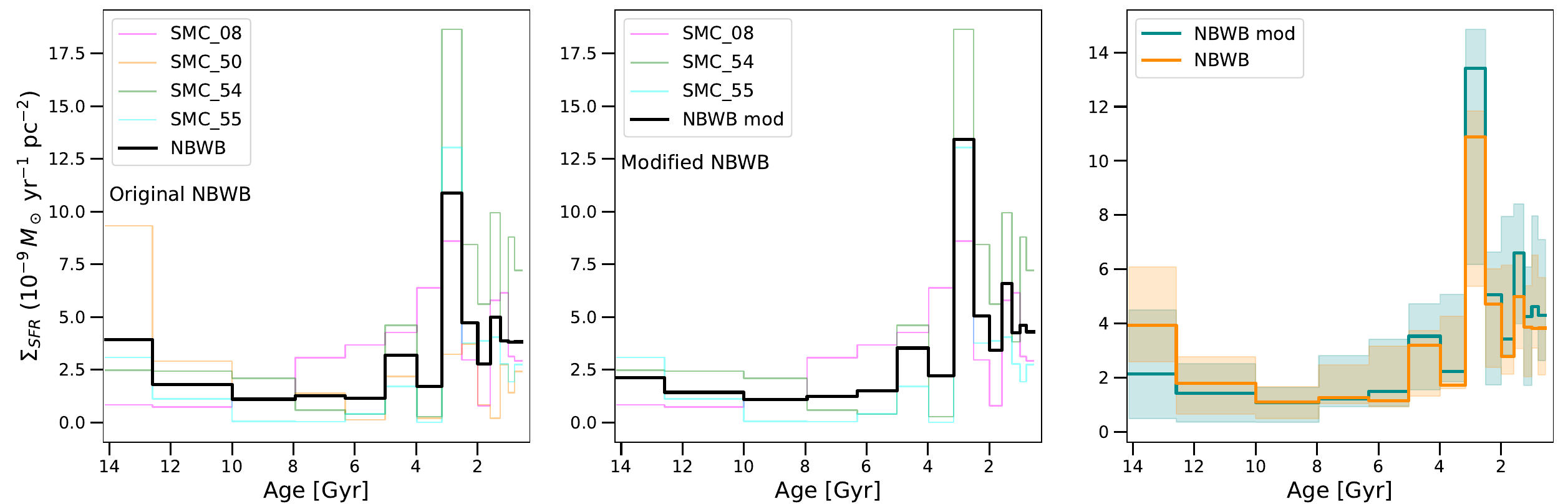}
	\caption{Area normalized SFH, $\Sigma_{SFH}$, for two possible combinations of fields in the NBWB sub-region, demonstrating how the choice of field combination did not impact the shape of the combined $\Sigma_{SFH}$ nor the location of the burst. \textbf{Left}: $\Sigma_{SFH}$ for individual files in the NBWB sub-region (colored lines) and the $\Sigma_{SFH}$ of the combined NBWB sub-region (black line). \textbf{Right}:  Same plot as left if SMC$\_{50}$  were omitted from the combined NBWB sub-region $\Sigma_{SFH}$. \textbf{Right}: The original NWBWB $\Sigma_{SFH}$ and modified NBWB $\Sigma_{SFH}$ overplotted with uncertainties. The inclusion or exclusion of this field does not change the shape or location of the burst signal in the $\Sigma_{SFH}$, only the magnitude of the SFR. This test was not performed in all groups, but rather for groups where it was unclear whether an individual field's $\Sigma_{SFH}$ would significantly change the signal of the burst in the combined solution. } 
        \label{fig:fieldcombination}
\end{figure*}

% \startlongtable
% \begin{deluxetable*}{lcccccc}
% \tabletypesize{\scriptsize}
% \tablecaption{Best-Fit Values for Scylla Fields \label{table:2}}
% \tablehead{\colhead{} & \colhead{} & \colhead{SMC} & \colhead{} & \colhead{} & \colhead{LMC} & \colhead{} \\ \colhead{} & \colhead{$(m-M)_0$ [mag]} & \colhead{Av [mag]} & \colhead{dAv [mag]} & \colhead{$(m-M)_0$ [mag]} & \colhead{Av [mag]} & \colhead{dAv [mag]}}
% \startdata
% Minimum & 18.86 & 0.0 & 0.0 & 18.29 & 0.0 & 0.0 \\
% Maximum & 19.26 & 0.2 & 0.4 & 18.79 & 0.5 & 3.4 \\
% Average & 19.05 & 0.12 & 0.15 & 18.53 &  0.28 & 0.34 \\ 
% \enddata
% \tablecomments{Minimum, maximum, and average best-fit values of the distance modulus $(m-M)_0$, foreground extinction ($Av$), and internal extinction ($dAv$). A complete list of values can be found in \citep{Cohen2024b} for the SMC and \citep{Cohen2024a} for the LMC.}
% \label{table:2}
% \end{deluxetable*}

\citet{Cohen2024b} demonstrated that our recovery method is robust for a range of line-of-sight distributions that exist in the SMC. This was done by recovering the SFHs of synthetically generated mock stellar populations with different line-of-sight distance distributions (these included a single fixed distance, two flat distributions, two Gaussian distributions, and a double Gaussian to model the bimodal structure of the SMC's wing). The recovery metrics (the lookback times to form $75\%$ and $90\%$ of stellar mass) were almost always recovered to within their uncertainties.

% The average best-fit values for all fields used in this analysis in the SMC and the LMC can be found in Table \ref{table:2}.   

The total uncertainties for each SFH fit were calculated, taking into account random and systematic uncertainties. Random uncertainties were calculated with a hybrid Markov chain Monte Carlo method \citep{Dolphin2013}, and systematic uncertainties were calculated by performing 50 Monte Carlo iterations that shift the observed CMD in $T_{eff}$ and $M_{bol}$ and then refitting the CMD \citep{Dolphin2012}. 

\subsection{Time Binning in the SFH Solutions} \label{sec:Timebin3.1}

We conducted time-binning tests to optimize the balance between temporal resolution and the statistical reliability of SFR measurements. Increasing the time resolution with finer binning has the benefit of more accurately constraining the duration of bursts, but comes at the cost of increasing the uncertainties in each of the bins. Coarser binning produces smaller uncertainties for two reasons. First, decreasing the number of time bins gives the model more flexibility, which means a wider range of SFRs in each bin can still produce a good fit to the data. Second, finer bins are less constrained by the data, so the SFR can vary more before it noticeably worsens the quality of the fit \citep{dolphin2002}. This is also entangled with the issue of covariance, wherein finer time binning increases the covariance between bins by decreasing the impact of moving star formation into a neighboring bin. The goal of our time-binning testing was to find a balance between a finer time resolution and reduced uncertainties that would still allow us to identify and characterize bursts.

% Our goal in testing was to optimize for the finest time bins and the lowest uncertainties for the star formation rates (SFRs) in each bin.

To determine the optimal time bin sizes for our SFH fits, we generated synthetic data covering the parameter space of the Scylla observations. We used MATCH to generate synthetic photometric catalogs and then recovered the SFHs ten times with MATCH to find the variance of the measured SFRs per bin. For the input SFH, we assumed a constant SFR of $ \simeq  1\times10^{-5}$  M$_{\odot}$ yr$^{-1}$, with three bursts superimposed with the following timing and intensity: $1$ – $1.1$ Gyr, $4.0$ - $4.5$ Gyr and $6.3$ - $7.0$ Gyr with SFR $3\times10^{-5}$  M$_{\odot}$ yr$^{-1}$, $2\times10^{-5}$  M$_{\odot}$ yr$^{-1}$ and $2\times10^{-5}$  M$_{\odot}$ yr$^{-1}$, respectively. These values were chosen to create a synthetic stellar population with approximately the same present-day stellar mass found in a typical Scylla field ($\sim 10^5 M_\odot$). Finally, the catalogs were injected with an observational noise model generated using the ASTs of representative fields within the two galaxies. We define representative as a field with an A(v), dA(v), and (m-M)$_0$ within $0.05$ dex of the mean value of that parameter in each galaxy. These fields were SMC$\_{20}$ and LMC$\_4$. 

We explored 21 different time-binning schemes, including those used in previous work such as the VMC survey, archival HST programs, and other Scylla analysis \citep{rubele_2015, weisz2013, Cohen2024a, Cohen2024b}. Based on these tests, we selected the binning scheme with the lowest variance among the ten MATCH solutions for each field. To further evaluate its performance, we generated synthetic photometry with input SFHs that included bursts shorter in duration than the width of the time bins used in the fitting process. This setup allowed us to assess how well MATCH recovers short-duration bursts when they fall within broader time bins where their signal could be diluted by lower star formation activity during the rest of the bin.

Figure~\ref{fig:testing} illustrates the results from fitting the SMC synthetic photometry using our final binning scheme. The left panel shows the ten SFH realizations, while the right panel presents the average SFR in each bin, with the square root of the variance overlaid in blue and the input SFH shown in black. We recovered the SFR in the oldest bin (6.3–8.0 Gyr), middle bin (4.0–4.5 Gyr), and youngest bin (1–1.1 Gyr) within 2, 1, and 2.5 standard deviations of the input values, respectively. The total stellar mass agreed with the input within 2$\%$. We found similar results for a synthetic catalog generated with noise characteristics from a representative field in the LMC.

% The total stellar masses were measured to be $1.6\times$, $1.9\times$, and $1.2\times$ the input values in the oldest, middle, and youngest bin, respectively.

Based on this testing, we selected the time-binning scheme listed in Table~\ref{table:cumulativeSFH} as the one that best minimized uncertainty while preserving the temporal resolution needed to identify and characterize bursts.

\subsection{Grouping Fields} \label{sec:Fields3.2}

While increasing the time resolution can help reduce systematic uncertainties, low star counts in many of the individual Scylla fields can yield large statistical uncertainties. To reduce the statistical uncertainties, we combined the best-fit solutions of fields within defined sub-regions (defined further in the section below). The reduction of uncertainties per bin plateaued for large groupings of fields in both the SMC and the LMC. Specifically, we found that the greatest decrease in uncertainty from a group of fields occurred when we combined three to five, which included approximately $150$K $-$ $250$K stars. There was also minimal additional improvement when we combined six to seven total fields, (i.e., $300$K $-$ $350$K stars), and no further improvement when we combined eight or more total fields (i.e., $>400$K stars).

Although the choice of which fields to combine involves some subjectivity, we aimed to ensure that the SFH solutions derived from these grouped fields are representative and not overly sensitive to the exact grouping. To ensure robustness, we performed two independent checks. First, we visually inspected the SFHs of neighboring fields within each group to confirm that their shapes were similar; more details on the selection process for these groups are provided in Sections \ref{sec:SMC3.2.1} and \ref{sec:LMC3.2.2}. Second, we performed a jackknife test on the groupings to verify that removing individual fields would not cause overt changes in the combined SFH results.

\begin{figure*}[h!]
 	\centering
        \textbf{SFH of NBWB subregion with burst metrics applied}\par\medskip
\includegraphics[width=1.0\textwidth]{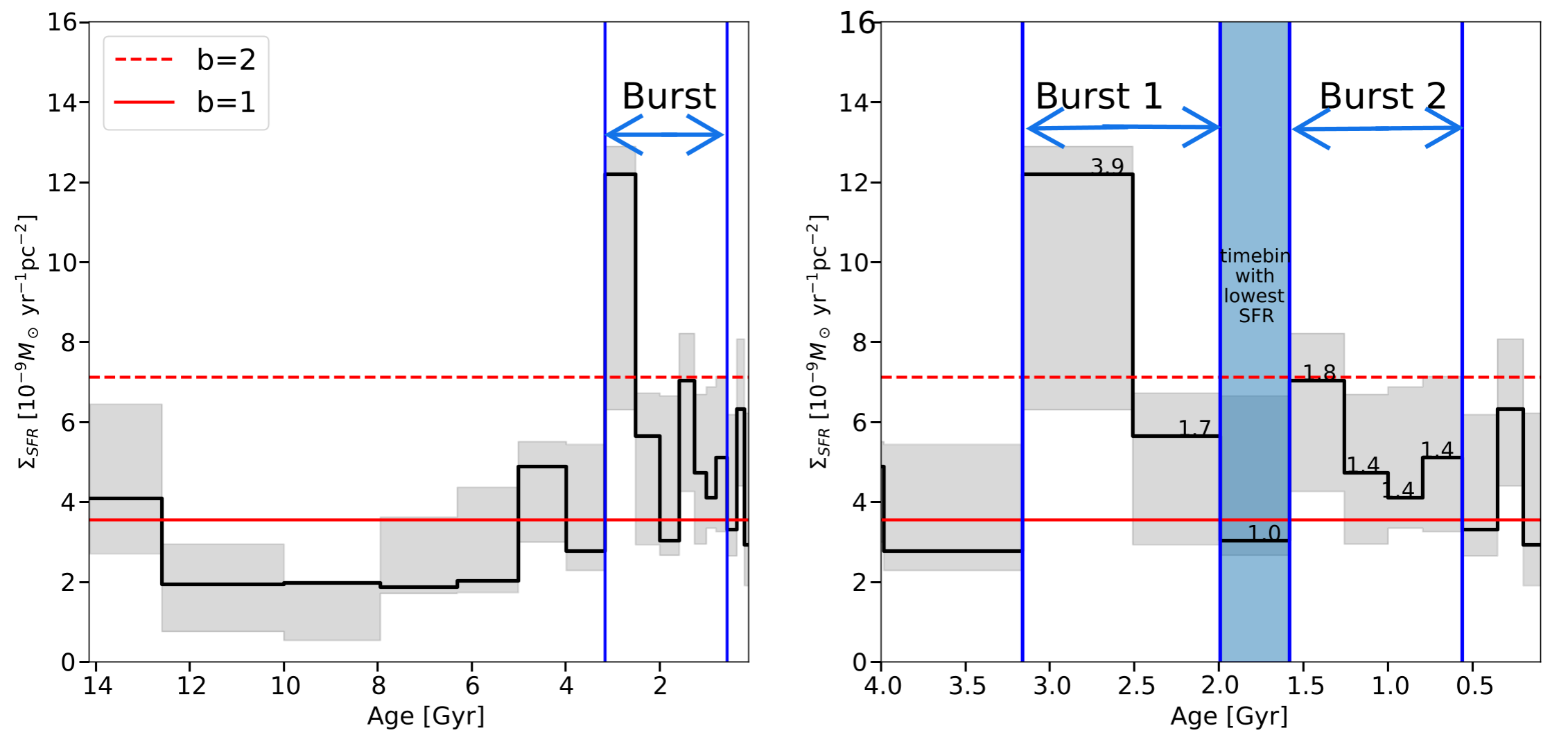}
	  \caption{The SFR(t) of the NBWB subregion, with $<SFR>$ and $2\times<SFR>$ plotted as solid and dotted red lines, respectively. In the left plot, the entire $\Sigma_{SFH}$ is shown with the duration of a burst, as defined by our metrics, demarcated by the two blue lines at $0.55$ Gyr and $3.3$ Gyr. In the right panel is the same $\Sigma_{SFH}$ shown only from $0-5$ Gyr. This panel illustrates how we use our secondary burst metric, which splits a burst into finer events if there is a b-value at least twice as great as the lowest b-value within the burst. The lowest b-value is highlighted in blue, and the \textit{b} for each bin within the burst is included above its respective bin. Since the bin with the lowest b-value within the burst has \textit{b} = 0.96 (plotted as 1.0), and the bin with the highest b-value within the burst has \textit{b} = 3.9, which is at least twice as large as the lowest b-value, the original burst is broken into two.}
  \label{fig:bvalue}
\end{figure*}

Figure \ref{fig:fieldcombination} presents an example of our jackknife test, which demonstrates that the inclusion or exclusion of a single field did not change the overall shape of our SFH nor the signal of the burst in the SFH. In the left panel are the SFHs for the four fields that comprise a grouping called the `NBWB subregion' and the combined SFH solution for the NBWB subregion (this is shown and Figure~\ref{fig:cmds} and defined more fully in Section~\ref{sec:SMC3.2.1}). In the right panel, we have excluded SMC$\_{50}$ from the final NBWB sub-region combined fit. The inclusion or exclusion of this field does not change the shape or location of the burst signal in the SFH; only the magnitude of the SFR. This test was not performed in all groups, but rather for groups where it was unclear whether an individual field's SFH would significantly change the signal of the burst in the combined solution.

\subsubsection{SMC} \label{sec:SMC3.2.1}

When selecting fields to combine, we were guided by previous studies of the kinematically, physically, and chemically associated regions in the SMC. We used the VIsible Soar photometry of star clusters in the tApii and Coxi HuguA (VISCACHA) survey \citep{Maia2019}. Regions are defined in \citet{dias2016, Dias2022} as the Center, Wing/Bridge, and North Bridge, all of which can be seen in Figure \ref{fig:maps}. Within regions, we grouped anywhere from three to eight fields into sub-regions. Fields were assigned to a subregion based on their proximity to one another and agreement between their individual SFHs. There were originally 45 fields in the SMC available at the time of this analysis, but 9 had to be excluded because they were too far from any of the sub-regions or too different than other potential sub-region constituents (differences may arise in nearby individual fields as a product of variations in dust and the stellar populations along the line of sight distance). The final subregions in the SMC are: Center 1 (C1), Center 2 (C2), North Bridge (NB), North Bridge/Wing Bridge (NB/WB), North Bridge/Center Bridge (NB/CB), Wing/Bridge (WB), Wing Bridge/North Bridge (WB/NB), and Wing Bridge/Center (WB/C).

% We considered a number of other methods for grouping the Scylla fields in the SMC. We tried combining fields based on their distance modulus, rather than their position in the plane of the sky, but doing so was not possible with comparable accuracy in the SMC due to the complicated morphology of the SMC \citep{almeida2023, nidever2017}. We also considered grouping fields based solely on their proximity to each other on the plane of the sky. This method posed an issue in the SMC due to the locations of the fields, many of which are in large clusters ($>8$). Combining all the fields in one of these clusters would not produce any further reduction in the uncertainties. By using the regions defined by VISCACHA, we were able to base our groupings in the SMC on the measurable, physical properties of star clusters inside the galaxy.

\subsubsection{LMC} \label{sec:LMC3.2.2}

The LMC’s regions were assigned cardinally. These are the Northeast, Northwest, Southeast, and Southwest. The Scylla program has a total of 45 fields with photometry in at least 2 bands in the LMC. After the initial inspection of the individual SFHs, we were left with 38 fields, as shown in Figure \ref{fig:maps}.

Seven of the fields were excluded by inspection because of their distance from one of the seven prospective sub-regions or because their SFH was significantly different than the rest of those in their prospective sub-region. The outliers were identified by inspection and proximity to pre-existing clusters of fields. 

\begin{figure*}
    \textbf{SFH history results for all sub-regions in the SMC and LMC}\par\medskip
  \subfloat
  {\includegraphics[width=.49\textwidth]{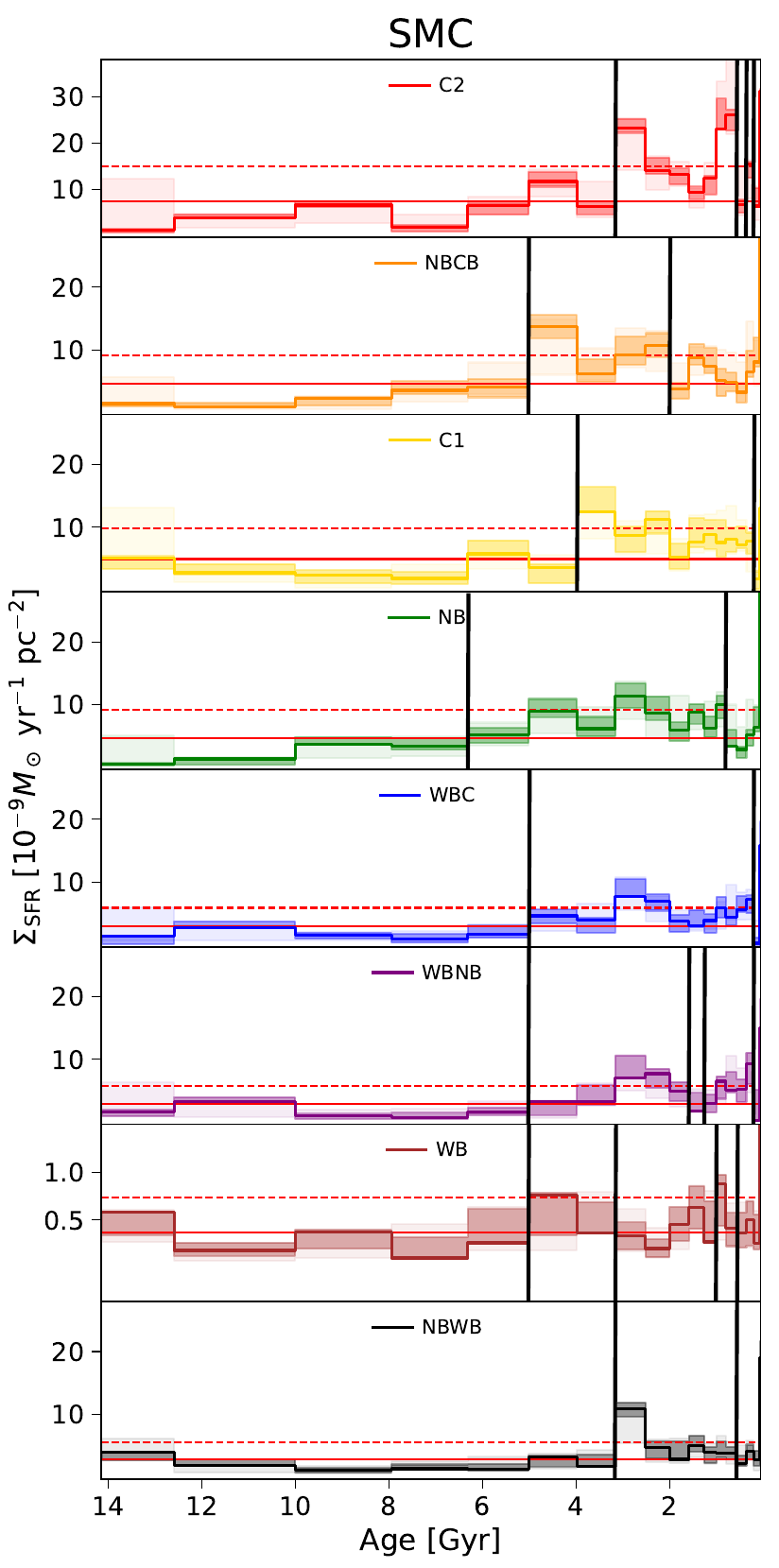}}\quad
  \subfloat
  {\includegraphics[width=.49\textwidth]{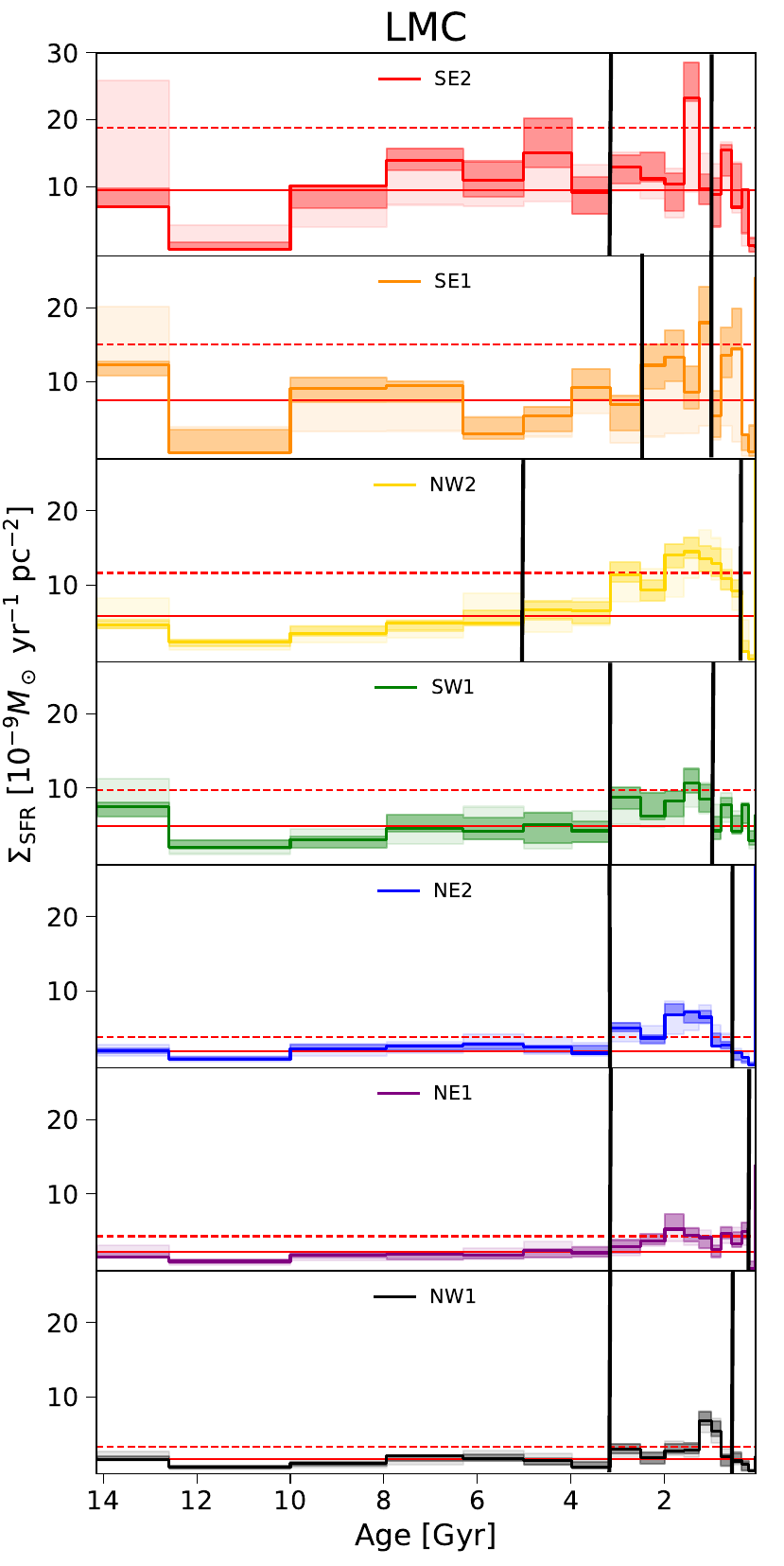}}\\
  \caption{$\Sigma_{SFR(t)}$ for eight sub-regions in the SMC and seven sub-regions in the LMC. Within the galaxies, the sub-regions are ordered from top to bottom, from areas with the highest to the lowest surface density. The SFHs are labeled with our broad burst metrics, with horizontal red lines showing b=1 and b=2, and vertical black lines demarcating the beginning and end of the bursts. The statistical uncertainties are plotted with the systematic uncertainties stacked on top with a lighter shading. Cumulative $\Sigma_{SFH}$s for each subregion are provided in the appendix. }
  \label{fig:stackedSFH}
\end{figure*}

\clearpage

\startlongtable
\begin{deluxetable*}{lcccccccc}
\tabletypesize{\scriptsize}
\tablecaption{SMC burst parameters $b = \mathrm{SFR}/\langle \mathrm{SFR} \rangle$ for each time bin in each region}
\tablehead{\colhead{log(Age)} & \colhead{C1} & \colhead{C2} & \colhead{WB} & \colhead{WBNB} & \colhead{WBC} & \colhead{NB} & \colhead{NBCB} & \colhead{NBWB} }
\startdata
6.60--7.20 & $2.39^{+0.33+0.36}_{-0.67-0.64}$ & $6.07^{+0.81+0.11}_{-0.47-1.45}$ & $26.30^{+5.49+1.99}_{-2.95-1.07}$ & $8.30^{+0.97+0.75}_{-1.78-2.08}$ & $4.64^{+0.30+1.17}_{-1.30-2.10}$ & $15.28^{+0.92+0.98}_{-2.52-3.32}$ & $17.12^{+0.97+1.52}_{-1.96-2.78}$ & $20.01^{+1.73+2.97}_{-1.99-1.95}$ \\
7.20--7.80 & $2.67^{+0.64+0.10}_{-0.52-1.61}$ & $2.82^{+0.15+0.38}_{-1.28-2.61}$ & $6.17^{+0.64+0.29}_{-3.69-0.52}$ & $5.27^{+1.66+0.21}_{-1.83-3.53}$ & $3.75^{+1.07+0.42}_{-0.54-4.47}$ & $10.56^{+1.28+0.44}_{-1.24-3.43}$ & $9.85^{+0.54+0.68}_{-1.88-2.96}$ & $6.86^{+1.07+0.00}_{-1.80-1.90}$ \\
7.80--8.30 & $0.36^{+0.24+0.07}_{-0.35-0.00}$ & $1.01^{+0.62+0.00}_{-0.05-0.38}$ & $0.69^{+0.82+0.12}_{-0.17-0.02}$ & $0.10^{+1.69+0.00}_{-0.05-0.00}$ & $1.48^{+0.26+0.99}_{-0.44-0.00}$ & $1.40^{+0.71+0.07}_{-0.19-0.21}$ & $1.46^{+0.69+0.11}_{-0.16-0.71}$ & $0.98^{+0.53+0.40}_{-0.36-0.00}$ \\
8.30--8.55 & $1.59^{+0.30+0.07}_{-0.43-0.60}$ & $2.40^{+0.12+0.17}_{-0.49-0.55}$ & $1.35^{+0.55+0.06}_{-0.73-0.07}$ & $3.26^{+0.67+0.18}_{-1.18-0.55}$ & $1.84^{+0.20+0.46}_{-0.62-0.56}$ & $1.11^{+0.21+1.03}_{-0.39-0.01}$ & $1.30^{+0.59+1.19}_{-0.17-0.06}$ & $1.45^{+0.41+0.12}_{-0.46-0.12}$ \\
8.55--8.75 & $1.47^{+0.60+0.25}_{-0.17-0.13}$ & $0.91^{+0.19+0.92}_{-0.27-0.06}$ & $0.99^{+0.60+0.27}_{-0.41-0.02}$ & $1.83^{+0.59+0.66}_{-0.72-0.06}$ & $1.95^{+0.61+0.39}_{-0.16-0.48}$ & $0.61^{+0.11+0.67}_{-0.29-0.01}$ & $0.63^{+0.26+0.79}_{-0.28-0.10}$ & $0.76^{+0.47+0.34}_{-0.15-0.07}$ \\
8.75--8.90 & $1.64^{+0.41+0.72}_{-0.42-0.13}$ & $3.86^{+0.26+1.59}_{-0.57-0.15}$ & $1.13^{+0.44+0.29}_{-0.53-0.03}$ & $1.78^{+1.03+0.94}_{-0.19-0.15}$ & $2.59^{+0.42+1.37}_{-0.38-0.00}$ & $0.73^{+0.57+0.87}_{-0.05-0.04}$ & $1.13^{+0.28+0.79}_{-0.25-0.11}$ & $1.38^{+0.36+0.23}_{-0.37-0.18}$ \\
8.90--9.00 & $1.54^{+0.74+0.33}_{-0.46-0.09}$ & $3.24^{+1.05+0.39}_{-0.14-0.25}$ & $2.38^{+0.35+0.09}_{-1.21-0.41}$ & $2.28^{+0.33+0.17}_{-1.00-0.35}$ & $1.48^{+0.47+0.00}_{-0.45-0.76}$ & $2.18^{+0.33+0.19}_{-0.51-0.32}$ & $1.00^{+0.34+0.76}_{-0.44-0.10}$ & $1.38^{+0.66+0.13}_{-0.28-0.00}$ \\
9.00--9.10 & $1.80^{+0.51+0.22}_{-0.49-0.11}$ & $1.58^{+0.10+0.39}_{-0.57-0.00}$ & $0.74^{+1.16+0.30}_{-0.08-0.03}$ & $1.06^{+0.49+0.28}_{-0.52-0.05}$ & $1.37^{+0.29+0.41}_{-0.55-0.00}$ & $1.35^{+0.42+0.53}_{-0.33-0.03}$ & $1.60^{+0.51+0.23}_{-0.22-0.06}$ & $1.40^{+0.34+0.23}_{-0.61-0.00}$ \\
9.10--9.20 & $1.57^{+0.79+0.11}_{-0.22-0.22}$ & $1.06^{+0.21+0.00}_{-0.23-0.25}$ & $1.72^{+0.61+0.05}_{-0.64-0.14}$ & $0.61^{+1.02+0.20}_{-0.05-0.10}$ & $1.42^{+0.67+0.17}_{-0.13-0.19}$ & $1.92^{+0.29+0.07}_{-0.43-0.24}$ & $1.93^{+0.41+0.05}_{-0.21-0.45}$ & $1.80^{+0.57+0.00}_{-0.40-0.41}$ \\
9.20--9.30 & $1.06^{+0.45+0.19}_{-0.38-0.10}$ & $1.90^{+0.22+0.22}_{-0.35-0.06}$ & $1.22^{+0.51+0.03}_{-0.48-0.23}$ & $1.72^{+0.53+0.08}_{-0.57-0.36}$ & $1.25^{+0.27+0.32}_{-0.48-0.14}$ & $1.29^{+0.29+0.92}_{-0.38-0.02}$ & $0.73^{+0.16+0.73}_{-0.29-0.05}$ & $1.00^{+1.06+0.00}_{-0.08-0.06}$ \\
9.30--9.40 & $2.29^{+0.33+0.13}_{-0.49-0.67}$ & $1.93^{+0.45+0.03}_{-0.15-0.47}$ & $0.55^{+0.27+0.33}_{-0.26-0.01}$ & $2.71^{+0.36+0.20}_{-0.84-0.62}$ & $2.01^{+0.33+0.13}_{-0.43-0.73}$ & $1.88^{+0.59+0.04}_{-0.27-0.62}$ & $2.37^{+0.37+0.24}_{-0.34-0.60}$ & $1.70^{+0.38+0.00}_{-0.65-0.14}$ \\
9.40--9.50 & $1.77^{+0.32+0.23}_{-0.57-0.11}$ & $3.35^{+0.35+0.15}_{-0.22-1.06}$ & $0.90^{+0.40+0.37}_{-0.38-0.07}$ & $2.47^{+1.27+0.07}_{-0.19-0.65}$ & $2.35^{+0.77+0.38}_{-0.15-0.67}$ & $2.47^{+0.46+0.14}_{-0.44-0.41}$ & $2.22^{+0.53+0.45}_{-0.29-0.21}$ & $3.93^{+0.45+0.00}_{-0.54-1.17}$ \\
9.50--9.60 & $2.54^{+0.83+0.07}_{-0.20-0.77}$ & $0.81^{+0.21+0.54}_{-0.25-0.07}$ & $1.00^{+0.86+0.29}_{-0.10-0.05}$ & $1.16^{+0.91+0.13}_{-0.22-0.12}$ & $1.71^{+0.15+0.76}_{-0.57-0.00}$ & $1.33^{+0.42+0.36}_{-0.27-0.04}$ & $1.17^{+0.42+0.50}_{-0.22-0.06}$ & $0.62^{+0.67+0.08}_{-0.06-0.11}$ \\
9.60--9.70 & $0.72^{+0.13+0.30}_{-0.51-0.01}$ & $1.25^{+0.33+0.07}_{-0.19-0.30}$ & $2.05^{+0.22+0.09}_{-1.14-0.44}$ & $1.14^{+0.09+0.12}_{-0.76-0.18}$ & $1.59^{+0.30+0.00}_{-0.34-0.54}$ & $1.95^{+0.43+0.05}_{-0.24-0.58}$ & $2.98^{+0.35+0.11}_{-0.36-1.37}$ & $1.15^{+0.10+0.14}_{-0.57-0.00}$ \\
9.70--9.80 & $1.17^{+0.45+0.03}_{-0.19-0.33}$ & $1.01^{+0.08+0.19}_{-0.32-0.00}$ & $0.71^{+0.95+0.08}_{-0.07-0.16}$ & $0.57^{+0.23+0.35}_{-0.18-0.02}$ & $0.71^{+0.39+0.19}_{-0.16-0.04}$ & $1.13^{+0.25+0.19}_{-0.31-0.10}$ & $0.94^{+0.23+0.64}_{-0.29-0.00}$ & $0.41^{+0.33+0.33}_{-0.09-0.01}$ \\
9.80--9.90 & $0.37^{+0.22+0.25}_{-0.17-0.02}$ & $0.24^{+0.07+0.30}_{-0.15-0.00}$ & $0.28^{+0.58+0.38}_{-0.03-0.01}$ & $0.25^{+0.27+0.21}_{-0.10-0.01}$ & $0.76^{+0.20+0.20}_{-0.17-0.07}$ & $0.71^{+0.30+0.07}_{-0.12-0.23}$ & $0.65^{+0.26+0.00}_{-0.11-0.31}$ & $0.46^{+0.25+0.25}_{-0.07-0.01}$ \\
9.90--10.00 & $0.47^{+0.17+0.01}_{-0.27-0.14}$ & $0.72^{+0.14+0.00}_{-0.08-0.46}$ & $1.02^{+0.12+0.04}_{-0.49-0.27}$ & $0.34^{+0.11+0.23}_{-0.16-0.02}$ & $0.42^{+0.11+0.09}_{-0.15-0.14}$ & $0.80^{+0.25+0.01}_{-0.05-0.46}$ & $0.65^{+0.05+0.05}_{-0.21-0.17}$ & $0.40^{+0.13+0.00}_{-0.17-0.24}$ \\
10.00--10.10 & $0.55^{+0.29+0.01}_{-0.08-0.24}$ & $0.61^{+0.13+0.00}_{-0.03-0.27}$ & $0.49^{+0.23+0.00}_{-0.16-0.15}$ & $1.15^{+0.25+0.06}_{-0.09-0.81}$ & $0.42^{+0.27+0.00}_{-0.03-0.69}$ & $0.28^{+0.02+0.06}_{-0.18-0.05}$ & $0.15^{+0.11+0.03}_{-0.01-0.08}$ & $0.65^{+0.36+0.00}_{-0.05-0.29}$ \\
10.10--10.15 & $1.03^{+0.10+1.55}_{-0.36-0.02}$ & $0.16^{+0.06+1.44}_{-0.07-0.00}$ & $1.59^{+0.16+0.08}_{-0.67-0.22}$ & $0.56^{+0.14+1.51}_{-0.26-0.01}$ & $0.69^{+0.04+1.54}_{-0.34-0.00}$ & $0.09^{+0.06+0.93}_{-0.03-0.00}$ & $0.28^{+0.03+0.91}_{-0.08-0.02}$ & $1.42^{+0.10+0.57}_{-0.50-0.00}$ \\
\enddata
\tablecomments{Birth rate parameters are listed with statistical and systematic uncertainties: $b^{+\mathrm{stat}+\mathrm{sys}}_{-\mathrm{stat}-\mathrm{sys}}$}

\label{table:statSMC}
\end{deluxetable*}

\startlongtable
\begin{deluxetable*}{lccccccc}
\tabletypesize{\scriptsize}
\tablecaption{LMC burst parameters $b = \mathrm{SFR}/\langle \mathrm{SFR} \rangle$ for each time bin in each region}
\tablehead{\colhead{log(Age)} & \colhead{SE2} & \colhead{SE1} & \colhead{NW2} & \colhead{SW1} & \colhead{NE2} & \colhead{NE1} & \colhead{NW1}}
\startdata
6.60--7.20 & $20.98^{+2.04+1.85}_{-1.62-2.70}$ & $26.47^{+2.49+2.28}_{-1.84-1.03}$ & $75.54^{+6.15+5.44}_{-4.93-5.71}$ & $5.88^{+0.58+3.56}_{-0.79-0.13}$ & $19.10^{+3.16+133.67}_{-1.34-1.49}$ & $28.02^{+2.37+12.96}_{-1.70-2.59}$ & $10.11^{+0.88+2.06}_{-1.05-0.62}$ \\
7.20--7.80 & $6.47^{+0.38+1.61}_{-2.33-0.19}$ & $12.82^{+1.42+0.20}_{-1.45-1.53}$ & $1.19^{+0.08+0.86}_{-0.97-0.09}$ & $6.02^{+0.26+0.24}_{-1.31-1.15}$ & $3.21^{+0.31+0.21}_{-0.90-2.34}$ & $0.10^{+0.29+0.00}_{-0.08-0.02}$ & $1.31^{+0.28+0.03}_{-0.69-0.47}$ \\
7.80--8.30 & $0.00^{+0.41+0.02}_{-0.00-0.00}$ & $0.00^{+0.17+0.00}_{-0.00-0.00}$ & $0.00^{+0.12+0.00}_{-0.00-0.00}$ & $0.00^{+0.09+0.00}_{-0.00-0.00}$ & $0.08^{+0.47+0.00}_{-0.08-0.00}$ & $0.13^{+0.12+0.01}_{-0.11-0.00}$ & $0.60^{+0.28+0.01}_{-0.12-0.11}$ \\
8.30--8.55 & $2.31^{+0.58+0.02}_{-0.49-0.43}$ & $0.41^{+0.04+0.15}_{-0.34-0.00}$ & $0.52^{+0.27+0.20}_{-0.47-0.05}$ & $0.18^{+0.24+0.49}_{-0.02-0.01}$ & $0.38^{+0.03+0.03}_{-0.28-0.11}$ & $1.01^{+0.06+0.09}_{-0.69-0.21}$ & $1.60^{+0.12+0.07}_{-0.53-0.15}$ \\
8.55--8.75 & $1.53^{+0.75+0.29}_{-0.23-0.02}$ & $0.88^{+0.05+0.15}_{-0.55-0.00}$ & $0.82^{+0.65+0.10}_{-0.15-0.16}$ & $1.58^{+0.26+0.02}_{-0.16-0.30}$ & $1.94^{+0.73+0.04}_{-0.14-1.57}$ & $0.74^{+0.69+0.03}_{-0.04-0.11}$ & $0.86^{+0.44+0.15}_{-0.10-0.02}$ \\
8.75--8.90 & $2.16^{+0.44+0.07}_{-0.30-0.34}$ & $0.73^{+0.24+0.34}_{-0.27-0.01}$ & $1.16^{+0.20+0.48}_{-0.71-0.11}$ & $1.86^{+0.22+0.50}_{-0.27-0.03}$ & $1.81^{+0.52+0.05}_{-0.42-0.87}$ & $1.64^{+0.13+0.15}_{-0.42-0.24}$ & $1.60^{+0.22+0.17}_{-0.36-0.15}$ \\
8.90--9.00 & $1.15^{+0.29+0.58}_{-0.46-0.01}$ & $0.84^{+0.90+0.20}_{-0.05-0.04}$ & $3.28^{+0.83+0.25}_{-0.47-0.57}$ & $2.21^{+0.37+0.26}_{-0.31-0.11}$ & $0.73^{+0.45+0.01}_{-0.40-0.05}$ & $0.94^{+0.26+0.24}_{-0.52-0.02}$ & $0.88^{+0.40+0.34}_{-0.24-0.01}$ \\
9.00--9.10 & $1.91^{+0.17+0.32}_{-0.67-0.05}$ & $2.61^{+0.48+0.19}_{-0.44-0.00}$ & $4.11^{+0.75+0.15}_{-0.48-0.59}$ & $2.33^{+0.32+0.36}_{-0.32-0.04}$ & $2.40^{+0.67+0.07}_{-0.43-1.46}$ & $1.03^{+0.24+0.34}_{-0.25-0.04}$ & $1.77^{+0.42+0.08}_{-0.38-0.07}$ \\
9.10--9.20 & $2.06^{+0.48+0.04}_{-0.42-0.23}$ & $2.77^{+0.19+0.07}_{-0.77-0.19}$ & $1.71^{+0.55+0.06}_{-0.32-0.30}$ & $2.49^{+0.35+0.04}_{-0.18-0.48}$ & $1.15^{+0.47+0.02}_{-0.32-0.45}$ & $2.47^{+0.59+0.09}_{-0.16-1.38}$ & $2.21^{+0.42+0.05}_{-0.32-0.40}$ \\
9.20--9.30 & $2.45^{+0.94+0.02}_{-0.15-0.65}$ & $2.12^{+0.78+0.00}_{-0.12-0.90}$ & $1.66^{+0.59+0.19}_{-0.21-0.21}$ & $2.42^{+0.27+0.05}_{-0.31-0.68}$ & $1.78^{+0.49+0.05}_{-0.46-0.95}$ & $1.10^{+0.18+0.12}_{-0.44-0.13}$ & $1.70^{+0.32+0.03}_{-0.44-0.28}$ \\
9.30--9.40 & $1.71^{+0.43+0.05}_{-0.25-0.12}$ & $1.57^{+0.25+0.40}_{-0.39-0.00}$ & $1.08^{+0.44+0.08}_{-0.47-0.03}$ & $1.61^{+0.24+0.27}_{-0.26-0.02}$ & $1.63^{+0.40+0.05}_{-0.44-0.86}$ & $1.18^{+0.43+0.03}_{-0.08-0.28}$ & $1.29^{+0.65+0.04}_{-0.14-0.21}$ \\
9.40--9.50 & $1.33^{+0.44+0.04}_{-0.39-0.18}$ & $1.70^{+0.39+0.00}_{-0.27-0.46}$ & $1.78^{+0.45+0.02}_{-0.39-0.40}$ & $1.97^{+0.29+0.03}_{-0.20-0.45}$ & $0.92^{+0.18+0.03}_{-0.47-0.23}$ & $1.37^{+0.21+0.11}_{-0.27-0.16}$ & $1.82^{+0.29+0.05}_{-0.36-0.44}$ \\
9.50--9.60 & $0.96^{+0.34+0.08}_{-0.26-0.03}$ & $0.47^{+0.54+0.01}_{-0.03-0.13}$ & $0.35^{+0.39+0.29}_{-0.14-0.01}$ & $1.12^{+0.22+0.09}_{-0.29-0.07}$ & $1.23^{+0.35+0.03}_{-0.23-0.27}$ & $0.97^{+0.25+0.21}_{-0.35-0.03}$ & $0.88^{+0.27+0.30}_{-0.33-0.01}$ \\
9.60--9.70 & $1.08^{+0.13+0.45}_{-0.42-0.01}$ & $0.90^{+0.13+0.40}_{-0.35-0.03}$ & $0.83^{+0.22+0.39}_{-0.34-0.02}$ & $1.14^{+0.20+0.03}_{-0.22-0.31}$ & $0.73^{+0.17+0.03}_{-0.29-0.10}$ & $1.60^{+0.56+0.04}_{-0.25-0.55}$ & $1.04^{+0.35+0.03}_{-0.52-0.16}$ \\
9.70--9.80 & $0.81^{+0.21+0.24}_{-0.22-0.03}$ & $1.16^{+0.09+0.44}_{-0.36-0.00}$ & $1.01^{+0.35+0.29}_{-0.19-0.03}$ & $0.83^{+0.30+0.41}_{-0.06-0.03}$ & $0.40^{+0.29+0.03}_{-0.09-0.02}$ & $1.16^{+0.31+0.04}_{-0.27-0.18}$ & $0.87^{+0.38+0.31}_{-0.25-0.01}$ \\
9.80--9.90 & $0.89^{+0.05+0.09}_{-0.33-0.08}$ & $0.95^{+0.12+0.10}_{-0.29-0.10}$ & $1.24^{+0.08+0.05}_{-0.45-0.26}$ & $0.83^{+0.07+0.03}_{-0.17-0.18}$ & $1.26^{+0.12+0.09}_{-0.30-0.53}$ & $1.47^{+0.22+0.08}_{-0.17-0.57}$ & $0.94^{+0.38+0.01}_{-0.11-0.33}$ \\
9.90--10.00 & $0.79^{+0.17+0.01}_{-0.12-0.20}$ & $0.92^{+0.29+0.00}_{-0.05-0.31}$ & $0.60^{+0.15+0.03}_{-0.17-0.14}$ & $0.60^{+0.16+0.03}_{-0.06-0.17}$ & $1.21^{+0.22+0.05}_{-0.27-0.52}$ & $1.07^{+0.06+0.10}_{-0.36-0.31}$ & $0.63^{+0.10+0.20}_{-0.23-0.02}$ \\
10.00--10.10 & $0.42^{+0.12+0.00}_{-0.11-0.10}$ & $0.30^{+0.15+0.00}_{-0.06-0.07}$ & $0.29^{+0.18+0.00}_{-0.05-0.12}$ & $0.40^{+0.06+0.00}_{-0.10-0.09}$ & $0.06^{+0.41+0.05}_{-0.00-0.01}$ & $0.06^{+0.12+0.27}_{-0.01-0.00}$ & $0.40^{+0.22+0.00}_{-0.05-0.13}$ \\
10.10--10.15 & $0.71^{+0.31+0.41}_{-0.04-0.03}$ & $0.80^{+0.15+0.16}_{-0.18-0.11}$ & $0.97^{+0.25+0.40}_{-0.09-0.07}$ & $0.78^{+0.12+0.51}_{-0.08-0.02}$ & $1.64^{+0.13+0.95}_{-0.22-0.09}$ & $0.74^{+0.30+1.71}_{-0.04-0.07}$ & $1.55^{+0.17+0.64}_{-0.28-0.04}$ \\
\enddata
\tablecomments{Birth rate parameters are listed with statistical and systematic uncertainties:$b^{+\mathrm{stat}+\mathrm{sys}}_{-\mathrm{stat}-\mathrm{sys}}$}

\label{table:statLMC}
\end{deluxetable*}

\clearpage

\noindent The distribution of the Scylla fields in the LMC were separated enough, and the SFHs of individual fields within those regions are similar enough, that we did not need to search for additional resources to define regions. Across the four regions, we defined seven sub-regions: North East 1 (NE1), North East 2 (NE2), North West 1 (NW1), North West 2 (NW2), South East 1 (SE1), South East 2 (SE2), and South West 1 (SW1). 

\section{Measuring Bursts} \label{sec:Bursts4}

To characterize a burst of star formation in the Clouds, we apply the birthrate parameter, $b = \frac{SFR}{<SFR>}$, as described by \citet{Kennicutt1983,Scalo86}, where $<SFR>$ defines the lifetime average SFR of the population. We adopted a threshold of $b = 2$ for identifying a burst and $b=1$ to demarcate the beginning and end of enhancement periods of star formation following approaches in previous studies e.g., \citet{Kennicutt2005, McQuinn2009, McQuinn2010}.

\begin{figure*}
 	\centering
        \textbf{Burst durations in the SMC and LMC and locations of the SMC local burst}\par\medskip
\includegraphics[width=0.315\textwidth]{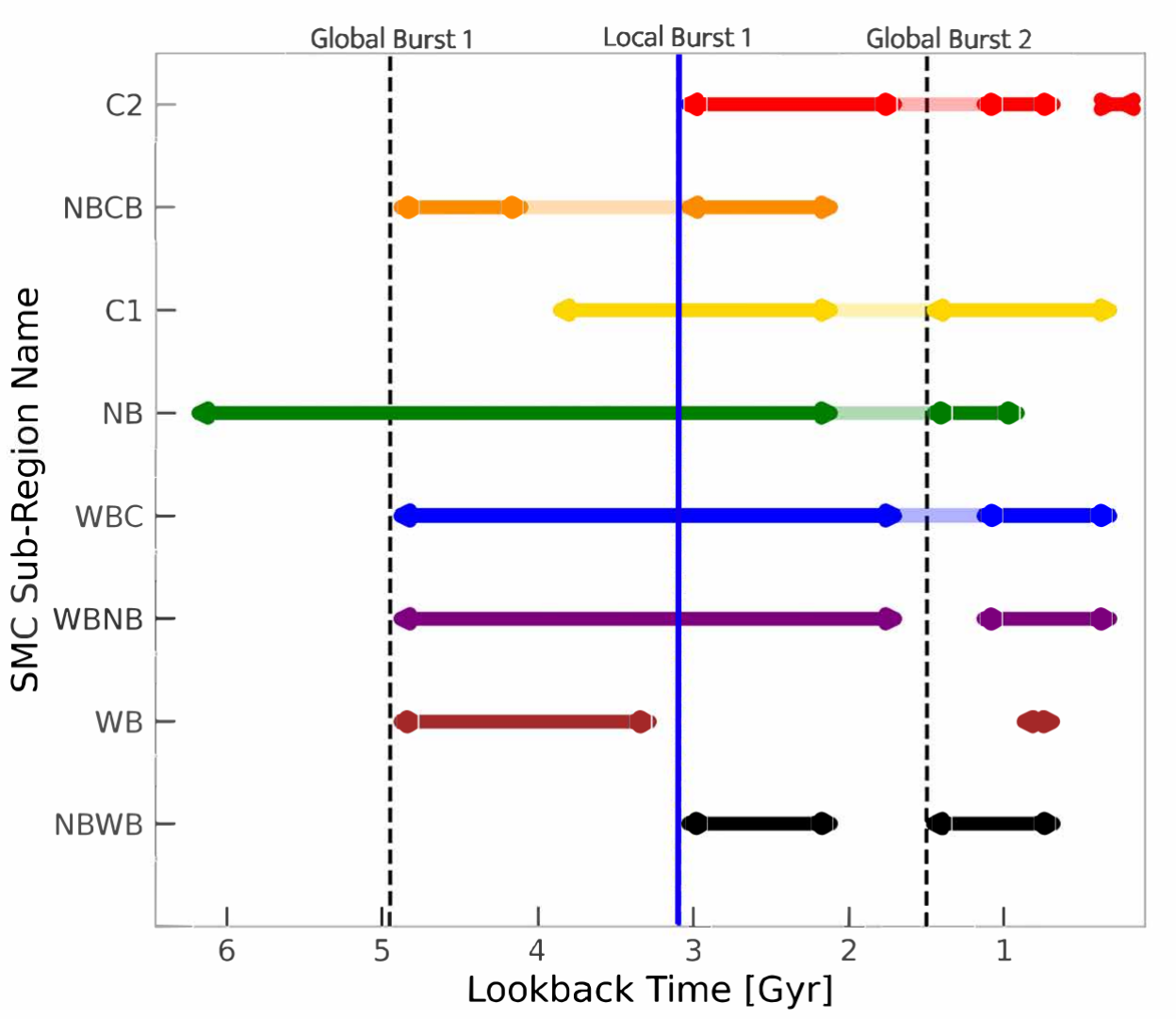}
\includegraphics[width=0.315\textwidth]{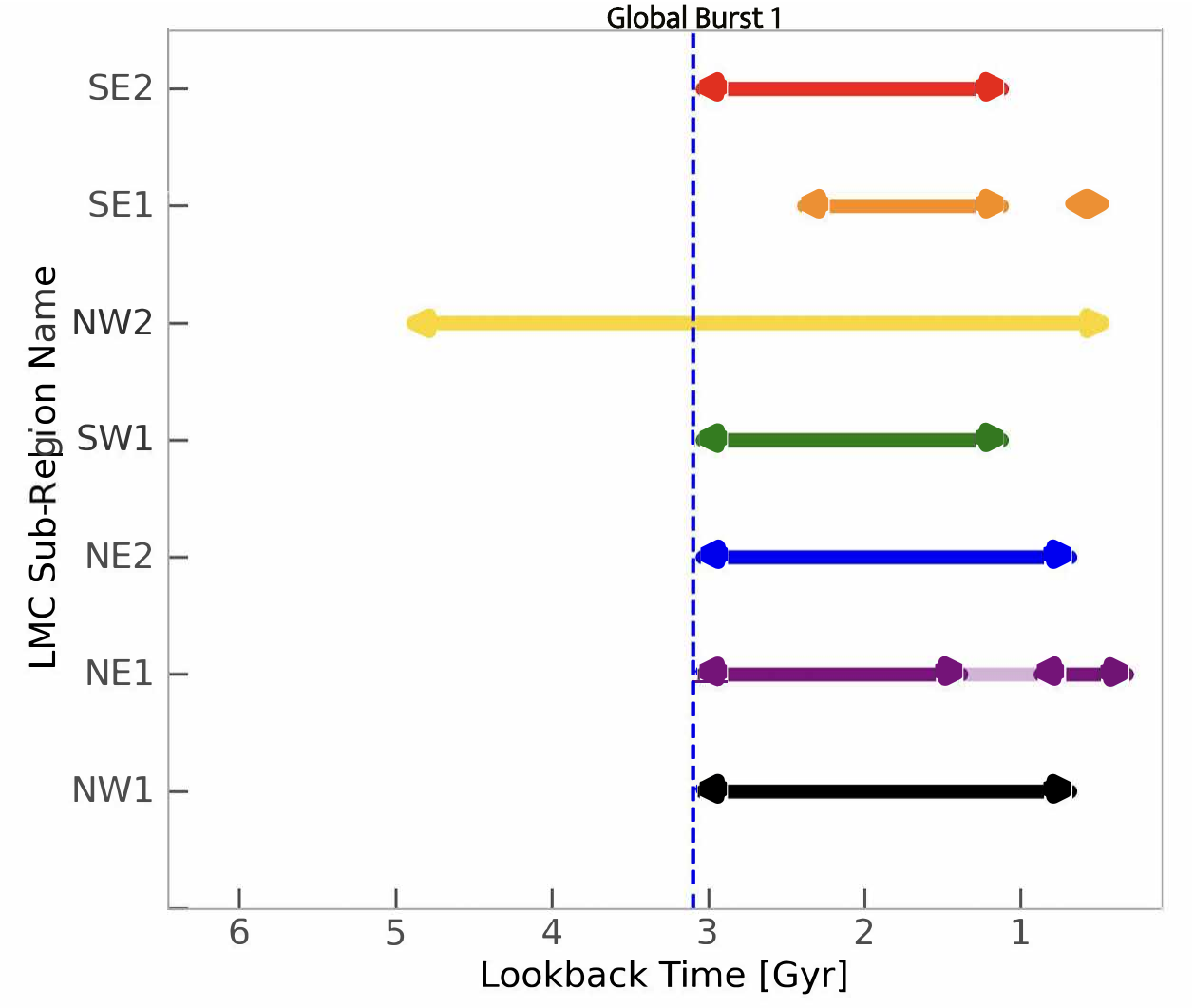}
\includegraphics[width=0.35\textwidth]{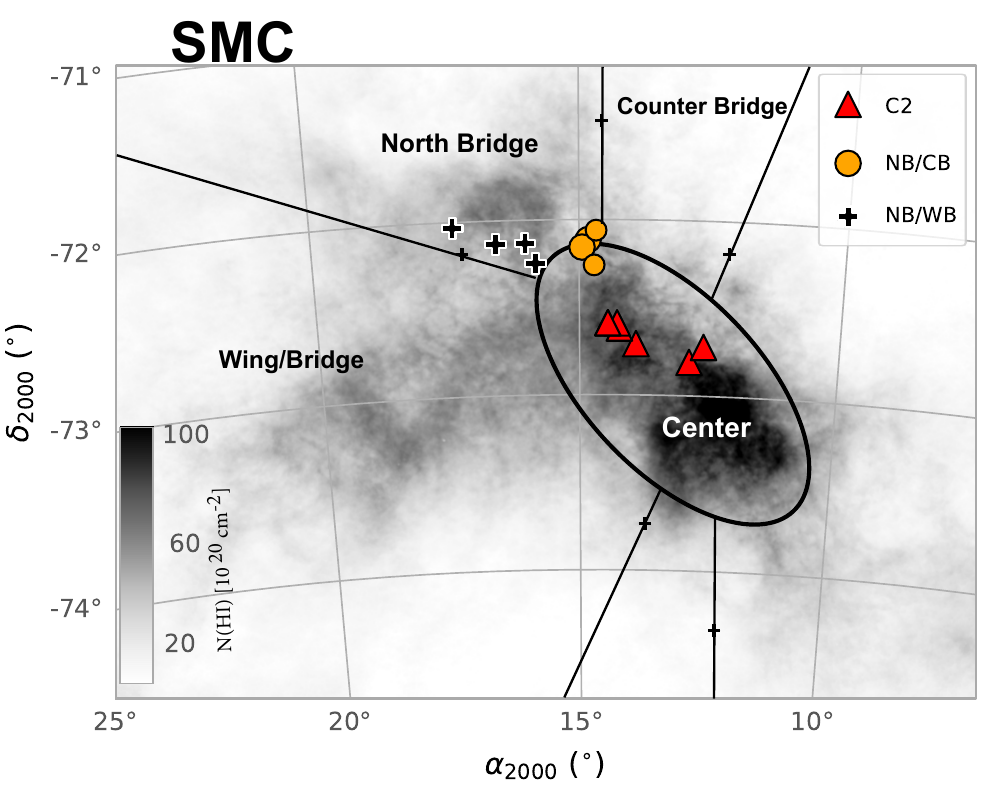}
      \caption{\textbf{Left}: The duration of the bursts as defined by our broad burst metric and modified burst metric for all sub-regions in the SMC.  The horizontal lines are colored by their stellar surface densities (which is the same coloring as their regions in Figure~\ref{fig:maps}). The solid horizontal lines are the burst durations defined by the fine burst metric, and a lightly shaded line connection indicates that the broader burst metric considers the bursts a single event. The vertical lines denote when global (dotted line) or local bursts (solid line) occur.  In the SMC there are two global bursts, at $5$ Gyr and $\sim 1.5$. Gyr, and a local burst at $\sim 3$ Gyr. 
      \textbf{Center}: The same as the left figure, but for sub-regions in the LMC.  In the LMC there is one global burst beginning $ \sim 3$ Gyr ago. 
      \textbf{Right}: The HI map of the SMC with the sub-regions of the Local Burst in the SMC labeled. These are the fields with the burst that coincides with the global burst in the LMC $\sim$ 3 Gyr ago. }
  \label{fig:burstplot}
\end{figure*}

\subsection{Characterizing finer features in bursts} \label{sec:Finebursts4.1}

When applying the birth rate parameter metric and adopting the burst threshold parameters, the bursts we measure can be long ( $> 1$ Gyr) in duration. An example of this can be found in the left-hand plot of Figure \ref{fig:bvalue}, in which we present the SFH of the NBWB sub-region where the burst lasts $\sim2.5$ Gyr. This SFH was area-normalized, performed on a field-by-field basis by summing over the area of all fields per sub-region, assuming a constant pixel/arcsec value of $1.4 \times 10^{-4}$, and dividing each sub-region's SFH by its final area (this is sometimes referred to as a specific SFH ($\Sigma_{SFH}$)).  

Within these long bursts, we identify noticeable variations in the SFR. Although our adopted starburst metric successfully identified the extended period of elevated star formation, the variable SFR within this burst is not captured by the metric. This is partially due to the dependence of \textit{b} on $<SFR>$ over the lifetime of galaxies, which will not effectively identify all variability in the SFR. Thus, we introduce a secondary metric to quantify and characterize changes in star formation activity within a burst. Specifically, we compare the time bins within the longer burst event against the lowest, non-edge \textit{b} within a burst.

We show an example of this secondary metric in Figure~\ref{fig:bvalue}. Because there is a time bin with a \textit{ b} at least two times greater than the lowest, we break the burst into two separate events. In the left panel, the entire SFH of NBWB is shown with the duration of a burst as defined by our broad burst metric. In the right-hand plot, the lowest SFR is highlighted in blue, and the \textit{b} for each bin within the burst is included above its respective bin. Since the bin with the lowest SFR within the burst has a \textit{b} $< 1$, and the bin with the highest SFR within the burst has a \textit{b} = 3.9, which is at least twice as large as the lowest SFR, the original burst is broken into two. This method will be referred to as our fine-burst metric.

\subsection{Identifying bursts of star formation in all sub-regions} \label{sec:Identify4.2}

In Figure \ref{fig:stackedSFH}, we present the SFHs of all sub-regions across both galaxies with the duration of the bursts according to our broad burst metric indicated by vertical black lines. The total uncertainties are plotted for each of the SFHs, and are further broken down into systematic and statistical components, indicated by the saturation of their shading (statistical uncertainties are shaded darker, systematic are shaded lighter). Tables \ref{table:statSMC} and \ref{table:statLMC} give the b-values for each timestep in the SMC and LMC, respectively with the statistical and systematic uncertainties given as $b^{+\mathrm{stat}+\mathrm{sys}}_{-\mathrm{stat}-\mathrm{sys}}$.

The left column of Figure \ref{fig:stackedSFH} contains the area-normalized SFHs for the SMC, the right column shows the area-normalized SFHs for the LMC. The fields are ordered and colored by their star formation surface densities. To enable direct comparison between regions, we fixed the SFR axis to be the same range for the majority of fields. Notable exceptions include C2 in the SMC and SE2 in the LMC, which exhibit higher SFRs, especially in the last 2 Gyrs, and the WB region in the SMC, which has an exceptionally low SFR. 

%the WB where we use a lower range, and C2 and SE2 where we use a higher range. The maximum SFR in the WB sub-region is one-tenth as large than the next lowest maximum in any region in the SMC. 

These plots are useful for direct intragalactic comparisons of the strength of the SFRs across all sub-regions. Comparing increases in the strength of the SFR across sub-regions is a useful first check in identifying whether bursts appear at similar times. Visual inspection alone shows that in the SMC, C1, C2, NBWB, NB, WBC, WBNB had peak SFRs around similar times of $2.5-3.0$ Gyr ago and $1.0-1.5$ Gyr ago. In the LMC, all sub-regions had increases in their SFRs around the same time ($\sim 3$ Gyrs). Additionally, in the LMC we note that the entire Southeast region (SE1, SE2) of the galaxy has the greatest amount of star formation of its four regions. It is worth noting that as these are ULYSES parallel fields, they may be preferentially clustered near star-forming regions and thus not fully representative of Southeast LMC generally. This aside, the results support findings from SMASH, which show that the Southern portion of the LMC experiences bursts more than double the intensity of the Northern portion of the galaxy for a duration of $\sim 0.5$ Gyrs $\sim 3$ Gyr ago  \citep{massana2022}.

\subsubsection{Measuring burst durations} \label{sec:MeasureBurst4.2.1}

We applied our burst metrics to the sub-regions in SMC and LMC, and report our measurements in Table \ref{table:SMCburst} with a visual representation in Figure \ref{fig:burstplot}. Figure \ref{fig:burstplot} highlights the duration of the bursts for all sub-regions to allow for intragalactic and intergalactic comparisons. A lighter shading is used to indicate that the burst has been identified as a single event when we only apply our broad burst metric (a burst must contain at least one bin of $b\geq 2$, begin with a bin of $b\geq 1$, end with a bin of $b\leq 1$), and as two separate events by our fine-burst metric (if a non-edge bin within a burst has $b\geq 2$, we break the burst into two separate events). These plots focus exclusively on the duration of the bursts and do not contain information about the change in star formation intensity over time, although that information is available in Figure \ref{fig:burstp20}.

% The SMC is on the left and the LMC is on the right, with the name of the sub-region on the y-axis and the timing and duration of their respective burst(s) on the x-axis.

Figure~\ref{fig:burstplot} shows that some bursts begin at nearly the same time across most sub-regions of each galaxy, while others appear only in a few areas. To capture these widespread features, we define `global burst events' as bursts that start at roughly the same time in $\ge$ half of sub-regions within a galaxy. Identifying these broader events allows for meaningful comparisons both between the two galaxies (see Section \ref{sec:MeasurDelay4.3}) and with dynamical models of their histories (see Section \ref{sec:Trigger5.1}). In Figure~\ref{fig:burstplot}, vertical dashed lines mark the start times of these global burst events.

In the SMC, we find two such global bursts. The most recent, which we call ``SMC Burst 2”, started around 1.5 Gyr ago and appears as a coeval gap in star formation in six out of eight sub-regions. The oldest, ``SMC Burst 1", started about 5 Gyr ago and is also present in four of the eight sub-regions. In contrast, the LMC shows only one global burst event, ``LMC Burst 1”, which began roughly 3 Gyr ago and is observed in six out of seven sub-regions.

Some sub-regions do not closely follow these global burst timings. We refer to those whose burst timings differ by several billion years as strong outliers; examples include the C1 and NB regions in the SMC, and NW2 in the LMC. Other sub-regions differ by less than 0.5 Gyr from the global burst timings, such as NBCB in the SMC and SE1 in the LMC, and are considered broadly consistent with the global bursts.

\begin{figure*}
 	\centering
        \textbf{Mass formed per burst across all sub-regions in the SMC and LMC}\par\medskip
\includegraphics[width=0.475\textwidth]{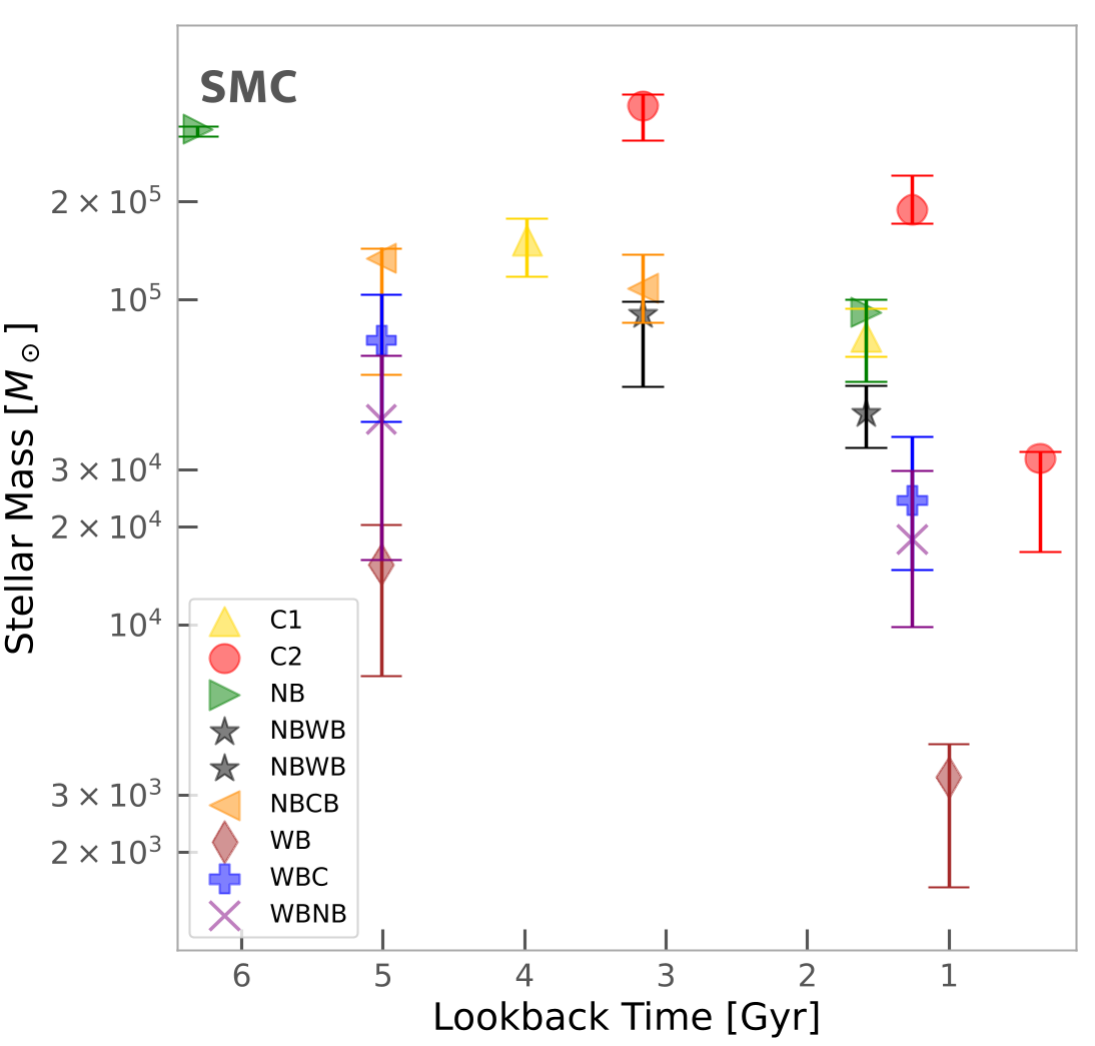}
\includegraphics[width=0.4425\textwidth]{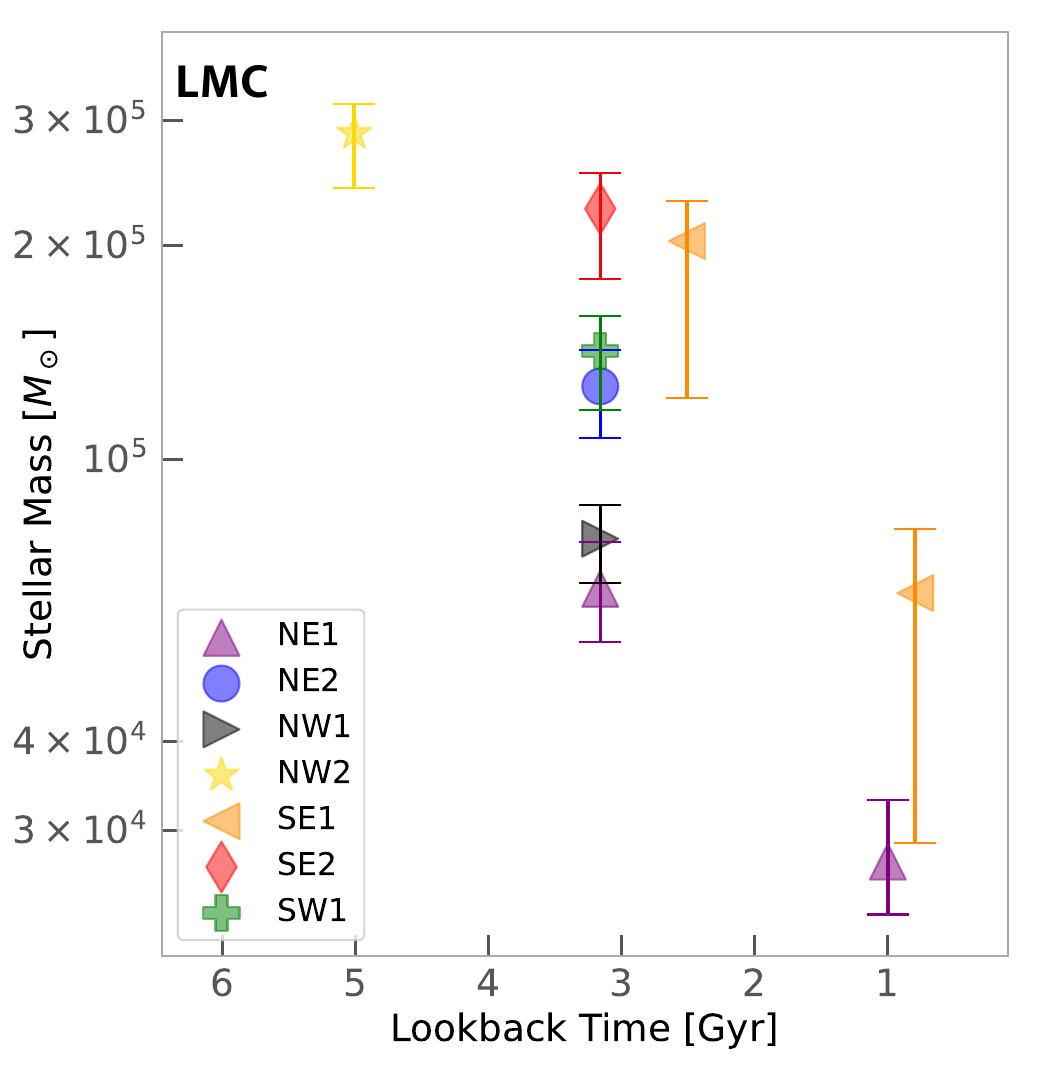}
	  \caption{The mass produced per burst according to our fine-burst metric (the solid-colored lines in Figure \ref{fig:burstplot}) with errors in the SMC and LMC. Markers are placed at the beginning of bursts. Note the y-axis scale for the SMC plot spans an order of magnitude more than the LMC plot.}
  \label{fig:mass}
\end{figure*}

We define `local burst events' as the time at which a burst begins across a minority of sub-regions in either galaxy. This metric is particularly useful for local bursts that occur in sub-regions with higher surface densities, as these bursts could be overpowering in a global SFH and be mistaken as an event that occurs across all parts of the galaxy. We find one such event in the SMC around 3 Gyr ago, which coincides with LMC Burst 1.

\subsubsection{Stellar mass formed in each burst event per sub-region} \label{sec:MeasureMass4.2.2}

Comparing the stellar mass formed during bursts across different sub-regions provides a way to assess how star formation activity varies spatially within the SMC and LMC. To make meaningful comparisons, we focus on bursts that occurred at similar times, which reduces biases from differing time resolutions and bin width, and ensures that we are analyzing physically comparable star formation events.

Figure \ref{fig:mass} presents the stellar mass formed within each burst, per sub-region, plotted against the burst onset time, with marker colors indicating the stellar surface density of each sub-region (these are the same as the coloring of the fields in Figure~\ref{fig:maps}). While the WB sub-region stands out as an outlier with both a significantly lower star formation rate and burst mass (see Section \ref{sec:WBdisagree}), the remaining bursts in the SMC and LMC span a similar mass range between approximately $10^{4.4}$ and $10^{5.5} M_{\odot}$. For instance, sub-regions in the SMC with bursts at the time of SMC Burst 1 (such as C1, WBC, and WBNB) produce similar stellar masses within the uncertainties, despite differences in surface density.

However, interpreting these comparisons requires accounting for stellar mixing: the gradual dispersal of stars from their birthplaces into the broader galaxy. Previous works estimate that in dwarf galaxies, this mixing occurs over timescales of roughly 100 to 350 Myr \citep{G2011}, with the SMC and LMC showing that substructure blends into the background population after about 75 and 175 Myr, respectively \citep{G08, G09}. This mixing limits our ability to associate older star formation events with their original sub-regions.

Despite these limitations, the similar stellar masses formed 3 Gyr ago in two distinct SMC sub-regions (NBCB and NBWB) and across the LMC, support the interpretation that bursts such as SMC Burst 1, SMC Burst 2, and LMC Burst 1 represent global star formation events. This finding implies that stellar populations in both galaxies are well mixed on billion-year timescales, with burst timing largely independent of stars’ present-day locations.

\subsection{Burst timing agreement within and between the galaxies} \label{sec:RegionalAgreement4.3.1}

In Figure~\ref{fig:cumulativeSFHs}, we compare the cumulative SFHs for different regions within and across both galaxies. Shown are the regional SFHs for the SMC (left) and the LMC (middle), and the combined SFH for all regions in both galaxies (right). The SMC regions are a modified version of those shown in Figure~\ref{fig:maps}. Specifically, we split the Wing/Bridge area into two components: the fields closest to the SMC center and the more isolated outer fields. This separation was made because the isolated Wing/Bridge fields show distinct SFHs and burst timing compared to the rest of the region. For clarity, these are treated as separate regions in this figure. The LMC regions follow the same definitions as in Figure~\ref{fig:maps}: North West, North East, South West, and South East.  Only statistical uncertainties are shown.

The cumulative SFHs of all sub-regions within the SMC are all quite similar, with the exception of the Wing/Bridge. The difference in the SFH of this section of the galaxy relative to the rest of the galaxy has 

 % We as we cannot be sure if a star's current position is where it was born in its galaxy,

% Given that the burst durations cannot be presented with errors for their SFRs, we include errors Table \ref{table:SMCburst}. Additionally, the errors for the mass produced per burst are directly proportional to the errors of their SFRs, becuase the stellar mass is calculated as the SFR$\times$time (given in the right column of Figure \ref{fig:burstplot}).

\newpage
\clearpage
\startlongtable
\begin{deluxetable*}{lllcccccc}
\tabletypesize{\scriptsize}
\tablecaption{Burst duration and mass produced per burst in SMC sub-regions \label{obstab}}
\tablehead{\colhead{Sub-region} & \colhead{Field Number} & \colhead{Field I.D.} & \colhead{Burst 1} & \colhead{Burst 1 Mass} & \colhead{Burst 2} & \colhead{Burst 2 Mass} & \colhead{Burst 3} & \colhead{Burst 3 Mass}\\ \colhead{} & \colhead{} & \colhead{} &  \colhead{[Gyr]} & \colhead{[M$_{\odot}$]} & \colhead{[Gyr]} & \colhead{[M$_{\odot}$]} & \colhead{[Gyr]} & \colhead{[M$_{\odot}$]}}
\startdata
 & SMC$\_${9} & 15891$\_$SMC-3587ne-10112  &  &  &  \\
 & SMC$\_${10} & 15891$\_$SMC-3149ne-12269  &  &  &  \\
North Bridge & SMC\_${16}$ & 15891\_SMC-3435ne-13258 &  6.3 - 2.0 & $3.4_{-5.6e4}^{+4.9e4} \times 10^{5}$ & 1.6 - 0.8 & $9.2_{-1.3e4}^{+1.3e4} \times 10^{4}$ & &  \\
 & SMC$\_${20} & 15891$\_$SMC-3370ne-13459 &  &  &  \\
 & SMC$\_${21} & 15891$\_$SMC-3104ne-13781 &  &  &  \\
 \hline
 & SMC$\_${12} & 15891$\_$SMC-3025ne-13499 &  &  &  \\
 % & SMC$\_${13} & 15891$\_$SMC-2983ne-12972 & 18.96 & (9.7,$1.34e5$) & (9.5,$1.08e5$) & -- \\
North Bridge/ & SMC\_${19}$ & 15891\_SMC-3032ne-13306 & 5.0 - 3.2 & $1.34_{-7.54e4}^{+1.01e4} \times 10^{5}$ & 3.2 - 2.0 & $1.9_{-2.3e4}^{+2.9e4} \times 10^{5}$ \\
Counter Bridge & SMC$\_${23} & 15891$\_$SMC-2584ne-14274 &  &  &  \\
 & SMC$\_${37} & 16235$\_$SMC-3154ne-32442 &  &  &  \\
 \hline
 % & SMC$\_${6} & 15891$\_$SMC-3956ne-9632 & 18.86 &  &  &  \\
 & SMC$\_${8} & 15891$\_$SMC-3955ne-9818 &  &  &  &  & & \\
North Bridge/ & SMC\_${50}$ & 16786\_SMC-4451ne-16362 & 3.2 - 2.0 & $9.0_{-3.6e4}^{+8.7e3} \times 10^{4}$ & 1.3-0.6 & $4.5_{-9.6e3}^{+9.7e3} \times 10^{4}$  \\
Wing/Bridge & SMC$\_${54} & 16786$\_$SMC-3529ne-15172 &  &  &  \\
 & SMC$\_${55} & 16786$\_$SMC-5409ne-15524 &  &  &  \\
 \hline
 & SMC$\_${11} & 15891$\_$SMC-8743se-11371 &  &  &  \\
 & SMC$\_${22} & 15891$\_$SMC-9034se-13316 &  &  &  \\
Wing/Bridge & SMC\_${28}$ & 16235\_SMC-10336se-14099 & 5.0 - 3.2 & $1.5_{-8.4e3}^{+4.9e3} \times 10^{4}$ & 1.0 - 0.6 & $3.4_{-1.8e3}^{+9.1e2} \times 10^{4}$\\
 & SMC$\_${38} & 16235$\_$SMC-8151se-32530 &  &  &  \\
 & SMC$\_${47} & 16786$\_$SMC-9277se-14900 &  &  &  \\
 & SMC$\_${52} & 16786$\_$SMC-9946se-16175 &  &  &  \\
 \hline
 & SMC$\_${5} & 15891$\_$SMC-3514se-8584 &  &  &  \\
Wing/Bridge/ & SMC\_${18}$ & 15891\_SMC-3029ne-13288 & 5.0 - 1.6 & $4.3_{-3.3e4}^{+2.9e4} \times 10^{4}$ & 1.3 - 0.2 & $1.8_{-9.4e4}^{+1.4e4} \times 10^{4}$ & & \\
Center & SMC$\_${34} & 16235$\_$SMC-2728ne-28918 &  &  &  \\
 & SMC$\_${53} & 16786$\_$SMC-4745se-7610 &  &  &  \\
 \hline
 % & SMC$\_${14} & 15891$\_$SMC-3669ne-13972 & 18.96 &  &  &  \\
Wing/Bridge/ & SMC\_${29}$ & 16235\_SMC-3870ne-14647 & 5.0 - 1.6 & $7.5_{-2.7e4}^{+2.4e4} \times 10^{4}$ & 1.3 - 0.2 & $2.4_{-8.5e3}^{+1.1e4} \times 10^{4}$ & & \\
North Bridge & SMC$\_${43} & 16235$\_$SMC-4996ne-34726 &  &  &  \\
 & SMC$\_${46} & 16235$\_$SMC-4450ne-32733 &  &  &  \\
 \hline
 & SMC$\_${2} & 15891$\_$SMC-2750ne-8567 &  &  &  \\
Center 1 & SMC\_${30}$ & 16235\_SMC-2259ne-15609 &  4.0 - 2.0 & $1.52_{-3.5e4}^{+2.5e4} \times 10^{5}$ & 1.6 - 0.2 & $7.7_{-9.8e3}^{+1.7e4} \times 10^{4}$ & & \\
 % & SMC$\_${31} & 16235$\_$SMC-2272ne-15308 & 19.11 &  &  &  \\
 & SMC$\_${33} & 16235$\_$SMC-2167ne-18821 &  &  &  \\
 \hline
 & SMC$\_${17} & 15891$\_$SMC-1588ne-12105 &  &  &  \\
 & SMC$\_${25} & 16235$\_$SMC-879ne-11082 &  &  &  \\
  & SMC$\_${40} & 16235$\_$SMC-1339ne-33009 &  &  &  \\
Center 2 & SMC\_${41}$ & 16235\_SMC-286sw-34349 & 3.2 - 1.6 & $4.0_{-8.6e4}^{+3.3e4} \times 10^{5}$ & 1.3 - 0.6 & $1.9_{-1.8e4}^{+5.2e4} \times 10^{5}$ & 0.4 - 0.2 & $3.3_{-3.5e4}^{+2.5e4} \times 10^{4}$\\
 & SMC$\_${42} & 16235$\_$SMC-1443ne-34945 &  &  &  \\
 & SMC$\_${45} & 15891$\_$SMC-641nw-12753 &   &  &  \\
\enddata
\tablecomments{A list of all the sub-regions in the SMC. For each sub-region we list the Scylla name file and field number of the constituent fields. Each field also has a list of burst timings, the mass produced per burst.}

\label{table:SMCburst}
\end{deluxetable*}

\clearpage

\newpage

\clearpage

\startlongtable
\begin{deluxetable*}{lllcccc}
\tabletypesize{\scriptsize}
\tablecaption{Burst duration and mass produced per burst in LMC sub-regions \label{obstab}}
\tablehead{\colhead{Sub-region} & \colhead{Field Number} & \colhead{Field I.D.} & \colhead{Burst 1} & \colhead{Burst 1 Mass} & \colhead{Burst 2} & \colhead{Burst 2 Mass}\\ \colhead{} & \colhead{} & \colhead{} &  \colhead{[Gyr]} & \colhead{[M$_{\odot}$]} & \colhead{[Gyr]} & \colhead{[M$_{\odot}$]}}
\startdata
 & LMC$\_${10} & 15891$\_$LMC-9446ne-10765 &  \\
 & LMC$\_${17} & 15891$\_$LMC-11384ne-12295 &  \\
Northeast 1 & LMC$\_{40}$ & 16235$\_$LMC-10728ne-8437 & 3.2 - 1.3 & $6.6_{-1.1e4}^{+1.1e4} \times 10^{4}$ & 1.0 - 0.2 & $2.7_{-4.3e3}^{+6.0e3} \times 10^{4}$ \\
 & LMC$\_${49} & 16786$\_$LMC-12311ne-5715 &  \\
 & LMC$\_${50} & 16786$\_$LMC-12141ne-5771 &  \\
 & LMC$\_${57} & 16786$\_$LMC-13556ne-15380 &    \\
 \hline
 & LMC$\_${2} & 15891$\_$LMC-9617ne-5147 &   \\
Northeast 2 & LMC$\_${15} & 15891$\_$LMC-7454ne-11865 & 3.2 - 0.6  & $1.3_{-2.0e4}^{+1.6e4} \times 10^{5}$ &  \\
 % & LMC$\_${30} & 16235$\_$LMC-9740nw-7508 & 18.59 &  &  &  \\
 % & LMC$\_${37} & 16235$\_$LMC-14421nw-26822 & 18.54 &  &  &  \\
 & LMC$\_${45} & 16786$\_$LMC-10253ne-6545 &  \\
 \hline
 & LMC$\_${1} & 15891$\_$LMC-15629nw-4948 &  \\
 & LMC$\_${8} & 15891$\_$LMC-15413nw-9621 & \\
Northwest 1 & LMC$\_{32}$ & 16235$\_$LMC-7623sw-22524 & 3.2 - 0.6  & $7.7_{-1.0e4}^{+8.9e4} \times 10^{4}$ &  \\
 & LMC$\_${41} & 16235$\_$LMC-17892nw-9532 &    \\
 & LMC$\_${42} & 16786$\_$LMC-15342nw-2460 &  \\
 \hline
 & LMC$\_${9} & 15891$\_$LMC-12315nw-11221 &  \\
 & LMC$\_${16} & 15891$\_$LMC-11456nw-12627 &   \\
 & LMC$\_${20} & 15891$\_$LMC-9679nw-13399 &  \\
Northwest 2 & LMC$\_{24}$ & 15891$\_$LMC-9690nw-13623 &  5.0 - 0.4  & $2.9_{-4.7e4}^{+2.9e4} \times 10^{5}$ & -- \\
 & LMC$\_${35} & 16235$\_$LMC-8599nw-23221 &   \\
 & LMC$\_${39} & 16786$\_$LMC-10028nw-33586 &    \\
 & LMC$\_${59} & 16786$\_$LMC-12269nw-24827 &   \\
 \hline
 & LMC$\_${11} & 15891$\_$LMC-5389ne-11134 &   \\
 & LMC$\_${27} & 16235$\_$LMC-4958ne-31479 &   \\
Southeast 1 & LMC$\_{48}$ & 16786$\_$LMC-5127ne-3118 &  2.5 - 1.0  & $2.0_{-8.1e4}^{+2.8e4} \times 10^{5}$ & 0.79 - 0.35  & $6.48_{-3.60e4}^{+1.50e4} \times 10^{4}$\\
 & LMC$\_${54} & 16786$\_$LMC-5943ne-9430 &  \\
 & LMC$\_${55} & 16786$\_$LMC-5850ne-10777 & \\
 & LMC$\_${56} & 16786$\_$LMC-5199ne-23482 &  \\
 \hline
 & LMC$\_${4} & 15891$\_$LMC-3610se-7920 &   \\
 & LMC$\_${5} & 15891$\_$LMC-5442ne-8000 &   \\
 Southeast 2 & LMC$\_{7}$ & 15891$\_$LMC-5619ne-9411 &  3.2 - 1.0  & $2.3_{-4.6e4}^{+2.8e4} \times 10^{5}$ \\
 % & LMC$\_${22} & 15891$\_$LMC-5421ne-12728 & 18.54 &  &  &  \\
 & LMC$\_${36} & 16235$\_$LMC-4763ne-26440 &  \\
 & LMC$\_${53} & 16786$\_$LMC-5045ne-18484 &  \\
 \hline
 & LMC$\_${21} & 15891$\_$LMC-8532sw-13647 &   \\
 & LMC$\_${29} & 16235$\_$LMC-5812sw-7744 &  \\
Southwest 1 & LMC$\_{32}$ & 16235$\_$LMC-7623sw-22524 &  3.2 - 1.0 & $1.4_{-2.5e4}^{+1.7e4} \times 10^{5}$ &  \\
 & LMC$\_${33} & 16235$\_$LMC-7234sw-22225 &  \\
 & LMC$\_${43} & 16786$\_$LMC-6222sw-15490 &   \\
\enddata
\tablecomments{A list of all the sub-regions in the LMC. For each sub-region we list the Scylla name file and field number of the constituent fields. Each field also has a list of burst timings, the mass produced per burst.}

\label{table:LMCbursts}
\end{deluxetable*}

\clearpage

\noindent been well characterized in other works \citep{nidever2014, dias2016, Cohen2024b}, and our cumulative SFHs agree well with these previous studies. Within the LMC, we find differences in the cumulative SFHs of the Northern and Southern areas.

The North East and North West regions agree with each other, and the South East and South West regions agree with each other. However, the South assembles a higher percentage of its mass earlier, with a large increase $\simeq10$ Gyr ago, while the North experiences more constant growth until around $\simeq3$ Gyr ago, at which point all regions have a steep and similar increase. When comparing these larger regions, the Wing/Bridge's cumulative SFH falls within the errors of the cumulative SFHs of the South East and South West regions of the LMC (discussed further in Section \ref{sec:WBdisagree}). 

\subsubsection{Measuring delays in Burst Timing between the galaxies} \label{sec:MeasurDelay4.3}

We compared the beginning of the bursts across both galaxies using their cumulative SFHs with the aim of better characterizing the onset of Burst 1 in the LMC and SMC. In the right panel of Figure \ref{fig:cumulativeSFHs}, we present the cumulative SFHs of the SMC and LMC with the total systematic and statistical uncertainties. The cumulative SFHs of the two galaxies are different from $\sim 8$ to $\sim2.5$ Gyr ago even within uncertainties. The SMC experienced an accelerated growth period from $\sim6$ Gyr ago to $\sim2.5$ Gyr ago, at which point the LMC and SMC had assembled roughly the same fraction of their masses, $\sim 75\%$.

The timing and intensity of the bursts noted in Figure \ref{fig:burstplot} can, of course, be seen as changes in slope in the cumulative SFH plots in Figure \ref{fig:cumulativeSFHs}. We quantify the degree of the enhancement by measuring the change in the slopes of the cumulative distributions. Specifically, a positive (negative) slope indicates that the SFR is increasing (decreasing), while a flat slope indicates that little star formation is occurring; the slope value constrains the intensity.  Table \ref{table:cumulativeSFH} includes the values of the SFR, slope of the cumulative SFH, and direction of change in the slope of the cumulative SFH at each time interval in our SFH solutions.

Specifically, we note that the SMC experienced its largest slope change from $5.0 - 4.0$ Gyr ago and $3.2 - 2.5$ Gyr ago, excluding the two most recent time bins, which were not the focus of this burst analysis. The beginning of this increase coincides with SMC Burst 1. The slope of the cumulative SFH in the SMC doubles from $5.0 - 4$ Gyr ago and the slope more than doubles from $3.2 - 2.5$ Gyr ago. This indicates a significant increase in the SFR intensity, corresponding to $98\%$ and $110\%$ increases, respectively. An enhancement that roughly doubles the SFR is consistent with relatively close interacting dwarf pairs, as observed in \cite{Stierwalt2015}. 

In the LMC, the slope of the cumulative SFH increases from 3.2 - 2.5 Gyr ago, a $46\%$ increase. This corresponds with LMC Burst 1, and is the LMC's largest increase in slope within the timeframe of interest.

% Additionally, we note that the Northern regions of the LMC more closely follow the rate of increase in star formation of the North Bridge, Center/Wing Bridge, and Central regions of the SMC and the Southern regions of the LMC. Observations from SMASH have found correlated SMC-LMC SFHs in the Northern region of the LMC and the entirety of the SMC \citep{massana2022}. Our results are in line with these observations. 

The SMC and LMC experienced their largest increases in SFR during the beginning of SMC Burst 1 and LMC Burst 1. If these events were related, i.e., if the burstiness in both galaxies was a product of the same event, then an increase in SFR in the SMC began in response to that trigger before the LMC and also responded more strongly than the LMC. This is discussed further in Section \ref{sec:Trigger5.1}.

\section{Trigger Mechanism for Burst} \label{sec:Trigger5.1}

Due to the widespread global agreement between bursts within the galaxies and the occurrence of bursts between the galaxies, the data indicate that bursts across both galaxies could share a trigger as the impetus for their star formation. \citet{massana2022} draws a similar conclusion by identifying bursty periods in the Clouds to assert that the galaxies have been mutually influencing each other for at least 3.5 Gyr. A likely candidate for this trigger mechanism is a dynamical interaction between the galaxies, which has been shown to drive increases in star formation activity in galaxies over a range of masses. In a study of 1500 nearby low- and intermediate-mass galaxies ($10^8$ - $10^{11} M_{\odot}$), \citet{Knapen_2015} found that the area-normalized SFR is enhanced in interacting galaxies by up to $\sim 2\times$ for merging systems, although this does change as a function of mass. \citet{Sun_2020} found that for low-mass, nearby pairs between $0.25 \le M_{\rm neighbor}/M_{\rm candidate} \le 4$  for $M = 10^8 - 10^{10} M_{\odot}$, the SFR relative to isolated control galaxies was enhanced $\sim 1.75\times$ near the inner regions and decreased with increasing galactic radii. Previous work done by the TNT program used observations to characterize how interactions between paired dwarfs affect the SFH of the component galaxies \citep{Stierwalt2015}. \citet{Stierwalt2015} found a SFR enhancement of ~$2 $-$ 2.75 \times$ the mean SFR for members of non-isolated dwarf galaxy pairs (those that are found within 1.5 Mpc of a massive host) separated by roughly 10 kpc.  The dwarf pairs in the TNT sample are slightly less massive than the SMC-LMC binary, but the MC system still fits within the overall trend of the data.

\begin{deluxetable*}{ll|ccc|ccc}
\tablecaption{SFRs \& Cumulative SFH slopes per time bin in the S/LMC \label{table:cumulativeSFH}}
\tablehead{
\colhead{} & \colhead{} & \multicolumn{3}{c}{\textbf{SMC}} & \multicolumn{3}{c}{\textbf{LMC}} \\
\colhead{Time Step} & \colhead{Time Step} & \colhead{SFR} & \colhead{Slope} & \colhead{Change of Slope} & 
\colhead{SFR} & \colhead{Slope} & \colhead{Change of Slope} \\
\colhead{log(Age)} & \colhead{Age [Gyr]} & \colhead{[$M_{\odot}$ yr$^{-1}$]} & \colhead{[$M_{\odot}$ yr$^{-1}$]} & \colhead{} &
\colhead{[$M_{\odot}$ yr$^{-1}$]} & \colhead{[$M_{\odot}$ yr$^{-1}$]} & \colhead{}
}
\startdata
10.15 - 10.10 & 14.13 - 12.59 & $1.52_{-3.52e-5}^{+3.59e-4} \times 10^{-4}$ & 0.032 & 0 & $3.06_{-3.68e-7}^{+2.77e-4} \times 10^{-4}$ & 0.078 & + \\
10.10 - 10.0  & 12.59 - 10.0  & $1.52_{-6.73e-5}^{+2.43e-5} \times 10^{-4}$ & 0.032 & 0 & $5.51_{-3.00e-8}^{+5.29e-5} \times 10^{-5}$ & 0.014 & - \\
10.0 - 9.9    & 10.0 - 7.94   & $2.23_{-9.33e-5}^{+9.78e-6} \times 10^{-4}$ & 0.048 & + & $2.71_{-1.19e-4}^{+3.60e-7} \times 10^{-4}$ & 0.069 & + \\
9.9 - 9.8     & 7.94 - 6.31   & $1.56_{-1.65e-5}^{+8.12e-5} \times 10^{-4}$ & 0.033 & - & $3.22_{-1.28e-4}^{+1.08e-5} \times 10^{-4}$ & 0.082 & + \\
9.8 - 9.7     & 6.31 - 5.01   & $2.89_{-2.92e-5}^{+8.00e-5} \times 10^{-4}$ & 0.062 & + & $2.37_{-2.55e-5}^{+6.03e-5} \times 10^{-4}$ & 0.060 & - \\
9.7 - 9.6     & 5.01 - 3.98   & $5.78_{-1.86e-4}^{+2.07e-5} \times 10^{-4}$ & 0.123 & + & $2.96_{-8.10e-5}^{+3.20e-5} \times 10^{-4}$ & 0.075 & + \\
9.6 - 9.5     & 3.98 - 3.16   & $4.17_{-2.47e-5}^{+1.21e-4} \times 10^{-4}$ & 0.089 & - & $2.72_{-3.62e-5}^{+5.10e-5} \times 10^{-4}$ & 0.069 & - \\
9.5 - 9.4     & 3.16 - 2.51   & $8.75_{-2.12e-4}^{+6.96e-5} \times 10^{-4}$ & 0.187 & + & $3.94_{-1.17e-4}^{+1.35e-5} \times 10^{-4}$ & 0.101 & + \\
9.4 - 9.3     & 2.51 - 2.0    & $6.71_{-1.66e-4}^{+4.62e-5} \times 10^{-4}$ & 0.143 & - & $3.94_{-1.08e-4}^{+5.68e-5} \times 10^{-4}$ & 0.100 & 0 \\
9.3 - 9.2     & 2.0 - 1.58    & $4.28_{-5.19e-5}^{+4.28e-5} \times 10^{-4}$ & 0.091 & - & $4.61_{-1.70e-4}^{+4.37e-5} \times 10^{-4}$ & 0.118 & + \\
9.2 - 9.1     & 1.58 - 1.26   & $5.10_{-7.88e-5}^{+5.75e-5} \times 10^{-4}$ & 0.109 & + & $5.88_{-2.22e-4}^{+6.88e-5} \times 10^{-4}$ & 0.150 & + \\
9.1 - 9.0     & 1.26 - 1.0    & $5.06_{-6.50e-5}^{+1.08e-4} \times 10^{-4}$ & 0.108 & 0 & $5.42_{-1.03e-4}^{+5.22e-5} \times 10^{-4}$ & 0.139 & - \\
9.0 - 8.90     & 1.00 - 0.79  & $6.78_{-2.37e-5}^{+2.00e-4} \times 10^{-4}$ & 0.245 & - & $3.22_{-4.87e-5}^{+9.09e-5} \times 10^{-5}$ & 0.090 & - \\
8.90 - 8.75     & 0.79 - 0.56  & $6.79_{-5.63e-5}^{+3.32e-4} \times 10^{-4}$ & 0.145 & - & $4.52_{-1.24e-4}^{+1.77e-5} \times 10^{-5}$ & 0.115 & + \\
8.75 - 8.55     & 0.56 - 0.35  & $3.51_{-3.95e-5}^{+2.56e-4} \times 10^{-4}$ & 0.075 & - & $3.70_{-1.02e-4}^{+1.35e-4} \times 10^{-5}$ & 0.094 & - \\
8.55 - 8.3     & 0.35 - 0.2  & $5.69_{-8.23e-5}^{+8.13e-4} \times 10^{-4}$ & 0.121 & + & $2.37_{-1.05e-4}^{+0.00e-0} \times 10^{-5}$ & 0.061 & - \\
8.3 - 7.8     & 0.20 - 0.063  & $3.28_{-7.11e-5}^{+1.05e-4} \times 10^{-4}$ & 0.069 & - & $3.90_{-1.26e-5}^{+3.33e-5} \times 10^{-5}$ & 0.010 & - \\
7.8 - 7.2     & 0.063 - 0.016 & $2.04_{-1.20e-3}^{+6.20e-7} \times 10^{-3}$ & 0.436 & + & $9.04_{-3.85e-4}^{+3.34e-7} \times 10^{-4}$ & 0.231 & + \\
7.2 - 6.6     & 0.016 - 0.004 & $3.41_{-5.96e-4}^{+0.0e-0} \times 10^{-3}$ & 0.733 & + & $6.24_{-2.20e-7}^{+1.19e-2} \times 10^{-3}$ & 1.592 & + \\
\enddata

\tablecomments{The SFR per timestep, the slope per timestep of the cumulative SFH, and an indication of whether the change in slope
from the previous timestep to the current was positive, negative, or zero for the SMC on the left and the LMC on the right.}
\end{deluxetable*}

%\startlongtable
\begin{deluxetable*}{lccccccccccc}
\tabletypesize{\scriptsize}
\tablecaption{Timing of Dynamic Events \label{table:p20}}
\tablehead{\colhead{} & \colhead{MW Mass} &  \colhead{Peri 1} & \colhead{Peri 1 range} & \colhead{Peri 2} & \colhead{Peri 2 range} & \colhead{SMC Infall} & \colhead{SMC Infall range} & \colhead{LMC Infall} & \colhead{LMC Infall range}
\\ \colhead{} & \colhead{M$_{\odot}$} & \colhead{Gyr} & \colhead{Gyr} & \colhead{Gyr} & \colhead{Gyr} & \colhead{Gyr} & \colhead{Gyr} & \colhead{Gyr} & \colhead{Gyr} }
\startdata
Model 1 & $1 \times 10^{12}$ & 1.4$_{-0.25}^{+0.79}$ & [1.15, 2.19] & 0.16$_{-0.03}^{+0.06}$ & [0.13, 0.22] & 1.59$_{-0.17}^{+0.25}$ & [1.42, 1.84] & 1.42$_{-0.22}^{+0.19}$ & [1.2, 1.61]\\
Model 2 & $1.5 \times 10^{12}$ & 0.16$_{-0.03}^{+0.05}$ & [0.13, 0.21] & 1.34$_{-0.29}^{+3.12}$ & [1.05, 4.46] & 5.04$_{-0.65}^{+0.42}$& [4.39, 5.46]  & 5.31$_{-0.58}^{+0.40}$ & [4.73, 5.71]\\
\enddata
\tablecomments{The timing of the dynamical interactions between the SMC, LMC, and Milky Way from \citet{patel2020}. The penultimate SMC-LMC pericenter, Peri 1, the ultimate SMC-LMC pericenter, Peri 2, the SMC's infall into the Milky Way's virial radius, SMC Infall, and LMC's infall into the Milky Way's virial radius, LMC Infall, are listed with one-sigma errors for Model 1 in the top row and Model 2 in the bottom row. The Milky Way masses used for each model are listed in the first column of each table. Fiducial values were used for the LMC and SMC masses. These are $1.8 \times 10^{10}$ M$_{\odot}$ and $5 \times 10^9$ $M_{\odot}$, respectively. }

\end{deluxetable*}

\subsection{Dynamical explanations for global bursts in the Clouds} \label{sec:Modeling5.2}

\begin{figure*}
 	\centering
        \textbf{Cumulative SFHs for the entire SMC and LMC, and their regions}\par\medskip
\includegraphics[width=1.0\textwidth]{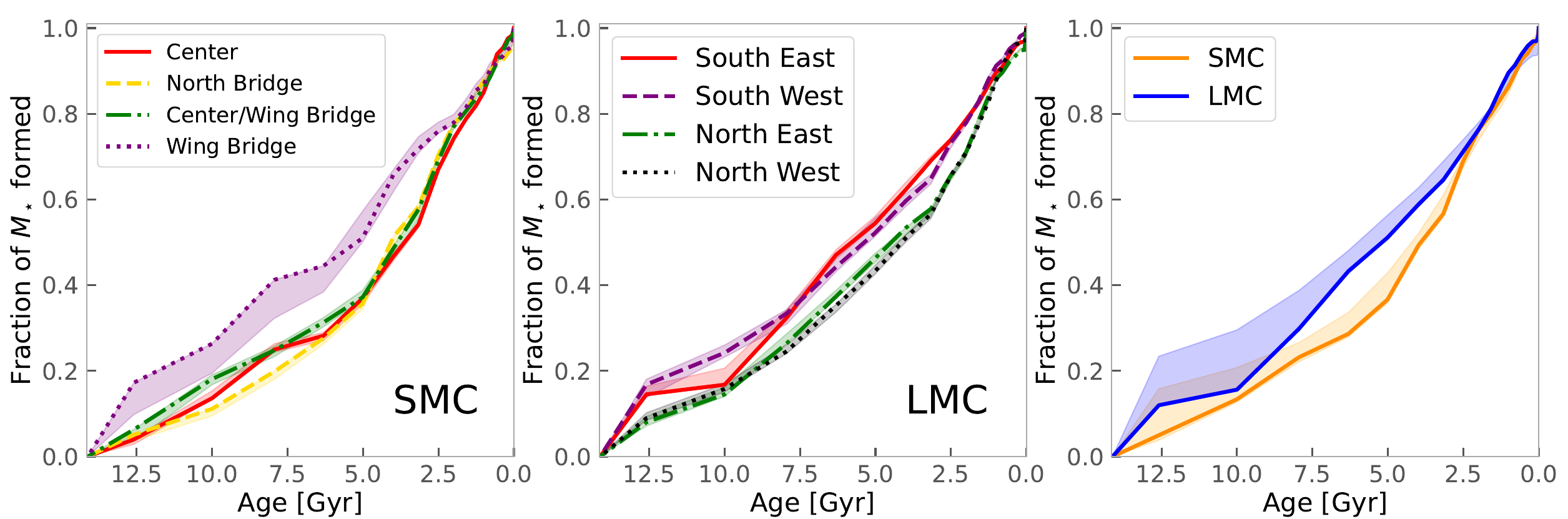}
	  \caption{\textbf{Left}: Cumulative SFHs for four regions in the SMC. The Wing/Bridge region of the SMC accrued its mass more quickly than the other three until $\sim 2.5 $ Gyr ago. \textbf{Central}: Cumulative SFHs for the four regions in the LMC. The South of the LMC accrued its mass more quickly than the North, despite the South containing the higher surface density regions. \textbf{Right}: The SFHs of all combined fields for each galaxy. The slope of the SMC’s cumulative SFH begins to steepen around 6 Gyr ago such that by 2.5 Gyr ago the SMC and LMC have accrued roughly the same amount of mass, $\sim 75\%$. These trends also indicate that, assuming the increase in SF around the time of the burst was due to a dynamical interaction, the SMC responds before the LMC to this disruption.}
  \label{fig:cumulativeSFHs}
\end{figure*}

There are several possible dynamical interactions that could account for SMC Burst 1 and LMC Burst 1, such as infall of the galaxies into the virial radius of the Milky Way or a pericentric passage between the galaxies at some point in their history as a binary pair. The timing of these events depends on the model used to determine the interaction history. Work from \citet{Cohen2024b} posits that infall into the virial radius of the Milky Way may not be a strong regulator of star formation in the Clouds, based on the constancy of the outside-in radial gradient over many Gyr observed in the SMC. For example, the study found discrepancies in star formation toward the Wing/Bridge compared to the outer non-Wing fields, with the former, showing a marked increase approximately 3–6 Gyr ago. This spatial variation in star formation suggests that environmental effects related to the Milky Way's halo, such as ram pressure stripping, might influence the distribution of young stars. However, ram pressure stripping seems an unlikely cause for the lack of young stars in the outer SMC based on modeling by \citet{besla2012}, which indicates that the Clouds likely did not enter the Milky Way’s virial radius until 1–2 Gyr ago. This timing is too recent to explain the earlier star formation patterns.

We explore the case of a dynamical trigger for star formation by comparing our results with modeling from \citet{patel2020} (hereafter P20). P20 uses proper motions from \citet{zivick2018} for the SMC and \citet{kallivayalil2013} for the LMC to model several combinations of the LMC-SMC-MW system orbits over the last 6 Gyr using backward integration in rigid galaxy potentials. To compare to our SFHs, we used two of P20’s models, which employ different halo masses for the MW (MW$_1$ = $1 \times 10^{12}$ M$_{\odot}$, MW$_2$ = $1.5 \times 10^{12}$ M$_{\odot}$ for models 1 and 2, respectively) and a fiducial value for the LMC and SMC ($1.8 \times 10^{10}$ M$_{\odot}$ and $5 \times 10^9$ $M_{\odot}$, respectively). The difference in MW mass can change the predicted timing of important dynamical events by Gyrs. For example, the larger MW mass corresponding to model 2 predicts the time of the infall of the SMC in the virial radius of the MW occurred $\sim 5$ Gyrs ago, while the lower MW mass corresponding to model 1 predicts the infall of the LMC in the MW occurred $\sim 1.6$ Gyrs ago (see Table \ref{table:p20}). This is similarly true for other dynamical events such as the two most recent pericenters of the Clouds with each other, and the SMC's and LMC’s infall into the virial radius of the MW.

\subsubsection{Comparing burst timing across regions to model predictions} \label{sec:BurstModelComparison5.3}

To compare regional burst timing to the modeled orbital histories, we examined the alignment of burst onset in each sub-region with key dynamical events (infall and pericentric passages) from Models 1 and 2 of \citet{patel2020}. 

\begin{figure*}
 	\centering
        \textbf{Comparison of a sub-region's SFH against dynamical models of the LMC/SMC/MW system}\par\medskip
\includegraphics[width=0.75\textwidth]{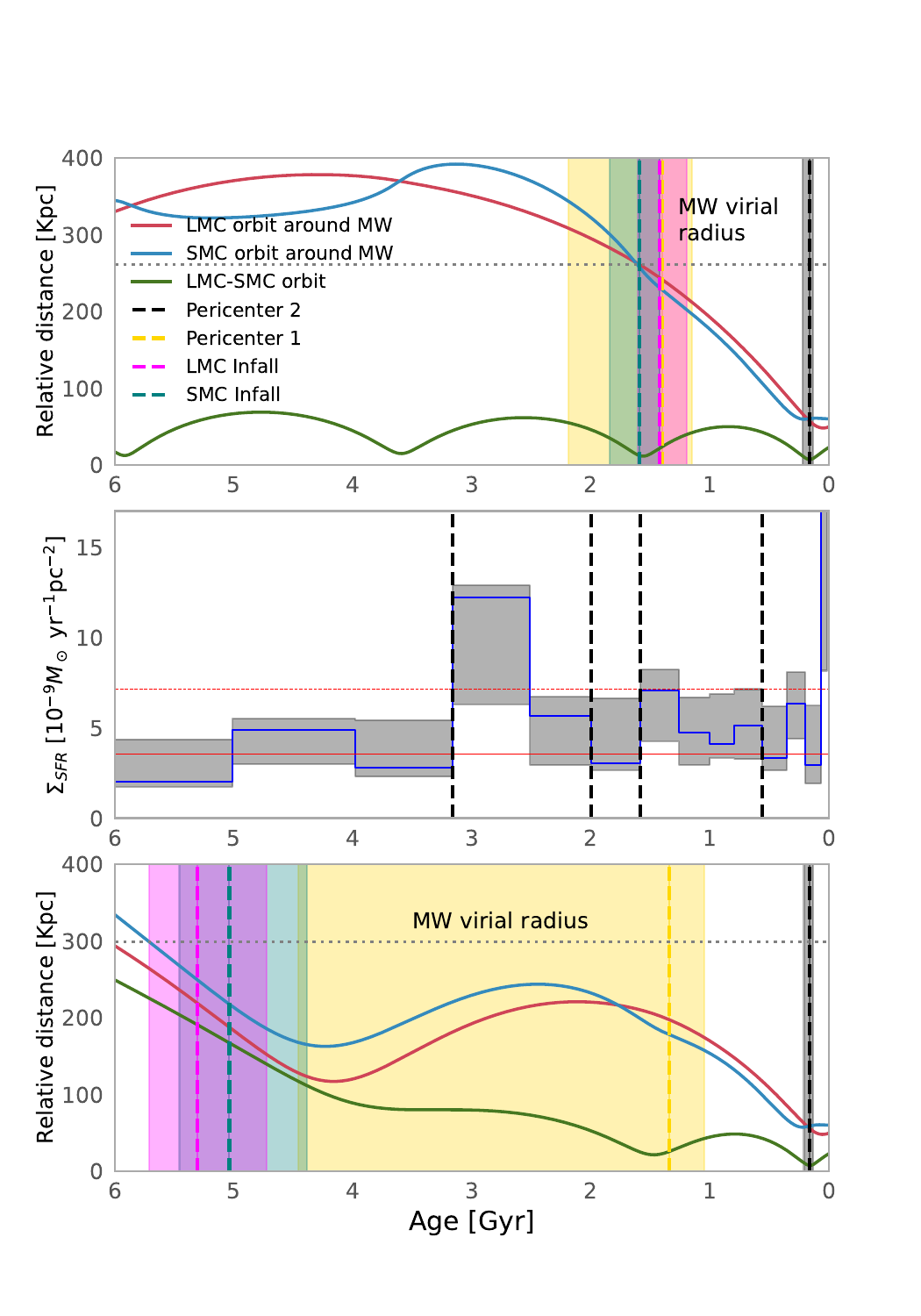}
	  \caption{We present the NBWB sub-region’s SFH compared to two dynamical models from \citet{patel2020} (P20) of the MW-LMC-SMC system over the last 6 Gyr. The top plot is modeled with a less massive MW, $1\times10^{12}$ M$_{\odot}$, and the bottom plot is the more massive MW $1.5\times 10^{12}$ M$_{\odot}$. The LMC and SMC are modeled in both cases with masses of $1.8\times10^{10}$ M$_{\odot}$  and $5\times10^9$ M$_{\odot}$, respectively. Overplotted on both models are vertical lines corresponding to dynamical events including the SMC’s infall into the MW virial radius, the LMC’s infall into the MW virial radius, the most recent SMC-LMC pericentric passage, and the penultimate SMC-LMC pericentric passage. In the center of the figure is the NBWB sub-region $\Sigma_{SFH}$ is plotted with its $<SFH>$, $2*<SFH>$ and the boundaries of its burst.}
  \label{fig:p20compare}
\end{figure*}

Figure~\ref{fig:p20compare} illustrates this comparison for one SMC sub-region (NBWB), showing the timing of the LMC and SMC’s infall into the Milky Way’s virial radius, as well as the most recent and penultimate pericentric passages. Vertical lines mark these events alongside the SFH and burst durations, allowing us to visually assess possible dynamical triggers. We focus on the two most recent pericentric passages, since Model 2 predicts only two within the last 6 Gyr compared to four in Model 1. Table~\ref{table:p20} summarizes the timing of these interactions with uncertainties, which are dependent on the proper motion, line of sight velocities and distance of the LMC and SMC.

The main takeaway from the SMC analysis is that Model 1 aligns the timing of infall to the MW and the penultimate pericenter closely with the onset of SMC Burst 1 in the global SFH and several sub-regions, suggesting these events may have triggered enhanced star formation. However, Model 1 does not explain bursts older than 1.5 Gyr, nor does it account for variation in burst timing across older sub-regions. In contrast, Model 2 places the SMC infall to the MW prior to burst onset in most sub-regions, with the notable exception of the NB region, which shows elevated star formation roughly 2 Gyr before the adjacent NBCB region. This spatially close yet temporally offset behavior is striking and may reflect uncertainties in the model or limitations from the age resolution of the SFH.

For the LMC, Model 1 generally aligns the timing of infall to the MW and the penultimate pericenter with the peak star formation in most sub-regions, although the NW2 sub-region shows elevated star formation beginning about 5 Gyr ago, significantly earlier than others. Model 2 predicts an earlier infall time that precedes this early NW2 burst onset, but other regions show bursts beginning roughly 1.5 Gyr later. Thus, Model 2 may better capture some early activity but still does not fully explain the spatial and temporal burst distribution.

Applying this approach to all sub-regions in both galaxies reveals a complex picture where both models offer possible dynamical triggers for bursts, but no single model fully explains the timing and variation observed. For example, in the NBWB sub-region, both models suggest that the first pericentric passage or galaxy infall events could have triggered bursts within their respective timing uncertainties. To further evaluate the causes of the global bursts identified in the SMC and LMC, we consider the duration and intensity of star formation in all sub-regions relative to these dynamical interactions, as discussed in Section~\ref{sec:Discussion5}. A supplementary plot presenting results for all sub-regions is included in the Appendix (Figure~\ref{fig:AMRs}).

% \clearpage

\section{Discussion} \label{sec:Discussion5}

We have presented measurements of bursts in the SMC and LMC by applying our broad and fine burst metrics (Section \ref{sec:Bursts4}). Applying these metrics, as shown in Figure \ref{fig:burstplot} and Figure \ref{fig:cumulativeSFHs}, we report two main epochs of bursts in the SMC, SMC Bursts 1 and 2, and one main epoch of bursts in the LMC, LMC Burst 1, as well as a localized but strong burst in the SMC. The first onset of bursts in most sub-regions in the SMC began around $5$ Gyr ago. Although a single sub-region in the LMC, NW2, began bursting around this time as well, most sub-regions in the LMC began bursting only $3$ Gyr ago. 

Our findings are broadly consistent with previous measurements of enhanced star formation in the Magellanic Clouds based on VMC, SMASH, and archival HST surveys. Across these studies, bursts are generally found to have occurred between $\sim$1.5–5 Gyr ago in both galaxies, with most surveys identifying at least two major episodes of elevated star formation in each system.

However, notable differences emerge in the precise timing and duration of these events. For example, \citet{rubele_2015}, using VMC data, identifies enhancements in the Northern LMC around $2$–$3$ Gyr ago, and in the SMC up to $5$ Gyr ago. SMASH-based analyses by \citet{rubele2018, mazzi2021, ruizlara2020} report correlated activity in the SMC and Northern Arm of the LMC around $\sim3$ Gyr ago. Meanwhile, \citet{weisz2013} finds peaks in the SMC at $\sim4.5$ and $9$ Gyr, and the LMC’s largest enhancement at $\sim3.5$ Gyr. Our results agree with the general timing of burst activity, but, as discussed in Section~\ref{sec:Bursts4}, differ in the specific onset times, particularly in the LMC.

Overall, we find one instance of bursts that occur synchronously across the Magellanic Clouds prior to $0.5$ Gyr ago, where we focus our burst analysis. This corresponds to a localized burst in the SMC and a global burst in the LMC around 3 Gyr ago. Aside from this, the timing and location of bursts differ significantly between the galaxies. For instance, we detect enhanced star formation in the northern LMC before 3.5 Gyr ago, while in the SMC, we observe sustained activity extending up to 5 Gyr ago.

These results differ from those reported by the SMASH survey, which finds an enhancement in the southern LMC around 5 Gyr ago, comparable in SFR to the peak activity in the SMC, but does not report the same northern LMC activity seen here. Additionally, we find lower SFRs in the SMC at ages older than ~3.5 Gyr compared to SMASH. In contrast, our burst timing shows stronger agreement with the VMC survey and with results from \citet{weisz2013}, though some differences remain in the exact duration and spatial extent of the identified enhancements. These discrepancies can arise from differences in assumed evolutionary models and the SFH fitting codes.

\subsection{Agreement with dynamical modeling} \label{sec:CompareModel5.3}

\begin{figure*}[ht]
\centering
\textbf{SMC and LMC sub-region burst durations vs.\ dynamical timings}\par\medskip
  \subfloat[][SMC burst intensity, Model 1]{\includegraphics[width=.45\textwidth]{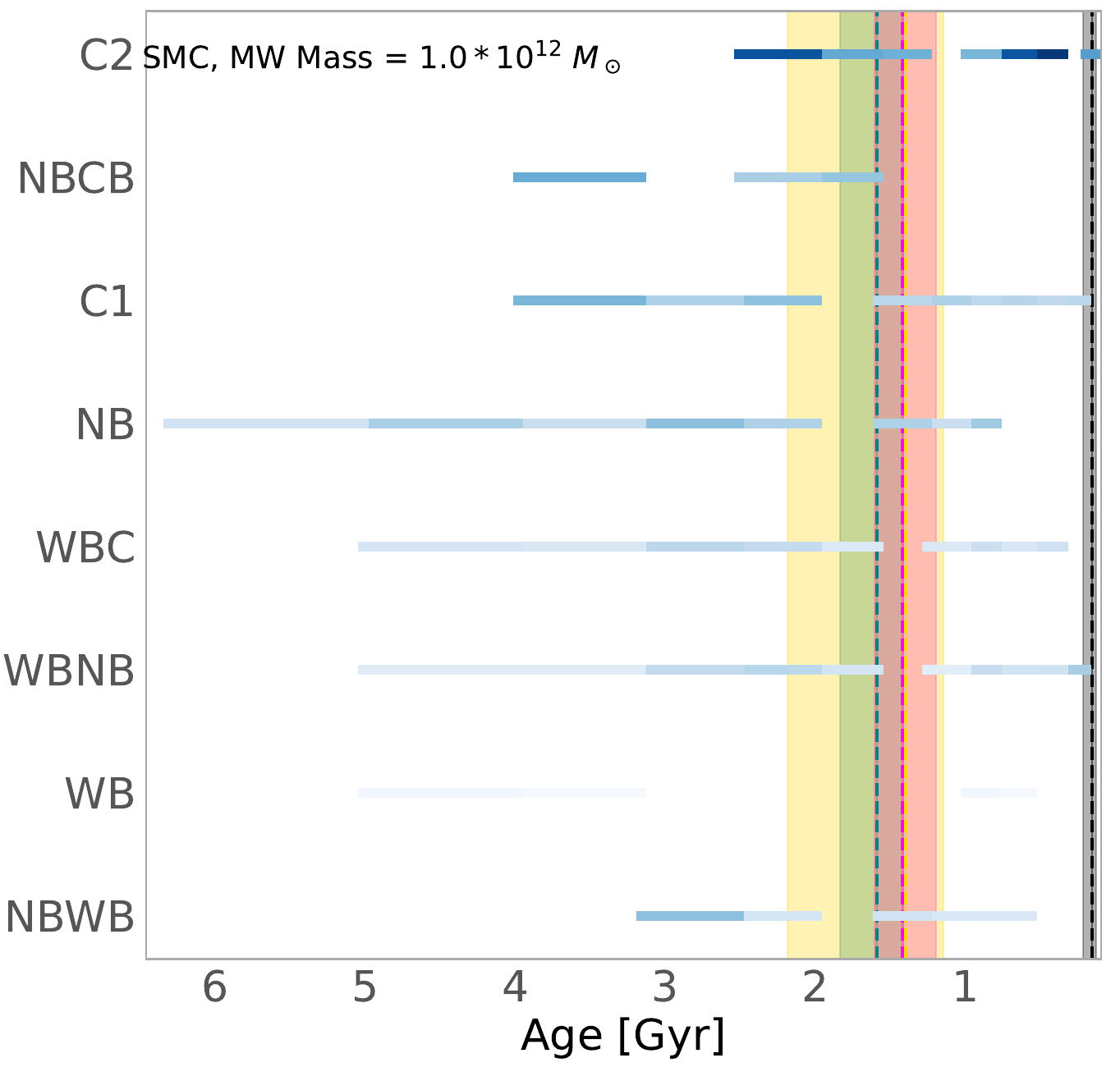}}\quad
  \subfloat[][SMC burst intensity, Model 2]{\includegraphics[width=.53\textwidth]{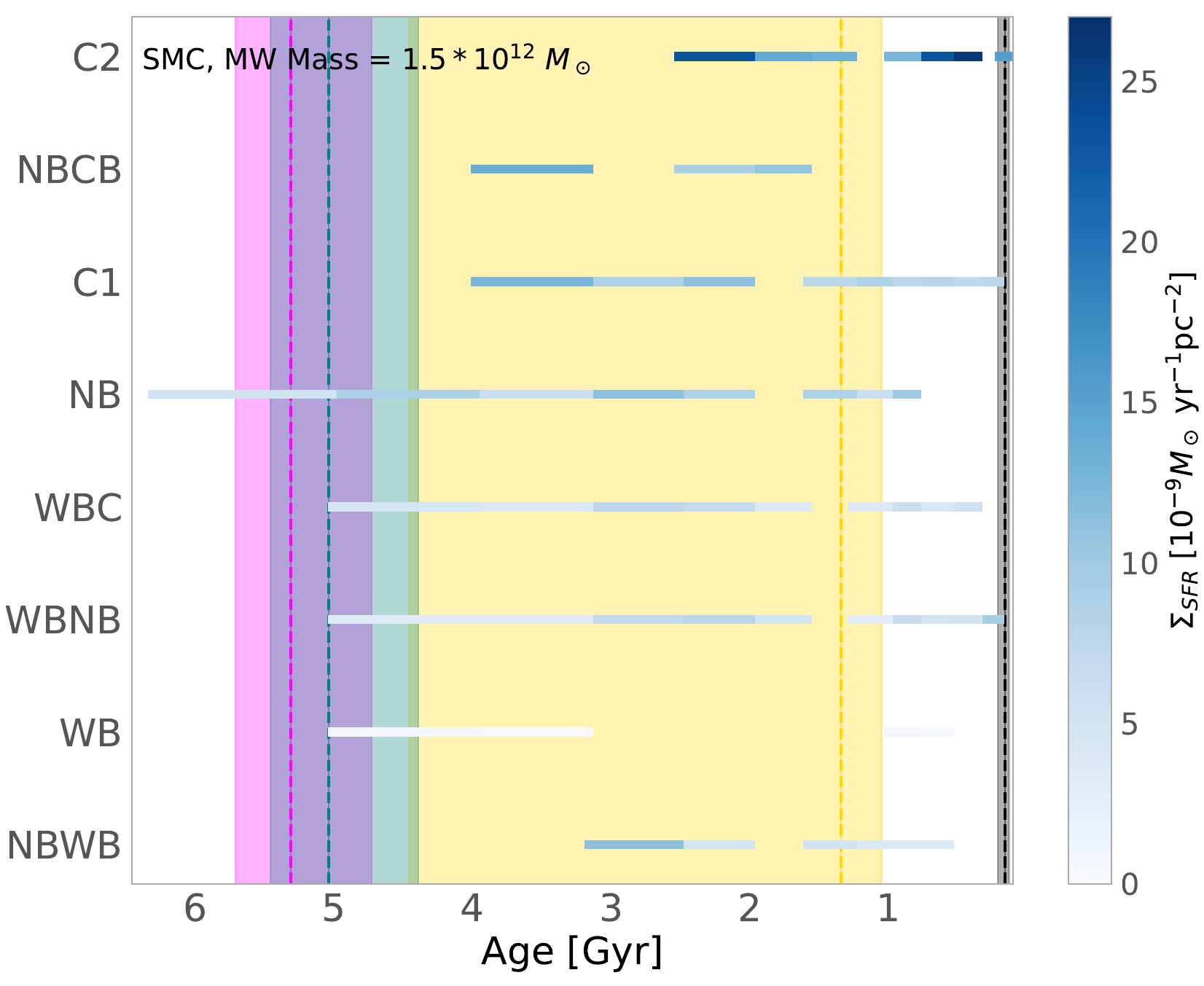}}\\
  \centering
  \subfloat[][LMC burst intensity, Model 1]{\includegraphics[width=.455\textwidth]{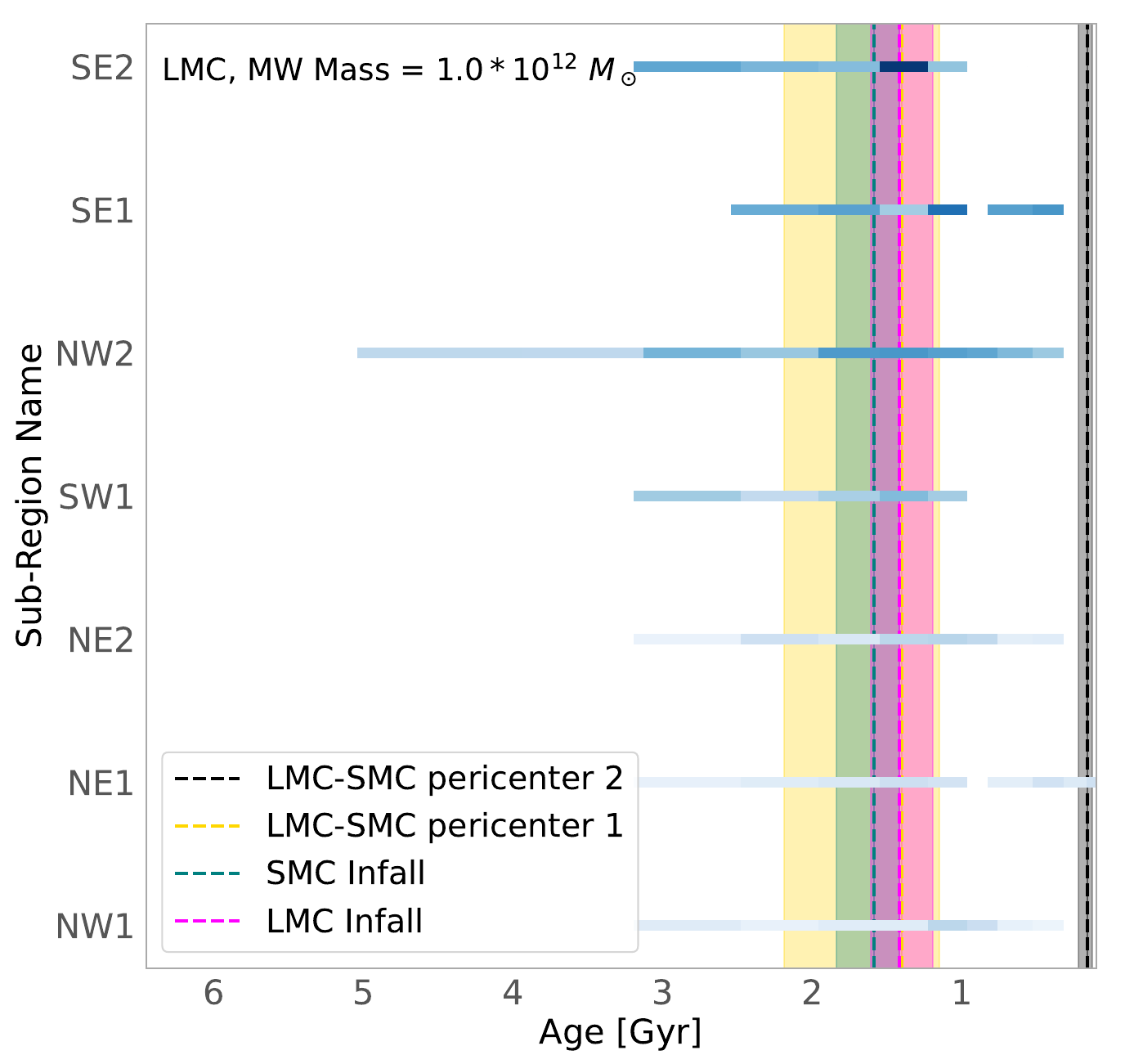}}\quad
  \subfloat[][LMC burst intensity, Model 2]{\includegraphics[width=.525\textwidth]{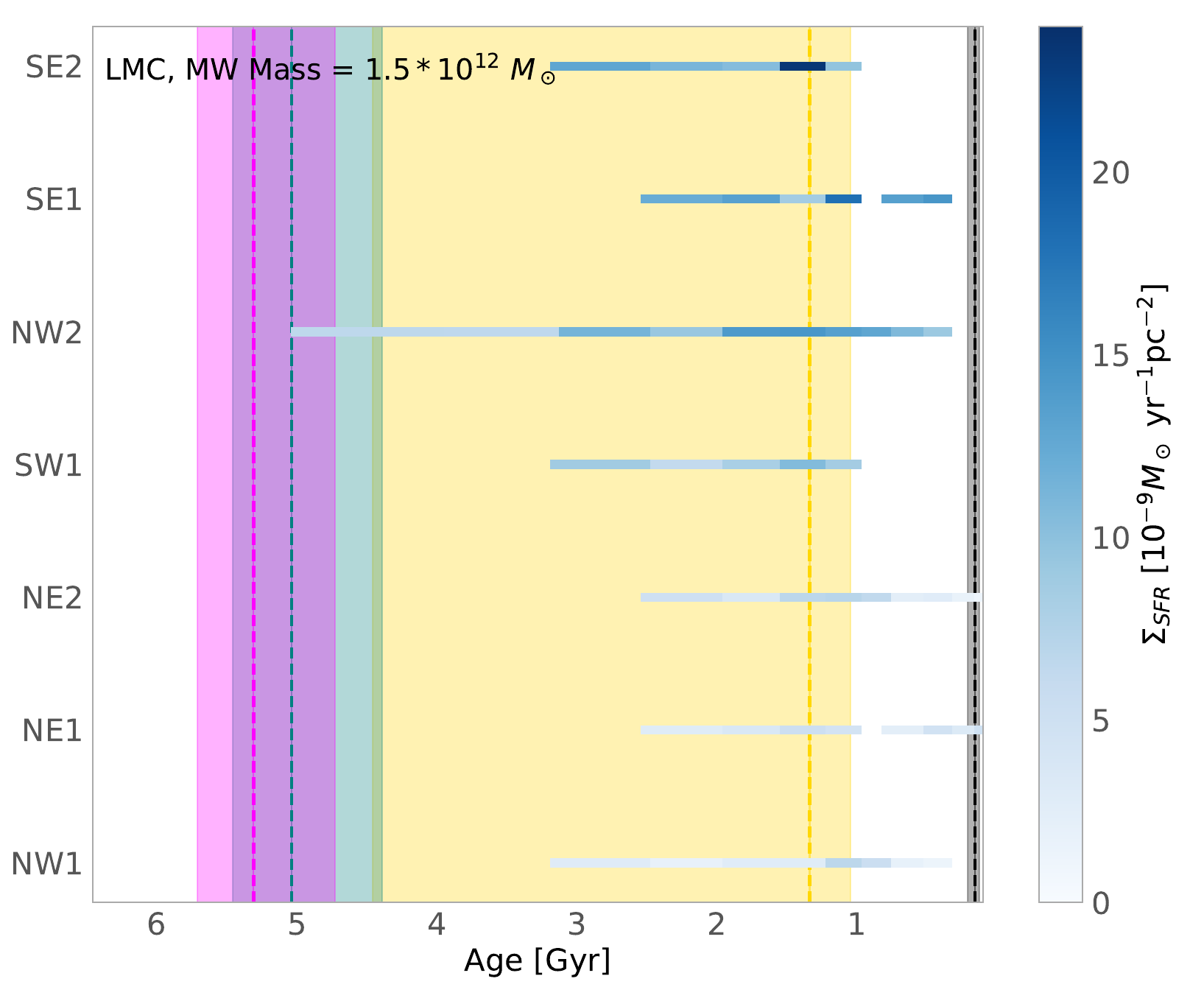}}
  \caption{The same plot as Figure \ref{fig:burstplot} colored by intensity of the burst and with dynamical modeling results overlaid. The intensity of the burst is given by the shading of the line, which corresponds to the $\Sigma_{SFR}$ noted in the colorbar. The timing of four dynamical events with errors are given vertically. The top row shows the SMC's sub-regions and dynamical timings calculated with Model 1 on the left and Model 2 on the right. The same is shown for the LMC in the bottom row. The most noticeable difference between the models is the timing of the infall of the SMC and LMC into the virial radius of the MW and the width of the errors on the first pericentric passage. While the comparison of the burst timings to dynamical models does not provide a conclusive dynamical trigger for all detected bursts, we do find that the penultimate pericentric passage in both models offers an explanation for SMC global burst 1.}
  \label{fig:burstp20}
\end{figure*}

To explore dynamical scenarios that may consistently explains the observed burst timings, we compared both the duration and intensity of our burst measurements to two orbital models of the Magellanic Clouds from P20. By incorporating both the duration and the intensity of bursts, we assessed whether the observed SFR enhancements align with expectations for interaction-driven events between dwarf galaxies. Figure~\ref{fig:burstp20} provides a broad overview of burst strength and timing across all sub-regions, alongside the predicted epochs of the four most significant potential dynamical triggers from both P20 models. The model-dependent timing of events can lead to different interpretations of which interaction triggers a given burst. For example, SMC Burst 2 could plausibly be linked to either infall or the penultimate pericentric passage in Model 1, where those events are closely spaced in time. In contrast, Model 2 separates these events by several billions of years, implying that different bursts may trace different dynamical encounters depending on the model.

Based on our findings, it is not possible to favor one dynamical model over another.  The discrepancy in which dynamical event triggers bursts as a function of the model indicates that neither model provides a fully consistent explanation for the timing of bursts across all sub-regions, particularly for those older than 1.5 Gyr. 

Despite the comparison not revealing a clear correlation with the timing of the interactions with the onset of the observed bursts in most cases, this study highlights how dynamical models could be adjusted for the Clouds going forward. For instance, a more precise Milky Way mass estimate and the inclusion of time-evolving gravitational potentials could help quantify the different types of bursts (e.g., bursts that occur from triggered interactions, secular evolution, etc.). This would be useful for comparison with SFHs like those in this study, and would help determine which bursts might be a product of interaction.

\subsection{The SMC responds more strongly to dynamical interactions than the LMC}

Our results suggest that the SMC experiences more intense bursts of star formation than the LMC during periods of interaction. In particular, the penultimate pericentric passage between the Clouds coincides with a resurgence of star formation in both galaxies, but the associated SFR increase is more pronounced in the SMC. This difference may reflect the lower total mass and shallower potential well of the SMC, which make it more susceptible to tidal disturbances. Additionally, if the SMC retained higher reservoirs of cold gas prior to the interaction, it may have been more primed for a burst. These factors likely contributed to the SMC's heightened sensitivity to the same dynamical trigger compared to the LMC.

That lower-mass galaxies may exhibit greater star formation enhancements in dwarf-dwarf interactions is supported by previous studies. Observationally, \citet{Subramanian_2023} found that some gas-rich dwarf-dwarf paired galaxies with stellar masses of $10^7$–$10^8~M_\odot$ can show SFR enhancements up to 3.5$\times$ higher than their isolated counterparts. Although not observing exclusively dwarf-dwarf pairs, \citet{Subramanian_2023} found that lower mass, interacting dwarf galaxies ($10^7$–$10^8~M_\odot$) can have greater star formation enhancements than more massive interacting dwarfs.  While the LMC and SMC are more massive than these examples, the mass difference between them may still drive differential responses, with the lower-mass SMC undergoing more substantial star formation enhancements. This mass dependence may help explain the larger burst signature that we observe in the SMC during shared dynamical triggers with the LMC.

\begin{figure*}
 	\centering
        \textbf{Cumulative SFHs and AMRs for regions in the SMC and LMC}\par\medskip
\includegraphics[width=1.0\textwidth]{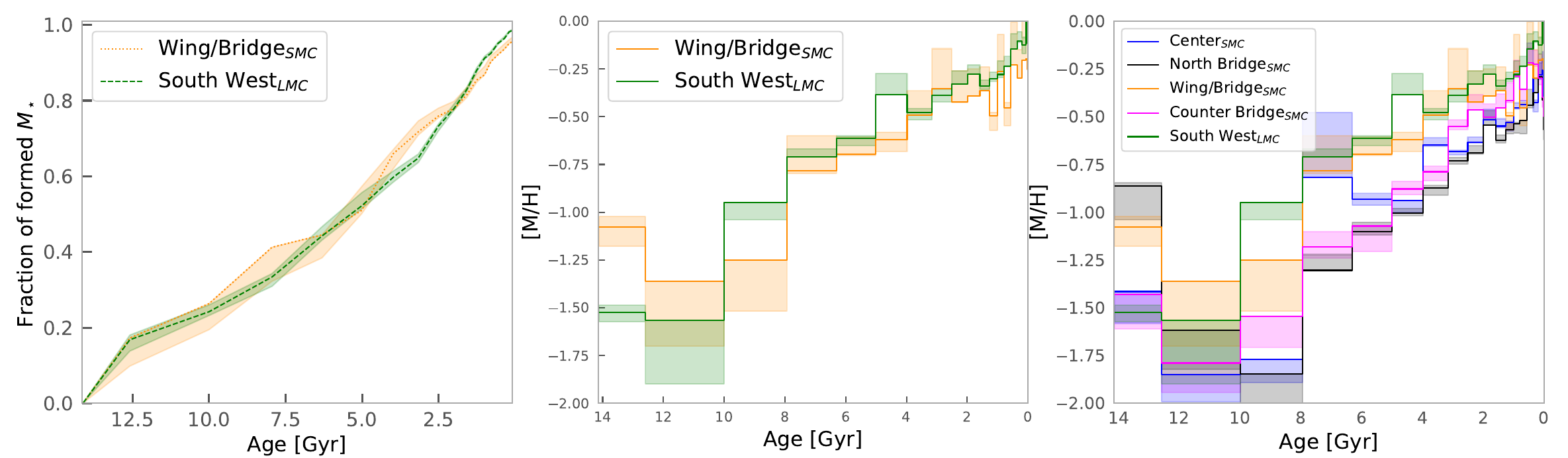}
	  \caption{\textbf{Left}: A comparison of the cumulative SFHs and AMRs of the Wing/Bridge (SMC) and South West (LMC). \textbf{Center}: The AMRs of the Wing/Bridge region and the LMC's South West region. \textbf{Right}: The AMRs of the LMC's South West region compared to all regions in the SMC. The AMRs show a similar result to the cumulative SFHs, especially from 7 Gyr to 1 Gyr ago, when the AMR of the Wing/Bridge looks is most similar to the South West of the LMC than the other regions of the SMC. }
  \label{fig:AMRs}
\end{figure*}

\subsection{Wing/Bridge disagreement}
\label{sec:WBdisagree}

When comparing the cumulative SFHs of regions across the Magellanic Clouds, we found that the Wing/Bridge region of the SMC assembles its stellar mass more similarly to the southern LMC than to any other part of the SMC (see Section \ref{sec:RegionalAgreement4.3.1}). To verify these results, we also examined the age-metallicity relations (AMRs) of these regions to determine if the metal composition of the regions showed similar trends. These were measured via CMD fitting, similar to our SFHs. Figure \ref{fig:AMRs} shows that, like the cumulative SFH, the AMR of the Wing/Bridge aligns more closely with that of the LMC Southwest than with other regions of the SMC, particularly between 7 Gyr and 1 Gyr ago, when the other AMRs of the SMC begin to diverge significantly.

Studies have proposed a tidal origin for stars in the Wing/Bridge region. Using data from the Apache Point Observatory Galactic Evolution Experiment 
(APOGEE), \citet{almeida2023} find that stars towards the eastern periphery of the SMC resemble those in the central SMC in both composition and kinematics, concluding they were likely stripped from the center by interactions with the LMC. \citet{Oliveira2023}, using star cluster data from the VISCACHA survey, similarly identifies intermediate-age, metal-poor Bridge clusters with distinct dips in their AMRs around 1.5–1.0 Gyr ago and again at 200–150 Myr, epochs that coincide with known interaction events and global bursts in the SMC \citep{Besla2010, Fox2016}.

Our results echo those of \citet{Oliveira2023}, showing metallicity dips in the Wing/Bridge AMR at approximately 1.75 Gyr, 1.5 Gyr, and 500 Myr ago (Figure \ref{fig:AMRs}, middle panel). These dips align with global burst activity in the SMC and with the timing of the most recent close dynamical encounters. For example, \citet{choi2022} reports a recent collision between the SMC and the LMC within the last 200 Myr, with an impact parameter $\le$ 10 kpc. This interaction could account for both the burst in star formation and the distinct chemical signature of the Wing/Bridge region, linking its formation to a specific dynamical event.

Further evidence suggests the Wing/Bridge's higher metallicity may also be partly due to material stripped from the LMC. While the Wing/Bridge is more metal-rich than the SMC overall \citep{rubele2018, Oliveira2023}, it remains less enriched than the most metal-rich LMC clusters, which are about 0.3 dex higher in metallicity \citep{parisis2009}. If gas was stripped from the southwest LMC during the last close encounter, possibly as the SMC passed through the LMC bar, this could explain the unusual enrichment of the Wing/Bridge. Additional chemical abundance comparisons between stellar populations in the Wing/Bridge and the LMC would help test this scenario. This could perhaps be addressed with the HI data from GASKAP and surveys like The Magellanic Edges Survey \citep{Cullinane2020}.

\section{Summary}\label{sec:Conclusions}

Using 74 Hubble Space Telescope (HST) pointings from homogeneous two-filter photometry (F475W and F814W) of the pure parallel survey, Scylla, we measured the SFHs of 36 fields in the SMC and 38 fields in the LMC to identify and characterize bursts of star formation across both galaxies. Our main results are summarized below:

\begin{enumerate}
    \item We defined a fine-burst metric to differentiate features within broader bursts of star formation, building on the framework of \citet{Kennicutt2005, McQuinn2009, McQuinn2010}.
    
    \item We grouped fields within each galaxy into spatial sub-regions (see Fig.~\ref{fig:maps}), combined their SFH solutions (see Fig.~\ref{fig:stackedSFH}), and used these groupings (seven in the LMC and eight in the SMC) to identify distinct bursts and measure the stellar mass formed in each (see Section ~\ref{sec:MeasureBurst4.2.1} and Fig.~\ref{fig:burstplot}).
    
    \item We identify two global bursts (bursts that occur across $\ge$ half of sub-regions) of star formation in the SMC at 1.5 Gyr and 5 Gyr ago and one global burst in the LMC, 3 Gyr ago. We identify one local burst (bursts that occur across a minority of sub-regions) in the SMC 3 Gyr across the Central and North Bridge regions. The local burst in the SMC and the global burst in the LMC are the only burst events that coincide between the galaxies.
    
        \item By comparing both $\Sigma_{SFH}$ and cumulative SFHs, we find that the timing of bursts is delayed between the SMC and LMC. If a shared dynamical event triggered star formation in both galaxies, the SMC responded earlier than the LMC ($\sim$ 5 Gyr ago in the SMC compared to $\sim$ 3 Gyr in the LMC).

    \item We compared our SFH results to two dynamical models from \citet{patel2020}, which simulate the LMC–SMC–MW orbital interactions over the past $\sim6$ Gyr using different masses for the Milky Way. While both models predict interactions capable of triggering bursts, our analysis did not definitively favor one model over the other (see Figs.~\ref{fig:cumulativeSFHs} and \ref{fig:p20compare}).
    
    \item Finally, we analyzed the cumulative SFHs and AMRs across the SMC and LMC and found that the Wing/Bridge region of the SMC is more similar—in both chemical enrichment and SFH over the past $\sim$7 Gyr — to the southwestern LMC than to other parts of the SMC (see Fig.~\ref{fig:AMRs}). 
\end{enumerate}

These results suggest that characterizing bursts in the Clouds as synchronous may need to be reconsidered. In particular, the consistent delay in LMC burst onset relative to the SMC supports a scenario where both galaxies are affected by shared interactions, but with differing responses.

Based on the observed similarity between the Wing/Bridge region of the SMC and the southwestern LMC, we posit that the Wing/Bridge includes stars stripped from the outer LMC. This interpretation contrasts with the APOGEE results and conclusions by \citet{almeida2023}, which suggest that the Wing/Bridge stars originated from the inner SMC, based on radial velocity measurements and chemical similarity to the central SMC. Our conclusions drawn from the analysis of the AMR and SFH of the Wing/Bridge may require further study. 

\section{Acknowledgements}

C.B. thanks Ivanna Escala for the helpful discussions of AMRs in the LMC and SMC. The authors also acknowledge the anonymous referee for their constructive comments. Some of the data in this article were obtained via the Mikulski Archive for Space Telescopes (MAST), which is supported by the NASA Office of Space Science via grant NAG5-7584 and the Space Telescope Science Institute, supported under NASA contract NAS 5-26555. This research was supported, in part, by the following grants:  GO-HST-15891, GO-HST-16235, and GO-HST-16786. The specific observations analyzed in this paper can be found via\dataset[doi: 10.17909/8ads-wn75]{https://doi.org/10.17909/8ads-wn75}. P.Y.M.J. acknowledges support by “Young Scientists and Postdoctoral Scholars” by the Bulgarian Ministry of Education and Science. C.B. also acknowledges Rutgers, the State University of New Jersey.

\facilities{HST (WFC3)}
\software{astropy \citep{astropy:2013, astropy:2018, astropy:2022}, matplotlib \citep{Hunter:2007}, numpy \citep{harris2020array}, emcee \citep{ForemanMackey2013}, Dolphot \citep{dolphin2000}, MATCH \citep{dolphin2002}}

% Although our results do not provide definitive evidence against the interpretation of APOGEE, they are consistent with the chemical abundance trends observed in the VISCACHA survey \citep{Oliveira2023}, which found dips in the AMR of Bridge clusters at times that corresponded to known dynamical interactions. Further detailed chemical abundance studies of clusters in the Wing/Bridge and southern LMC will be essential to distinguish between these scenarios.

\renewcommand\bibname{{References}}
\bibliography{ms}

\section{Appendix}

\begin{figure*}[h!]
\centering
\textbf{SFHs for sub-regions in the SMC and LMC with modeled dynamical timings}\par\medskip
  \subfloat
  {\includegraphics[width=.49\textwidth]{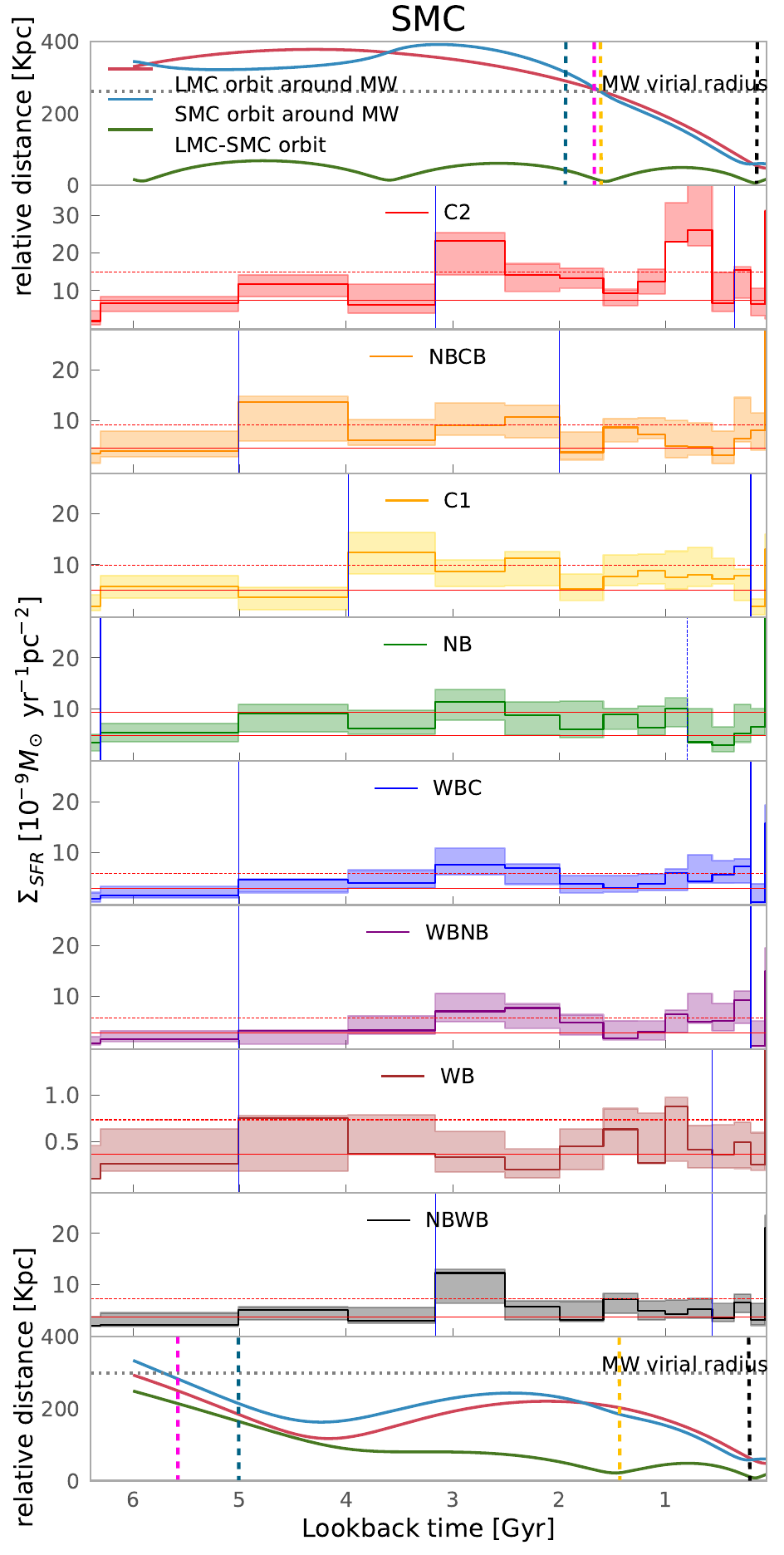}}\quad
  \subfloat
  {\includegraphics[width=.49\textwidth]{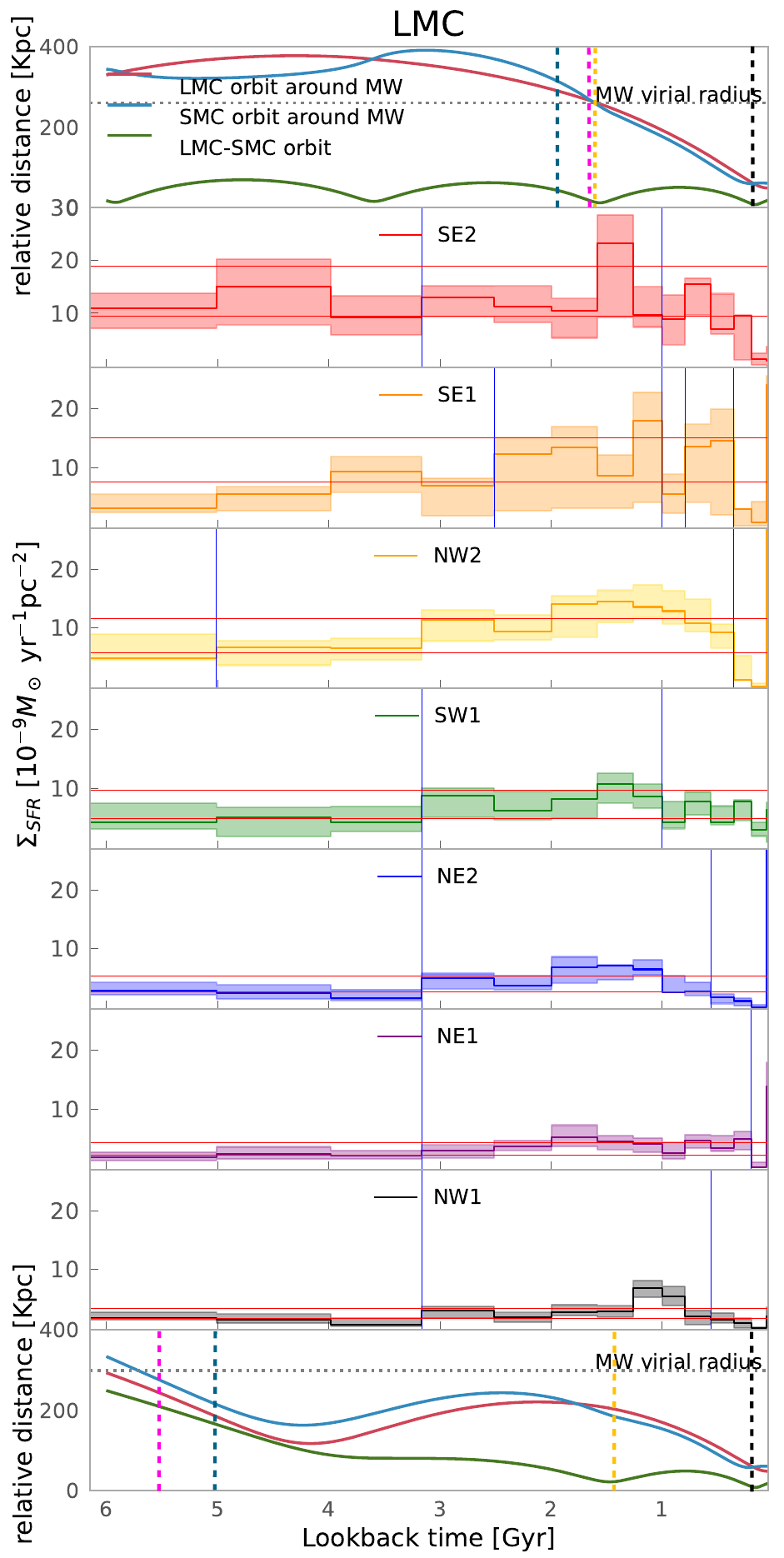}}\\
  \caption{A combination of Figure \ref{fig:stackedSFH} with the dynamical models from  Figure \ref{fig:p20compare} to compare the SFHs of all sub-regions to the dynamical timings we measure in P20's dynamical models, where the top row have timings from model 1 and the bottom row have timings from model 2. Note the dynamical timing of the penultimate pericentric passage (yellow), most recent pericentric passage (black), LMC's infall into the MW (pink), and SMC's infall into the MW (teal) have been indicated with dotted vertical lines in the top and bottom plots, identical to those given in Figure~\ref{fig:p20compare}. These designations have been excluded from the legend to improve readability of the plot. The plot also only gives the outter most bounds of the bursts. }
  \label{fig:dynam}
\end{figure*}

Figure \ref{fig:dynam} presents the SFHs of each sub-region in the LMC and SMC plotted against the key dynamical event timings from the two models proposed in \citet{patel2020}. Vertical dashed lines indicate the approximate timing of major interactions, including pericentric passages and infall into the Milky Way’s halo. Overlaying the SFHs with these dynamical markers allows for a visual assessment of whether bursts of star formation may have been temporally correlated with interaction events. This appendix figure provides additional support for the comparisons discussed in Section~\ref{sec:Discussion5}, offering readers a more granular look at the temporal alignment between star formation activity and dynamical triggers.

\begin{figure*}
 	\centering
        \textbf{Cumulative SFHs for all sub-regions in the SMC}\par\medskip
\includegraphics[width=1.0\textwidth]{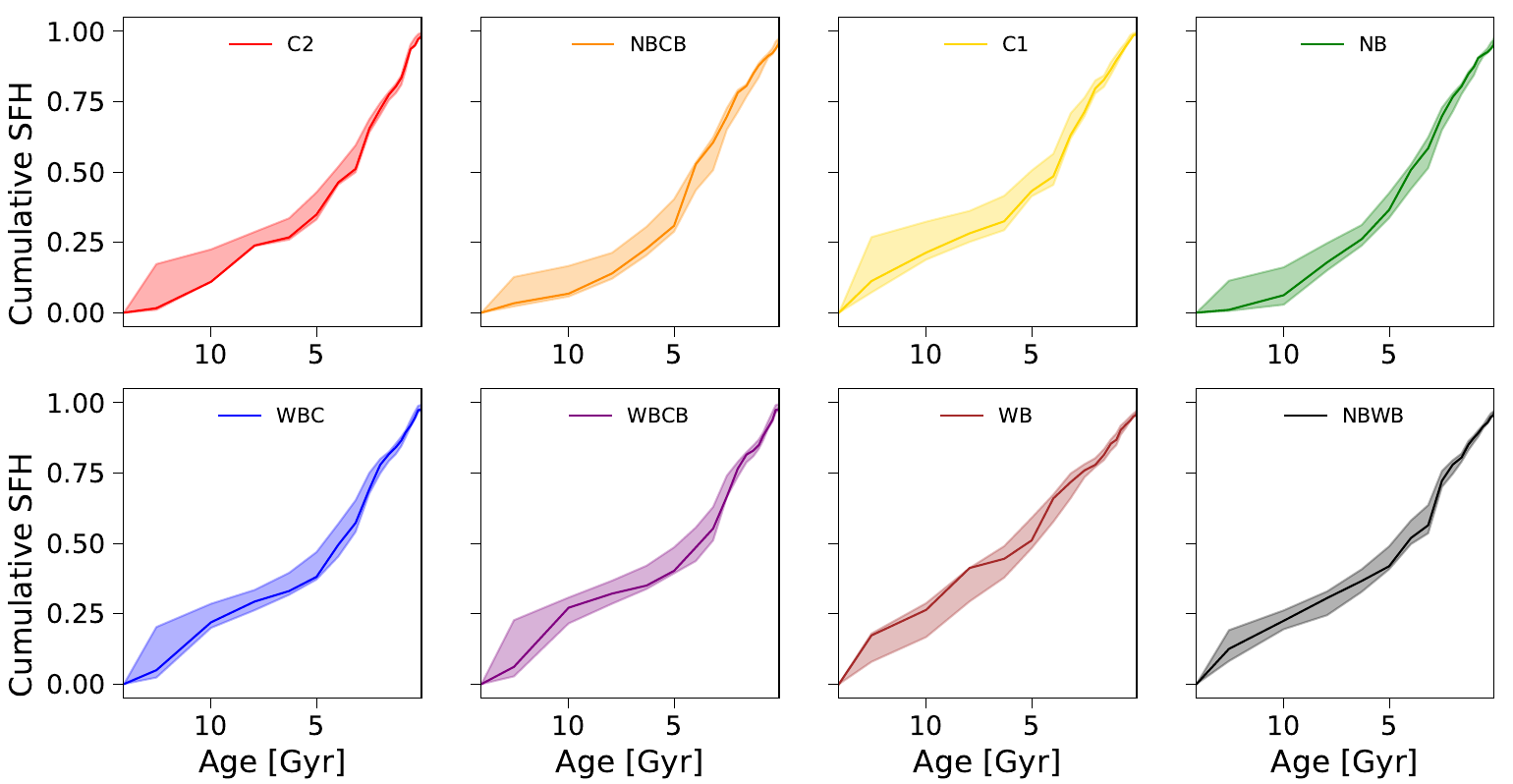}
	  \caption{Cumulative SFHs for all SMC sub-regions across both galaxies. Galaxies are colored and ordered from highest to lowest stellar surface density. }
  \label{fig:cumulativesSMC}
\end{figure*}

\begin{figure*}
 	\centering
        \textbf{Cumulative SFHs for all sub-regions in the SMC and LMC}\par\medskip
\includegraphics[width=1.0\textwidth]{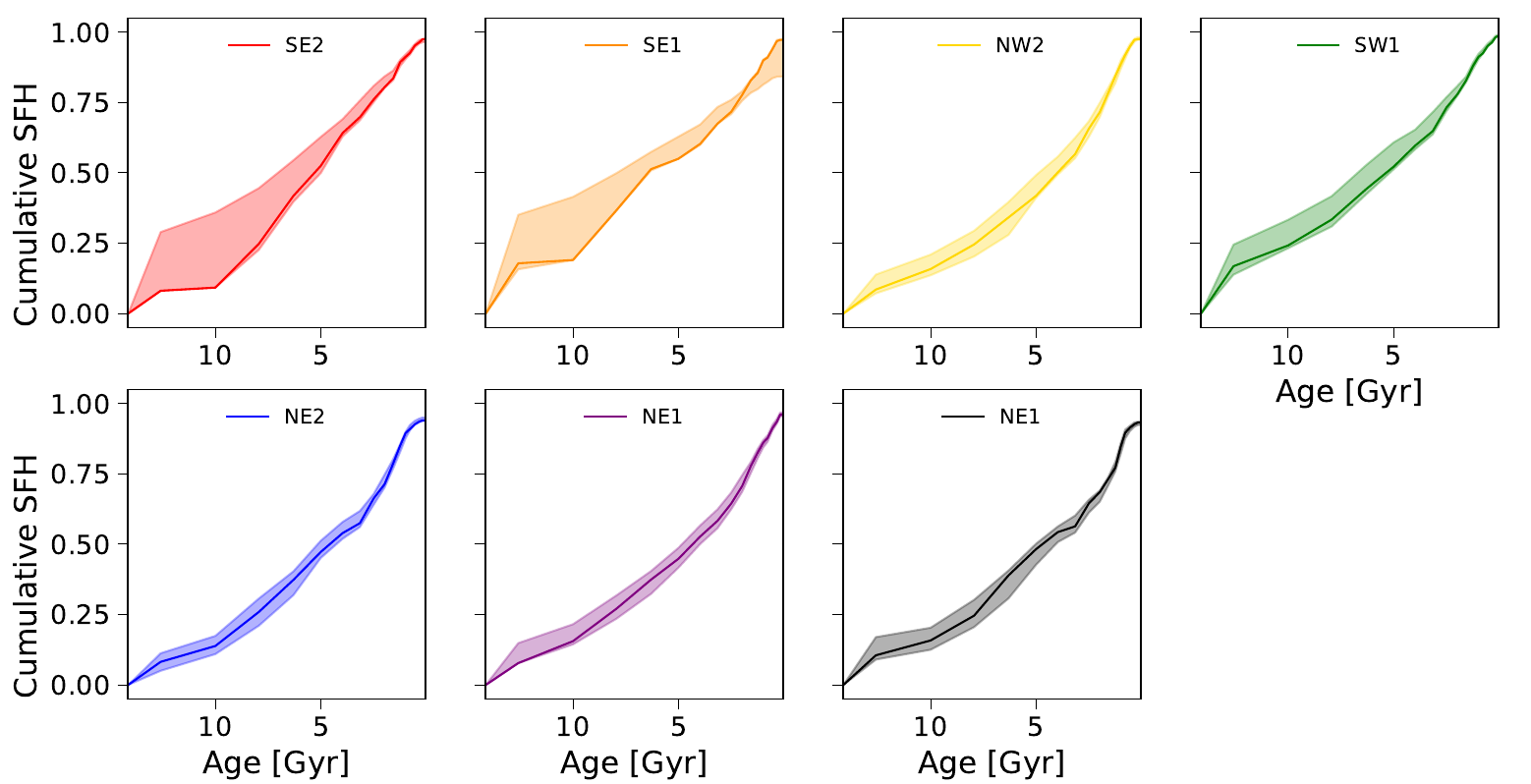}
	  \caption{Cumulative SFHs for all LMC sub-regions across both galaxies. Galaxies are colored and ordered from highest to lowest stellar surface density.}
  \label{fig:cumulativesLMC}
\end{figure*}

Figure \ref{fig:cumulativesSMC} and Figure \ref{fig:cumulativesLMC} display the cumulative SFHs for all individual sub-regions analyzed in the SMC and LMC. By showing the fractional stellar mass growth over cosmic time for each region, this plot provides a complementary view to the burst-focused analyses in the main text. Readers can use these cumulative SFHs to assess both the overall pace of stellar mass buildup and to compare evolutionary trends between regions. In particular, this figure illustrates the diversity of SFHs across the Clouds and contextualizes how certain regions deviate or conform to broader galactic trends, as discussed in Section~\ref{sec:Bursts4}.

\end{document}